\documentclass[final,3p,times]{elsarticle}

\usepackage{graphicx,color}
\usepackage{xcolor}
\usepackage{tcolorbox}
\usepackage{enumerate}
\usepackage[colorlinks]{hyperref}
\usepackage{natbib}
\usepackage{amsmath,amssymb,amsthm,bm}
\usepackage{multicol}
\usepackage[dvipsnames]{xcolor}
\usepackage{ulem}
\usepackage{tikz}
\usetikzlibrary{automata,positioning}
\usepackage{afterpage}
\usepackage{calc}
\usepackage{ifthen}
\usepackage{algorithm}
\usepackage{algpseudocode}
\usepackage{caption}
\usepackage{subcaption}
\usepackage{tabularx,ragged2e}
\usepackage{multirow}
\usepackage{booktabs}
\usepackage{comment}
\usepackage{colortbl}
\usepackage[title]{appendix}
\usepackage{cancel}
\usepackage{soul}
\usepackage{lineno}

\definecolor{colA}{HTML}{538FFF}
\definecolor{colB}{HTML}{B431E6}
\definecolor{colAf}{HTML}{FF5B58}
\definecolor{colBf}{HTML}{F7ED65}
\definecolor{colAn}{HTML}{28D2AB}
\definecolor{colBn}{HTML}{FCA207}
\definecolor{colAfs}{HTML}{F6CCF9}
\definecolor{colBfs}{HTML}{268189}
\definecolor{colK}{HTML}{2D1A77}

\journal{arXiv}

\begin{document}
\newcommand{\beq}{\begin{equation}}
\newcommand{\eeq}{\end{equation}}
\newcommand{\D}  {\displaystyle}
\newcommand{\sca}[1]{\mbox{\rm{#1}}{}}
\newcommand{\mat}[1]{\mbox{\bfseries #1}{}}
\newcommand{\scas}[1]{{\rm{#1}}{}}
\newcommand{\vecs}[1]{{\rm{\bfseries{#1}}}
\newcommand{\ten}[1]{\ten{#1}}   
\newcommand{\vecb}[1]{\ten{#1}}  
{}}


\newcommand{\CLMa}{$8.090\times10^{-4}$}
\newcommand{\CLMb}{$7.474\times10^{0}$}
\newcommand{\CLMaf}{$1.911\times10^{-3}$}
\newcommand{\CLMbf}{$2.206\times10^{1}$}
\newcommand{\CLMan}{$2.270\times10^{-4}$}
\newcommand{\CLMbn}{$3.480\times10^{1}$}
\newcommand{\CLMafs}{$3.139\times10^{0}$}
\newcommand{\CLMbfs}{$3.256\times10^{0}$}
\newcommand{\CLMK}{$1.000\times10^{-1}$}
\newcommand{\CRMa}{$8.090\times10^{-4}$}
\newcommand{\CRMb}{$7.474\times10^{0}$}
\newcommand{\CRMaf}{$1.911\times10^{-3}$}
\newcommand{\CRMbf}{$2.206\times10^{1}$}
\newcommand{\CRMan}{$2.270\times10^{-4}$}
\newcommand{\CRMbn}{$3.480\times10^{1}$}
\newcommand{\CRMafs}{$5.447\times10^{-4}$}
\newcommand{\CRMbfs}{$4.921\times10^{0}$}
\newcommand{\CRMK}{$10.000\times10^{-2}$}
\newcommand{\RHMa}{$8.090\times10^{-4}$}
\newcommand{\RHMb}{$7.474\times10^{0}$}
\newcommand{\RHMaf}{$1.911\times10^{-3}$}
\newcommand{\RHMbf}{$2.206\times10^{1}$}
\newcommand{\RHMan}{$2.270\times10^{-4}$}
\newcommand{\RHMbn}{$3.480\times10^{1}$}
\newcommand{\RHMafs}{$5.481\times10^{-4}$}
\newcommand{\RHMbfs}{$4.961\times10^{0}$}
\newcommand{\RHMK}{$1.000\times10^{-1}$}
\newcommand{\CLTa}{$8.090\times10^{-4}$}
\newcommand{\CLTb}{$7.474\times10^{0}$}
\newcommand{\CLTaf}{$1.911\times10^{-3}$}
\newcommand{\CLTbf}{$2.206\times10^{1}$}
\newcommand{\CLTan}{$2.270\times10^{-4}$}
\newcommand{\CLTbn}{$3.480\times10^{1}$}
\newcommand{\CLTafs}{$5.436\times10^{-4}$}
\newcommand{\CLTbfs}{$3.147\times10^{0}$}
\newcommand{\CLTK}{$1.000\times10^{-1}$}
\newcommand{\CRTa}{$8.090\times10^{-4}$}
\newcommand{\CRTb}{$7.474\times10^{0}$}
\newcommand{\CRTaf}{$1.911\times10^{-3}$}
\newcommand{\CRTbf}{$2.206\times10^{1}$}
\newcommand{\CRTan}{$2.270\times10^{-4}$}
\newcommand{\CRTbn}{$3.480\times10^{1}$}
\newcommand{\CRTafs}{$5.478\times10^{-4}$}
\newcommand{\CRTbfs}{$5.695\times10^{0}$}
\newcommand{\CRTK}{$1.000\times10^{-1}$}
\newcommand{\RHTa}{$8.090\times10^{-4}$}
\newcommand{\RHTb}{$7.474\times10^{0}$}
\newcommand{\RHTaf}{$1.911\times10^{-3}$}
\newcommand{\RHTbf}{$2.206\times10^{1}$}
\newcommand{\RHTan}{$2.270\times10^{-4}$}
\newcommand{\RHTbn}{$3.480\times10^{1}$}
\newcommand{\RHTafs}{$5.478\times10^{-4}$}
\newcommand{\RHTbfs}{$5.675\times10^{0}$}
\newcommand{\RHTK}{$1.000\times10^{-1}$}
\newcommand{\CLMsigmaa}{$1.494\times10^{-8}$}
\newcommand{\CLMsigmab}{$1.044\times10^{-4}$}
\newcommand{\CLMsigmaaf}{$2.502\times10^{-9}$}
\newcommand{\CLMsigmabf}{$1.835\times10^{-5}$}
\newcommand{\CLMsigmaan}{$1.337\times10^{-9}$}
\newcommand{\CLMsigmabn}{$4.554\times10^{-5}$}
\newcommand{\CLMsigmaafs}{$1.342\times10^{1}$}
\newcommand{\CLMsigmabfs}{$1.370\times10^{1}$}
\newcommand{\CLMsigmaK}{$1.965\times10^{-7}$}
\newcommand{\CRMsigmaa}{$1.507\times10^{-8}$}
\newcommand{\CRMsigmab}{$8.826\times10^{-5}$}
\newcommand{\CRMsigmaaf}{$2.659\times10^{-8}$}
\newcommand{\CRMsigmabf}{$1.972\times10^{-4}$}
\newcommand{\CRMsigmaan}{$1.143\times10^{-7}$}
\newcommand{\CRMsigmabn}{$3.477\times10^{-2}$}
\newcommand{\CRMsigmaafs}{$3.965\times10^{-5}$}
\newcommand{\CRMsigmabfs}{$4.445\times10^{0}$}
\newcommand{\CRMsigmaK}{$7.520\times10^{-7}$}
\newcommand{\RHMsigmaa}{$1.526\times10^{-8}$}
\newcommand{\RHMsigmab}{$9.443\times10^{-5}$}
\newcommand{\RHMsigmaaf}{$2.948\times10^{-8}$}
\newcommand{\RHMsigmabf}{$1.879\times10^{-4}$}
\newcommand{\RHMsigmaan}{$6.658\times10^{-9}$}
\newcommand{\RHMsigmabn}{$3.559\times10^{-4}$}
\newcommand{\RHMsigmaafs}{$4.295\times10^{-5}$}
\newcommand{\RHMsigmabfs}{$3.318\times10^{0}$}
\newcommand{\RHMsigmaK}{$6.438\times10^{-7}$}
\newcommand{\CLTsigmaa}{$1.415\times10^{-8}$}
\newcommand{\CLTsigmab}{$1.090\times10^{-4}$}
\newcommand{\CLTsigmaaf}{$3.981\times10^{-9}$}
\newcommand{\CLTsigmabf}{$1.491\times10^{-5}$}
\newcommand{\CLTsigmaan}{$1.059\times10^{-9}$}
\newcommand{\CLTsigmabn}{$4.139\times10^{-5}$}
\newcommand{\CLTsigmaafs}{$6.385\times10^{-5}$}
\newcommand{\CLTsigmabfs}{$1.209\times10^{1}$}
\newcommand{\CLTsigmaK}{$2.033\times10^{-7}$}
\newcommand{\CRTsigmaa}{$1.580\times10^{-8}$}
\newcommand{\CRTsigmab}{$1.072\times10^{-4}$}
\newcommand{\CRTsigmaaf}{$4.635\times10^{-8}$}
\newcommand{\CRTsigmabf}{$6.669\times10^{-4}$}
\newcommand{\CRTsigmaan}{$1.098\times10^{-7}$}
\newcommand{\CRTsigmabn}{$3.142\times10^{-2}$}
\newcommand{\CRTsigmaafs}{$3.896\times10^{-6}$}
\newcommand{\CRTsigmabfs}{$4.433\times10^{-2}$}
\newcommand{\CRTsigmaK}{$8.810\times10^{-7}$}
\newcommand{\RHTsigmaa}{$1.653\times10^{-8}$}
\newcommand{\RHTsigmab}{$1.119\times10^{-4}$}
\newcommand{\RHTsigmaaf}{$2.670\times10^{-8}$}
\newcommand{\RHTsigmabf}{$1.929\times10^{-4}$}
\newcommand{\RHTsigmaan}{$6.207\times10^{-9}$}
\newcommand{\RHTsigmabn}{$3.382\times10^{-4}$}
\newcommand{\RHTsigmaafs}{$1.010\times10^{-5}$}
\newcommand{\RHTsigmabfs}{$2.674\times10^{-1}$}
\newcommand{\RHTsigmaK}{$6.560\times10^{-7}$}
\newcommand{\CLTLNa}{$8.086\times10^{-4}$}
\newcommand{\CLTLNb}{$7.475\times10^{0}$}
\newcommand{\CLTLNaf}{$1.919\times10^{-3}$}
\newcommand{\CLTLNbf}{$2.202\times10^{1}$}
\newcommand{\CLTLNan}{$2.259\times10^{-4}$}
\newcommand{\CLTLNbn}{$3.484\times10^{1}$}
\newcommand{\CLTLNafs}{$2.901\times10^{-8}$}
\newcommand{\CLTLNbfs}{$3.047\times10^{0}$}
\newcommand{\CLTLNK}{$1.000\times10^{-1}$}
\newcommand{\CRTLNa}{$8.051\times10^{-4}$}
\newcommand{\CRTLNb}{$7.535\times10^{0}$}
\newcommand{\CRTLNaf}{$1.989\times10^{-3}$}
\newcommand{\CRTLNbf}{$2.135\times10^{1}$}
\newcommand{\CRTLNan}{$2.367\times10^{-4}$}
\newcommand{\CRTLNbn}{$3.136\times10^{1}$}
\newcommand{\CRTLNafs}{$2.590\times10^{-4}$}
\newcommand{\CRTLNbfs}{$9.088\times10^{0}$}
\newcommand{\CRTLNK}{$9.952\times10^{-2}$}
\newcommand{\RHTLNa}{$8.055\times10^{-4}$}
\newcommand{\RHTLNb}{$7.491\times10^{0}$}
\newcommand{\RHTLNaf}{$1.946\times10^{-3}$}
\newcommand{\RHTLNbf}{$2.183\times10^{1}$}
\newcommand{\RHTLNan}{$2.288\times10^{-4}$}
\newcommand{\RHTLNbn}{$3.479\times10^{1}$}
\newcommand{\RHTLNafs}{$4.273\times10^{-5}$}
\newcommand{\RHTLNbfs}{$2.492\times10^{1}$}
\newcommand{\RHTLNK}{$1.001\times10^{-1}$}
\newcommand{\CLTHNa}{$8.102\times10^{-4}$}
\newcommand{\CLTHNb}{$7.825\times10^{0}$}
\newcommand{\CLTHNaf}{$2.004\times10^{-3}$}
\newcommand{\CLTHNbf}{$2.129\times10^{1}$}
\newcommand{\CLTHNan}{$2.326\times10^{-4}$}
\newcommand{\CLTHNbn}{$3.378\times10^{1}$}
\newcommand{\CLTHNafs}{$2.185\times10^{-8}$}
\newcommand{\CLTHNbfs}{$2.926\times10^{0}$}
\newcommand{\CLTHNK}{$9.924\times10^{-2}$}
\newcommand{\CRTHNa}{$9.911\times10^{-4}$}
\newcommand{\CRTHNb}{$8.775\times10^{0}$}
\newcommand{\CRTHNaf}{$3.669\times10^{-3}$}
\newcommand{\CRTHNbf}{$9.787\times10^{0}$}
\newcommand{\CRTHNan}{$1.279\times10^{-8}$}
\newcommand{\CRTHNbn}{$7.070\times10^{-1}$}
\newcommand{\CRTHNafs}{$1.324\times10^{-8}$}
\newcommand{\CRTHNbfs}{$8.505\times10^{-1}$}
\newcommand{\CRTHNK}{$7.005\times10^{-2}$}
\newcommand{\RHTHNa}{$9.240\times10^{-4}$}
\newcommand{\RHTHNb}{$7.922\times10^{0}$}
\newcommand{\RHTHNaf}{$2.042\times10^{-3}$}
\newcommand{\RHTHNbf}{$2.087\times10^{1}$}
\newcommand{\RHTHNan}{$2.498\times10^{-4}$}
\newcommand{\RHTHNbn}{$3.406\times10^{1}$}
\newcommand{\RHTHNafs}{$1.582\times10^{-8}$}
\newcommand{\RHTHNbfs}{$3.055\times10^{0}$}
\newcommand{\RHTHNK}{$9.157\times10^{-2}$}
\newcommand{\CLTLNsigmaa}{$6.286\times10^{-8}$}
\newcommand{\CLTLNsigmab}{$4.969\times10^{-4}$}
\newcommand{\CLTLNsigmaaf}{$2.011\times10^{-8}$}
\newcommand{\CLTLNsigmabf}{$1.139\times10^{-4}$}
\newcommand{\CLTLNsigmaan}{$5.345\times10^{-9}$}
\newcommand{\CLTLNsigmabn}{$2.732\times10^{-4}$}
\newcommand{\CLTLNsigmaafs}{$3.916\times10^{-6}$}
\newcommand{\CLTLNsigmabfs}{$1.295\times10^{1}$}
\newcommand{\CLTLNsigmaK}{$9.225\times10^{-7}$}
\newcommand{\CRTLNsigmaa}{$5.740\times10^{-8}$}
\newcommand{\CRTLNsigmab}{$3.975\times10^{-4}$}
\newcommand{\CRTLNsigmaaf}{$1.834\times10^{-7}$}
\newcommand{\CRTLNsigmabf}{$2.209\times10^{-3}$}
\newcommand{\CRTLNsigmaan}{$3.721\times10^{-7}$}
\newcommand{\CRTLNsigmabn}{$1.098\times10^{-1}$}
\newcommand{\CRTLNsigmaafs}{$7.839\times10^{-6}$}
\newcommand{\CRTLNsigmabfs}{$1.624\times10^{-1}$}
\newcommand{\CRTLNsigmaK}{$2.781\times10^{-6}$}
\newcommand{\RHTLNsigmaa}{$5.530\times10^{-8}$}
\newcommand{\RHTLNsigmab}{$3.900\times10^{-4}$}
\newcommand{\RHTLNsigmaaf}{$8.448\times10^{-8}$}
\newcommand{\RHTLNsigmabf}{$6.359\times10^{-4}$}
\newcommand{\RHTLNsigmaan}{$2.430\times10^{-8}$}
\newcommand{\RHTLNsigmabn}{$1.305\times10^{-3}$}
\newcommand{\RHTLNsigmaafs}{$5.946\times10^{-6}$}
\newcommand{\RHTLNsigmabfs}{$1.060\times10^{0}$}
\newcommand{\RHTLNsigmaK}{$2.089\times10^{-6}$}
\newcommand{\CLTHNsigmaa}{$5.143\times10^{-7}$}
\newcommand{\CLTHNsigmab}{$4.260\times10^{-3}$}
\newcommand{\CLTHNsigmaaf}{$1.811\times10^{-7}$}
\newcommand{\CLTHNsigmabf}{$1.074\times10^{-3}$}
\newcommand{\CLTHNsigmaan}{$4.829\times10^{-8}$}
\newcommand{\CLTHNsigmabn}{$2.254\times10^{-3}$}
\newcommand{\CLTHNsigmaafs}{$3.022\times10^{-6}$}
\newcommand{\CLTHNsigmabfs}{$1.258\times10^{1}$}
\newcommand{\CLTHNsigmaK}{$8.517\times10^{-6}$}
\newcommand{\CRTHNsigmaa}{$3.082\times10^{-7}$}
\newcommand{\CRTHNsigmab}{$1.882\times10^{-3}$}
\newcommand{\CRTHNsigmaaf}{$1.930\times10^{-6}$}
\newcommand{\CRTHNsigmabf}{$1.491\times10^{-2}$}
\newcommand{\CRTHNsigmaan}{$4.575\times10^{-8}$}
\newcommand{\CRTHNsigmabn}{$4.072\times10^{0}$}
\newcommand{\CRTHNsigmaafs}{$1.813\times10^{-6}$}
\newcommand{\CRTHNsigmabfs}{$3.901\times10^{0}$}
\newcommand{\CRTHNsigmaK}{$2.025\times10^{-5}$}
\newcommand{\RHTHNsigmaa}{$4.174\times10^{-7}$}
\newcommand{\RHTHNsigmab}{$2.899\times10^{-3}$}
\newcommand{\RHTHNsigmaaf}{$7.679\times10^{-7}$}
\newcommand{\RHTHNsigmabf}{$5.549\times10^{-3}$}
\newcommand{\RHTHNsigmaan}{$2.148\times10^{-7}$}
\newcommand{\RHTHNsigmabn}{$1.052\times10^{-2}$}
\newcommand{\RHTHNsigmaafs}{$2.142\times10^{-6}$}
\newcommand{\RHTHNsigmabfs}{$1.148\times10^{1}$}
\newcommand{\RHTHNsigmaK}{$1.810\times10^{-5}$}
\newcommand{\CLTSNa}{$4.348\times10^{-4}$}
\newcommand{\CLTSNb}{$1.076\times10^{1}$}
\newcommand{\CLTSNaf}{$2.580\times10^{-3}$}
\newcommand{\CLTSNbf}{$1.920\times10^{1}$}
\newcommand{\CLTSNan}{$2.541\times10^{-4}$}
\newcommand{\CLTSNbn}{$3.257\times10^{1}$}
\newcommand{\CLTSNafs}{$5.688\times10^{-5}$}
\newcommand{\CLTSNbfs}{$2.856\times10^{0}$}
\newcommand{\CLTSNK}{$9.678\times10^{-2}$}
\newcommand{\CRTSNa}{$8.135\times10^{-4}$}
\newcommand{\CRTSNb}{$1.053\times10^{1}$}
\newcommand{\CRTSNaf}{$6.905\times10^{-3}$}
\newcommand{\CRTSNbf}{$1.265\times10^{-2}$}
\newcommand{\CRTSNan}{$1.714\times10^{-8}$}
\newcommand{\CRTSNbn}{$5.721\times10^{-1}$}
\newcommand{\CRTSNafs}{$8.329\times10^{-8}$}
\newcommand{\CRTSNbfs}{$4.417\times10^{-1}$}
\newcommand{\CRTSNK}{$5.278\times10^{-2}$}
\newcommand{\RHTSNa}{$5.520\times10^{-4}$}
\newcommand{\RHTSNb}{$9.863\times10^{0}$}
\newcommand{\RHTSNaf}{$1.262\times10^{-3}$}
\newcommand{\RHTSNbf}{$2.893\times10^{1}$}
\newcommand{\RHTSNan}{$3.668\times10^{-3}$}
\newcommand{\RHTSNbn}{$1.187\times10^{-4}$}
\newcommand{\RHTSNafs}{$1.889\times10^{-7}$}
\newcommand{\RHTSNbfs}{$1.867\times10^{0}$}
\newcommand{\RHTSNK}{$9.391\times10^{-2}$}
\newcommand{\CLTSNsigmaa}{$4.725\times10^{-7}$}
\newcommand{\CLTSNsigmab}{$7.082\times10^{-3}$}
\newcommand{\CLTSNsigmaaf}{$3.194\times10^{-7}$}
\newcommand{\CLTSNsigmabf}{$1.470\times10^{-3}$}
\newcommand{\CLTSNsigmaan}{$7.436\times10^{-8}$}
\newcommand{\CLTSNsigmabn}{$3.217\times10^{-3}$}
\newcommand{\CLTSNsigmaafs}{$2.880\times10^{-4}$}
\newcommand{\CLTSNsigmabfs}{$1.218\times10^{1}$}
\newcommand{\CLTSNsigmaK}{$1.194\times10^{-5}$}
\newcommand{\CRTSNsigmaa}{$2.584\times10^{-7}$}
\newcommand{\CRTSNsigmab}{$1.884\times10^{-3}$}
\newcommand{\CRTSNsigmaaf}{$3.074\times10^{-6}$}
\newcommand{\CRTSNsigmabf}{$1.399\times10^{-2}$}
\newcommand{\CRTSNsigmaan}{$6.089\times10^{-8}$}
\newcommand{\CRTSNsigmabn}{$3.227\times10^{0}$}
\newcommand{\CRTSNsigmaafs}{$8.836\times10^{-6}$}
\newcommand{\CRTSNsigmabfs}{$2.678\times10^{0}$}
\newcommand{\CRTSNsigmaK}{$2.176\times10^{-5}$}
\newcommand{\RHTSNsigmaa}{$1.626\times10^{-7}$}
\newcommand{\RHTSNsigmab}{$1.828\times10^{-3}$}
\newcommand{\RHTSNsigmaaf}{$2.374\times10^{-7}$}
\newcommand{\RHTSNsigmabf}{$2.462\times10^{-3}$}
\newcommand{\RHTSNsigmaan}{$1.222\times10^{-6}$}
\newcommand{\RHTSNsigmabn}{$1.810\times10^{-4}$}
\newcommand{\RHTSNsigmaafs}{$1.078\times10^{-5}$}
\newcommand{\RHTSNsigmabfs}{$7.454\times10^{0}$}
\newcommand{\RHTSNsigmaK}{$9.835\times10^{-6}$}
\newcommand{\CLMalphaVs}{$1.691\times10^{1}$}
\newcommand{\CLMbetaVs}{$7.215\times10^{7}$}
\newcommand{\CRMalphaVs}{$1.803\times10^{1}$}
\newcommand{\CRMbetaVs}{$7.008\times10^{7}$}
\newcommand{\RHMalphaVs}{$1.800\times10^{1}$}
\newcommand{\RHMbetaVs}{$6.978\times10^{7}$}
\newcommand{\CLTalphaVs}{$1.721\times10^{1}$}
\newcommand{\CLTbetaVs}{$7.393\times10^{7}$}
\newcommand{\CRTalphaVs}{$1.802\times10^{1}$}
\newcommand{\CRTbetaVs}{$7.506\times10^{7}$}
\newcommand{\RHTalphaVs}{$1.799\times10^{1}$}
\newcommand{\RHTbetaVs}{$7.491\times10^{7}$}
\newcommand{\CLMalphasigma}{$5.204\times10^{6}$}
\newcommand{\CLMbetasigma}{$1.920\times10^{1}$}
\newcommand{\CRMalphasigma}{$5.210\times10^{6}$}
\newcommand{\CRMbetasigma}{$1.897\times10^{1}$}
\newcommand{\RHMalphasigma}{$5.222\times10^{6}$}
\newcommand{\RHMbetasigma}{$1.898\times10^{1}$}
\newcommand{\CLTalphasigma}{$5.404\times10^{6}$}
\newcommand{\CLTbetasigma}{$1.850\times10^{1}$}
\newcommand{\CRTalphasigma}{$5.401\times10^{6}$}
\newcommand{\CRTbetasigma}{$1.821\times10^{1}$}
\newcommand{\RHTalphasigma}{$5.416\times10^{6}$}
\newcommand{\RHTbetasigma}{$1.831\times10^{1}$}
\newcommand{\CLTLNalphaVs}{$1.837\times10^{1}$}
\newcommand{\CLTLNbetaVs}{$1.957\times10^{7}$}
\newcommand{\CRTLNalphaVs}{$1.847\times10^{1}$}
\newcommand{\CRTLNbetaVs}{$2.068\times10^{7}$}
\newcommand{\RHTLNalphaVs}{$1.822\times10^{1}$}
\newcommand{\RHTLNbetaVs}{$2.439\times10^{7}$}
\newcommand{\CLTHNalphaVs}{$1.923\times10^{1}$}
\newcommand{\CLTHNbetaVs}{$2.245\times10^{6}$}
\newcommand{\CRTHNalphaVs}{$1.932\times10^{1}$}
\newcommand{\CRTHNbetaVs}{$1.090\times10^{6}$}
\newcommand{\RHTHNalphaVs}{$1.916\times10^{1}$}
\newcommand{\RHTHNbetaVs}{$2.668\times10^{6}$}
\newcommand{\CLTLNalphasigma}{$2.683\times10^{6}$}
\newcommand{\CLTLNbetasigma}{$3.691\times10^{1}$}
\newcommand{\CRTLNalphasigma}{$2.890\times10^{6}$}
\newcommand{\CRTLNbetasigma}{$3.411\times10^{1}$}
\newcommand{\RHTLNalphasigma}{$2.857\times10^{6}$}
\newcommand{\RHTLNbetasigma}{$3.453\times10^{1}$}
\newcommand{\CLTHNalphasigma}{$9.050\times10^{5}$}
\newcommand{\CLTHNbetasigma}{$1.100\times10^{2}$}
\newcommand{\CRTHNalphasigma}{$1.080\times10^{6}$}
\newcommand{\CRTHNbetasigma}{$9.200\times10^{1}$}
\newcommand{\RHTHNalphasigma}{$9.907\times10^{5}$}
\newcommand{\RHTHNbetasigma}{$1.004\times10^{2}$}
\newcommand{\CLTSNalphaVs}{$1.938\times10^{1}$}
\newcommand{\CLTSNbetaVs}{$1.515\times10^{6}$}
\newcommand{\CRTSNalphaVs}{$1.944\times10^{1}$}
\newcommand{\CRTSNbetaVs}{$7.690\times10^{5}$}
\newcommand{\RHTSNalphaVs}{$1.906\times10^{1}$}
\newcommand{\RHTSNbetaVs}{$3.695\times10^{6}$}
\newcommand{\CLTSNalphasigma}{$7.540\times10^{5}$}
\newcommand{\CLTSNbetasigma}{$1.323\times10^{2}$}
\newcommand{\CRTSNalphasigma}{$1.017\times10^{6}$}
\newcommand{\CRTSNbetasigma}{$9.776\times10^{1}$}
\newcommand{\RHTSNalphasigma}{$1.355\times10^{6}$}
\newcommand{\RHTSNbetasigma}{$7.322\times10^{1}$}
\newcommand{\CLMEa}{$-1.022\times10^{-5}$}
\newcommand{\CLMEb}{$-2.568\times10^{-5}$}
\newcommand{\CLMEaf}{$-8.278\times10^{-5}$}
\newcommand{\CLMEbf}{$-3.188\times10^{-5}$}
\newcommand{\CLMEan}{$1.158\times10^{-4}$}
\newcommand{\CLMEbn}{$2.496\times10^{-5}$}
\newcommand{\CLMEafs}{$-5.738\times10^{3}$}
\newcommand{\CLMEbfs}{$4.279\times10^{-1}$}
\newcommand{\CLMEK}{$-3.005\times10^{-5}$}
\newcommand{\CRMEa}{$4.572\times10^{-5}$}
\newcommand{\CRMEb}{$2.159\times10^{-5}$}
\newcommand{\CRMEaf}{$9.389\times10^{-6}$}
\newcommand{\CRMEbf}{$4.899\times10^{-5}$}
\newcommand{\CRMEan}{$-6.964\times10^{-5}$}
\newcommand{\CRMEbn}{$6.033\times10^{-5}$}
\newcommand{\CRMEafs}{$4.131\times10^{-3}$}
\newcommand{\CRMEbfs}{$1.354\times10^{-1}$}
\newcommand{\CRMEK}{$3.963\times10^{-5}$}
\newcommand{\RHMEa}{$-1.448\times10^{-5}$}
\newcommand{\RHMEb}{$-3.091\times10^{-5}$}
\newcommand{\RHMEaf}{$-6.745\times10^{-5}$}
\newcommand{\RHMEbf}{$-2.662\times10^{-5}$}
\newcommand{\RHMEan}{$5.965\times10^{-5}$}
\newcommand{\RHMEbn}{$-2.814\times10^{-5}$}
\newcommand{\RHMEafs}{$-1.945\times10^{-3}$}
\newcommand{\RHMEbfs}{$1.283\times10^{-1}$}
\newcommand{\RHMEK}{$-3.433\times10^{-5}$}
\newcommand{\CLTEa}{$7.019\times10^{-8}$}
\newcommand{\CLTEb}{$-1.492\times10^{-5}$}
\newcommand{\CLTEaf}{$-6.743\times10^{-5}$}
\newcommand{\CLTEbf}{$-1.735\times10^{-5}$}
\newcommand{\CLTEan}{$9.178\times10^{-5}$}
\newcommand{\CLTEbn}{$1.181\times10^{-6}$}
\newcommand{\CLTEafs}{$6.163\times10^{-3}$}
\newcommand{\CLTEbfs}{$4.471\times10^{-1}$}
\newcommand{\CLTEK}{$-1.350\times10^{-5}$}
\newcommand{\CRTEa}{$7.098\times10^{-7}$}
\newcommand{\CRTEb}{$-2.521\times10^{-5}$}
\newcommand{\CRTEaf}{$-9.161\times10^{-5}$}
\newcommand{\CRTEbf}{$-6.930\times10^{-5}$}
\newcommand{\CRTEan}{$-8.719\times10^{-5}$}
\newcommand{\CRTEbn}{$3.278\times10^{-5}$}
\newcommand{\CRTEafs}{$-1.459\times10^{-3}$}
\newcommand{\CRTEbfs}{$-7.901\times10^{-4}$}
\newcommand{\CRTEK}{$-8.489\times10^{-6}$}
\newcommand{\RHTEa}{$-1.826\times10^{-5}$}
\newcommand{\RHTEb}{$-3.542\times10^{-5}$}
\newcommand{\RHTEaf}{$-6.393\times10^{-5}$}
\newcommand{\RHTEbf}{$-2.107\times10^{-5}$}
\newcommand{\RHTEan}{$7.453\times10^{-5}$}
\newcommand{\RHTEbn}{$-1.752\times10^{-5}$}
\newcommand{\RHTEafs}{$-1.505\times10^{-3}$}
\newcommand{\RHTEbfs}{$2.882\times10^{-3}$}
\newcommand{\RHTEK}{$-3.858\times10^{-5}$}
\newcommand{\CLTLNEa}{$4.826\times10^{-4}$}
\newcommand{\CLTLNEb}{$-1.843\times10^{-4}$}
\newcommand{\CLTLNEaf}{$-4.237\times10^{-3}$}
\newcommand{\CLTLNEbf}{$1.938\times10^{-3}$}
\newcommand{\CLTLNEan}{$4.971\times10^{-3}$}
\newcommand{\CLTLNEbn}{$-1.187\times10^{-3}$}
\newcommand{\CLTLNEafs}{$9.999\times10^{-1}$}
\newcommand{\CLTLNEbfs}{$4.645\times10^{-1}$}
\newcommand{\CLTLNEK}{$-1.080\times10^{-4}$}
\newcommand{\CRTLNEa}{$4.764\times10^{-3}$}
\newcommand{\CRTLNEb}{$-8.210\times10^{-3}$}
\newcommand{\CRTLNEaf}{$-4.060\times10^{-2}$}
\newcommand{\CRTLNEbf}{$3.238\times10^{-2}$}
\newcommand{\CRTLNEan}{$-4.290\times10^{-2}$}
\newcommand{\CRTLNEbn}{$9.900\times10^{-2}$}
\newcommand{\CRTLNEafs}{$5.264\times10^{-1}$}
\newcommand{\CRTLNEbfs}{$-5.970\times10^{-1}$}
\newcommand{\CRTLNEK}{$4.751\times10^{-3}$}
\newcommand{\RHTLNEa}{$4.286\times10^{-3}$}
\newcommand{\RHTLNEb}{$-2.226\times10^{-3}$}
\newcommand{\RHTLNEaf}{$-1.806\times10^{-2}$}
\newcommand{\RHTLNEbf}{$1.070\times10^{-2}$}
\newcommand{\RHTLNEan}{$-7.797\times10^{-3}$}
\newcommand{\RHTLNEbn}{$2.294\times10^{-4}$}
\newcommand{\RHTLNEafs}{$9.219\times10^{-1}$}
\newcommand{\RHTLNEbfs}{$-3.379\times10^{0}$}
\newcommand{\RHTLNEK}{$-1.271\times10^{-3}$}
\newcommand{\CLTHNEa}{$-1.526\times10^{-3}$}
\newcommand{\CLTHNEb}{$-4.697\times10^{-2}$}
\newcommand{\CLTHNEaf}{$-4.859\times10^{-2}$}
\newcommand{\CLTHNEbf}{$3.506\times10^{-2}$}
\newcommand{\CLTHNEan}{$-2.456\times10^{-2}$}
\newcommand{\CLTHNEbn}{$2.947\times10^{-2}$}
\newcommand{\CLTHNEafs}{$10.000\times10^{-1}$}
\newcommand{\CLTHNEbfs}{$4.859\times10^{-1}$}
\newcommand{\CLTHNEK}{$7.621\times10^{-3}$}
\newcommand{\CRTHNEa}{$-2.251\times10^{-1}$}
\newcommand{\CRTHNEb}{$-1.741\times10^{-1}$}
\newcommand{\CRTHNEaf}{$-9.201\times10^{-1}$}
\newcommand{\CRTHNEbf}{$5.564\times10^{-1}$}
\newcommand{\CRTHNEan}{$9.999\times10^{-1}$}
\newcommand{\CRTHNEbn}{$9.797\times10^{-1}$}
\newcommand{\CRTHNEafs}{$10.000\times10^{-1}$}
\newcommand{\CRTHNEbfs}{$8.506\times10^{-1}$}
\newcommand{\CRTHNEK}{$2.995\times10^{-1}$}
\newcommand{\RHTHNEa}{$-1.421\times10^{-1}$}
\newcommand{\RHTHNEb}{$-5.992\times10^{-2}$}
\newcommand{\RHTHNEaf}{$-6.874\times10^{-2}$}
\newcommand{\RHTHNEbf}{$5.408\times10^{-2}$}
\newcommand{\RHTHNEan}{$-1.006\times10^{-1}$}
\newcommand{\RHTHNEbn}{$2.124\times10^{-2}$}
\newcommand{\RHTHNEafs}{$10.000\times10^{-1}$}
\newcommand{\RHTHNEbfs}{$4.632\times10^{-1}$}
\newcommand{\RHTHNEK}{$8.432\times10^{-2}$}
\newcommand{\CLTSNEa}{$4.626\times10^{-1}$}
\newcommand{\CLTSNEb}{$-4.398\times10^{-1}$}
\newcommand{\CLTSNEaf}{$-3.501\times10^{-1}$}
\newcommand{\CLTSNEbf}{$1.296\times10^{-1}$}
\newcommand{\CLTSNEan}{$-1.194\times10^{-1}$}
\newcommand{\CLTSNEbn}{$6.421\times10^{-2}$}
\newcommand{\CLTSNEafs}{$8.960\times10^{-1}$}
\newcommand{\CLTSNEbfs}{$4.981\times10^{-1}$}
\newcommand{\CLTSNEK}{$3.217\times10^{-2}$}
\newcommand{\CRTSNEa}{$-5.595\times10^{-3}$}
\newcommand{\CRTSNEb}{$-4.090\times10^{-1}$}
\newcommand{\CRTSNEaf}{$-2.613\times10^{0}$}
\newcommand{\CRTSNEbf}{$9.994\times10^{-1}$}
\newcommand{\CRTSNEan}{$9.999\times10^{-1}$}
\newcommand{\CRTSNEbn}{$9.836\times10^{-1}$}
\newcommand{\CRTSNEafs}{$9.998\times10^{-1}$}
\newcommand{\CRTSNEbfs}{$9.224\times10^{-1}$}
\newcommand{\CRTSNEK}{$4.722\times10^{-1}$}
\newcommand{\RHTSNEa}{$3.177\times10^{-1}$}
\newcommand{\RHTSNEb}{$-3.196\times10^{-1}$}
\newcommand{\RHTSNEaf}{$3.399\times10^{-1}$}
\newcommand{\RHTSNEbf}{$-3.111\times10^{-1}$}
\newcommand{\RHTSNEan}{$-1.516\times10^{1}$}
\newcommand{\RHTSNEbn}{$10.000\times10^{-1}$}
\newcommand{\RHTSNEafs}{$9.997\times10^{-1}$}
\newcommand{\RHTSNEbfs}{$6.719\times10^{-1}$}
\newcommand{\RHTSNEK}{$6.093\times10^{-2}$}
\newcommand{\CLMsa}{$-0.136$}
\newcommand{\CLMsb}{$0.088$}
\newcommand{\CLMsaf}{$-0.133$}
\newcommand{\CLMsbf}{$0.003$}
\newcommand{\CLMsan}{$0.007$}
\newcommand{\CLMsbn}{$-0.042$}
\newcommand{\CLMsafs}{\cellcolor{colAfs}$0.722$}
\newcommand{\CLMsbfs}{\cellcolor{colBfs}$0.847$}
\newcommand{\CLMsK}{$0.114$}
\newcommand{\CRMsa}{$-0.136$}
\newcommand{\CRMsb}{$0.088$}
\newcommand{\CRMsaf}{$-0.133$}
\newcommand{\CRMsbf}{$0.003$}
\newcommand{\CRMsan}{$0.007$}
\newcommand{\CRMsbn}{$-0.042$}
\newcommand{\CRMsafs}{$-0.104$}
\newcommand{\CRMsbfs}{$0.492$}
\newcommand{\CRMsK}{$0.114$}
\newcommand{\RHMsa}{$-0.136$}
\newcommand{\RHMsb}{$0.088$}
\newcommand{\RHMsaf}{$-0.133$}
\newcommand{\RHMsbf}{$0.003$}
\newcommand{\RHMsan}{$0.007$}
\newcommand{\RHMsbn}{$-0.042$}
\newcommand{\RHMsafs}{$-0.104$}
\newcommand{\RHMsbfs}{$0.337$}
\newcommand{\RHMsK}{$0.114$}
\newcommand{\CLTsa}{$-0.136$}
\newcommand{\CLTsb}{$0.088$}
\newcommand{\CLTsaf}{$-0.133$}
\newcommand{\CLTsbf}{$0.003$}
\newcommand{\CLTsan}{$0.007$}
\newcommand{\CLTsbn}{$-0.042$}
\newcommand{\CLTsafs}{$-0.104$}
\newcommand{\CLTsbfs}{\cellcolor{colBfs}$0.838$}
\newcommand{\CLTsK}{$0.114$}
\newcommand{\CRTsa}{$-0.136$}
\newcommand{\CRTsb}{$0.088$}
\newcommand{\CRTsaf}{$-0.133$}
\newcommand{\CRTsbf}{$0.003$}
\newcommand{\CRTsan}{$0.007$}
\newcommand{\CRTsbn}{$-0.042$}
\newcommand{\CRTsafs}{$-0.104$}
\newcommand{\CRTsbfs}{$-0.033$}
\newcommand{\CRTsK}{$0.114$}
\newcommand{\RHTsa}{$-0.136$}
\newcommand{\RHTsb}{$0.088$}
\newcommand{\RHTsaf}{$-0.133$}
\newcommand{\RHTsbf}{$0.003$}
\newcommand{\RHTsan}{$0.007$}
\newcommand{\RHTsbn}{$-0.042$}
\newcommand{\RHTsafs}{$-0.104$}
\newcommand{\RHTsbfs}{$-0.033$}
\newcommand{\RHTsK}{$0.114$}
\newcommand{\CLTLNsa}{$-0.136$}
\newcommand{\CLTLNsb}{$0.088$}
\newcommand{\CLTLNsaf}{$-0.133$}
\newcommand{\CLTLNsbf}{$0.003$}
\newcommand{\CLTLNsan}{$0.007$}
\newcommand{\CLTLNsbn}{$-0.042$}
\newcommand{\CLTLNsafs}{\cellcolor{colAfs}$0.799$}
\newcommand{\CLTLNsbfs}{\cellcolor{colBfs}$0.848$}
\newcommand{\CLTLNsK}{$0.114$}
\newcommand{\CRTLNsa}{$-0.136$}
\newcommand{\CRTLNsb}{$0.088$}
\newcommand{\CRTLNsaf}{$-0.133$}
\newcommand{\CRTLNsbf}{$0.003$}
\newcommand{\CRTLNsan}{$0.007$}
\newcommand{\CRTLNsbn}{$-0.042$}
\newcommand{\CRTLNsafs}{$-0.104$}
\newcommand{\CRTLNsbfs}{$-0.033$}
\newcommand{\CRTLNsK}{$0.114$}
\newcommand{\RHTLNsa}{$-0.136$}
\newcommand{\RHTLNsb}{$0.088$}
\newcommand{\RHTLNsaf}{$-0.133$}
\newcommand{\RHTLNsbf}{$0.003$}
\newcommand{\RHTLNsan}{$0.007$}
\newcommand{\RHTLNsbn}{$-0.042$}
\newcommand{\RHTLNsafs}{$-0.104$}
\newcommand{\RHTLNsbfs}{$-0.033$}
\newcommand{\RHTLNsK}{$0.114$}
\newcommand{\CLTHNsa}{$-0.136$}
\newcommand{\CLTHNsb}{$0.088$}
\newcommand{\CLTHNsaf}{$-0.133$}
\newcommand{\CLTHNsbf}{$0.003$}
\newcommand{\CLTHNsan}{$0.007$}
\newcommand{\CLTHNsbn}{$-0.042$}
\newcommand{\CLTHNsafs}{\cellcolor{colAfs}$0.799$}
\newcommand{\CLTHNsbfs}{\cellcolor{colBfs}$0.849$}
\newcommand{\CLTHNsK}{$0.114$}
\newcommand{\CRTHNsa}{$-0.136$}
\newcommand{\CRTHNsb}{$0.088$}
\newcommand{\CRTHNsaf}{$-0.133$}
\newcommand{\CRTHNsbf}{$0.003$}
\newcommand{\CRTHNsan}{\cellcolor{colAn}$0.933$}
\newcommand{\CRTHNsbn}{\cellcolor{colBn}$0.878$}
\newcommand{\CRTHNsafs}{\cellcolor{colAfs}$0.799$}
\newcommand{\CRTHNsbfs}{\cellcolor{colBfs}$0.855$}
\newcommand{\CRTHNsK}{$0.114$}
\newcommand{\RHTHNsa}{$-0.136$}
\newcommand{\RHTHNsb}{$0.088$}
\newcommand{\RHTHNsaf}{$-0.133$}
\newcommand{\RHTHNsbf}{$0.003$}
\newcommand{\RHTHNsan}{$0.007$}
\newcommand{\RHTHNsbn}{$-0.042$}
\newcommand{\RHTHNsafs}{\cellcolor{colAfs}$0.799$}
\newcommand{\RHTHNsbfs}{\cellcolor{colBfs}$0.836$}
\newcommand{\RHTHNsK}{$0.114$}
\newcommand{\CLTSNsa}{$-0.136$}
\newcommand{\CLTSNsb}{$0.088$}
\newcommand{\CLTSNsaf}{$-0.133$}
\newcommand{\CLTSNsbf}{$0.003$}
\newcommand{\CLTSNsan}{$0.007$}
\newcommand{\CLTSNsbn}{$-0.042$}
\newcommand{\CLTSNsafs}{\cellcolor{colAfs}$0.735$}
\newcommand{\CLTSNsbfs}{\cellcolor{colBfs}$0.848$}
\newcommand{\CLTSNsK}{$0.114$}
\newcommand{\CRTSNsa}{$-0.136$}
\newcommand{\CRTSNsb}{$0.088$}
\newcommand{\CRTSNsaf}{$-0.133$}
\newcommand{\CRTSNsbf}{\cellcolor{colBf}$0.611$}
\newcommand{\CRTSNsan}{\cellcolor{colAn}$0.932$}
\newcommand{\CRTSNsbn}{\cellcolor{colBn}$0.877$}
\newcommand{\CRTSNsafs}{\cellcolor{colAfs}$0.798$}
\newcommand{\CRTSNsbfs}{\cellcolor{colBfs}$0.875$}
\newcommand{\CRTSNsK}{$0.114$}
\newcommand{\RHTSNsa}{$-0.136$}
\newcommand{\RHTSNsb}{$0.088$}
\newcommand{\RHTSNsaf}{$-0.133$}
\newcommand{\RHTSNsbf}{$0.003$}
\newcommand{\RHTSNsan}{$0.007$}
\newcommand{\RHTSNsbn}{\cellcolor{colBn}$0.688$}
\newcommand{\RHTSNsafs}{\cellcolor{colAfs}$0.796$}
\newcommand{\RHTSNsbfs}{\cellcolor{colBfs}$0.842$}
\newcommand{\RHTSNsK}{$0.114$}


\newcommand{\CLMCSa}{$1000$}
\newcommand{\CLMCSb}{$1000$}
\newcommand{\CLMCSaf}{$1000$}
\newcommand{\CLMCSbf}{$1000$}
\newcommand{\CLMCSan}{$1000$}
\newcommand{\CLMCSbn}{$1000$}
\newcommand{\CLMCSafs}{$17000$}
\newcommand{\CLMCSbfs}{$17000$}
\newcommand{\CLMCSK}{$1000$}
\newcommand{\CRMCSa}{$1000$}
\newcommand{\CRMCSb}{$1000$}
\newcommand{\CRMCSaf}{$1000$}
\newcommand{\CRMCSbf}{$1000$}
\newcommand{\CRMCSan}{$1000$}
\newcommand{\CRMCSbn}{$1000$}
\newcommand{\CRMCSafs}{$4000$}
\newcommand{\CRMCSbfs}{$14000$}
\newcommand{\CRMCSK}{$1000$}
\newcommand{\RHMCSa}{$1000$}
\newcommand{\RHMCSb}{$1000$}
\newcommand{\RHMCSaf}{$1000$}
\newcommand{\RHMCSbf}{$1000$}
\newcommand{\RHMCSan}{$1000$}
\newcommand{\RHMCSbn}{$1000$}
\newcommand{\RHMCSafs}{$2000$}
\newcommand{\RHMCSbfs}{$14000$}
\newcommand{\RHMCSK}{$1000$}
\newcommand{\CLTCSa}{$1000$}
\newcommand{\CLTCSb}{$1000$}
\newcommand{\CLTCSaf}{$1000$}
\newcommand{\CLTCSbf}{$1000$}
\newcommand{\CLTCSan}{$1000$}
\newcommand{\CLTCSbn}{$1000$}
\newcommand{\CLTCSafs}{$5000$}
\newcommand{\CLTCSbfs}{$17000$}
\newcommand{\CLTCSK}{$1000$}
\newcommand{\CRTCSa}{$1000$}
\newcommand{\CRTCSb}{$1000$}
\newcommand{\CRTCSaf}{$1000$}
\newcommand{\CRTCSbf}{$1000$}
\newcommand{\CRTCSan}{$1000$}
\newcommand{\CRTCSbn}{$1000$}
\newcommand{\CRTCSafs}{$1000$}
\newcommand{\CRTCSbfs}{$1000$}
\newcommand{\CRTCSK}{$1000$}
\newcommand{\RHTCSa}{$1000$}
\newcommand{\RHTCSb}{$1000$}
\newcommand{\RHTCSaf}{$1000$}
\newcommand{\RHTCSbf}{$1000$}
\newcommand{\RHTCSan}{$1000$}
\newcommand{\RHTCSbn}{$1000$}
\newcommand{\RHTCSafs}{$1000$}
\newcommand{\RHTCSbfs}{$1000$}
\newcommand{\RHTCSK}{$1000$}
\newcommand{\CLTLNCSa}{$1000$}
\newcommand{\CLTLNCSb}{$1000$}
\newcommand{\CLTLNCSaf}{$1000$}
\newcommand{\CLTLNCSbf}{$1000$}
\newcommand{\CLTLNCSan}{$1000$}
\newcommand{\CLTLNCSbn}{$1000$}
\newcommand{\CLTLNCSbfs}{$17000$}
\newcommand{\CLTLNCSK}{$1000$}
\newcommand{\CRTLNCSa}{$1000$}
\newcommand{\CRTLNCSb}{$1000$}
\newcommand{\CRTLNCSaf}{$1000$}
\newcommand{\CRTLNCSbf}{$1000$}
\newcommand{\CRTLNCSan}{$1000$}
\newcommand{\CRTLNCSbn}{$1000$}
\newcommand{\CRTLNCSafs}{$1000$}
\newcommand{\CRTLNCSbfs}{$1000$}
\newcommand{\CRTLNCSK}{$1000$}
\newcommand{\RHTLNCSa}{$1000$}
\newcommand{\RHTLNCSb}{$1000$}
\newcommand{\RHTLNCSaf}{$1000$}
\newcommand{\RHTLNCSbf}{$1000$}
\newcommand{\RHTLNCSan}{$1000$}
\newcommand{\RHTLNCSbn}{$1000$}
\newcommand{\RHTLNCSafs}{$10000$}
\newcommand{\RHTLNCSbfs}{$9000$}
\newcommand{\RHTLNCSK}{$1000$}
\newcommand{\CLTHNCSa}{$1000$}
\newcommand{\CLTHNCSb}{$1000$}
\newcommand{\CLTHNCSaf}{$1000$}
\newcommand{\CLTHNCSbf}{$1000$}
\newcommand{\CLTHNCSan}{$1000$}
\newcommand{\CLTHNCSbn}{$1000$}
\newcommand{\CLTHNCSbfs}{$17000$}
\newcommand{\CLTHNCSK}{$1000$}
\newcommand{\CRTHNCSa}{$1000$}
\newcommand{\CRTHNCSb}{$1000$}
\newcommand{\CRTHNCSaf}{$1000$}
\newcommand{\CRTHNCSbf}{$1000$}
\newcommand{\CRTHNCSbn}{$16000$}
\newcommand{\CRTHNCSbfs}{$17000$}
\newcommand{\CRTHNCSK}{$1000$}
\newcommand{\RHTHNCSa}{$1000$}
\newcommand{\RHTHNCSb}{$1000$}
\newcommand{\RHTHNCSaf}{$1000$}
\newcommand{\RHTHNCSbf}{$1000$}
\newcommand{\RHTHNCSan}{$1000$}
\newcommand{\RHTHNCSbn}{$1000$}
\newcommand{\RHTHNCSafs}{$19000$}
\newcommand{\RHTHNCSbfs}{$17000$}
\newcommand{\RHTHNCSK}{$1000$}
\newcommand{\CLTSNCSa}{$1000$}
\newcommand{\CLTSNCSb}{$1000$}
\newcommand{\CLTSNCSaf}{$1000$}
\newcommand{\CLTSNCSbf}{$1000$}
\newcommand{\CLTSNCSan}{$1000$}
\newcommand{\CLTSNCSbn}{$1000$}
\newcommand{\CLTSNCSbfs}{$17000$}
\newcommand{\CLTSNCSK}{$1000$}
\newcommand{\CRTSNCSa}{$1000$}
\newcommand{\CRTSNCSb}{$1000$}
\newcommand{\CRTSNCSaf}{$1000$}
\newcommand{\CRTSNCSbf}{$15000$}
\newcommand{\CRTSNCSbn}{$16000$}
\newcommand{\CRTSNCSbfs}{$17000$}
\newcommand{\CRTSNCSK}{$1000$}
\newcommand{\RHTSNCSa}{$1000$}
\newcommand{\RHTSNCSb}{$1000$}
\newcommand{\RHTSNCSaf}{$1000$}
\newcommand{\RHTSNCSbf}{$1000$}
\newcommand{\RHTSNCSan}{$1000$}
\newcommand{\RHTSNCSbfs}{$17000$}
\newcommand{\RHTSNCSK}{$1000$}

\newif\ifrevision
\revisionfalse  

\ifrevision
\newenvironment{revision}{\color{red}}{}
\newcommand{\revisionst}[1]{{\color{red}\st{#1}}}
\else
\newenvironment{revision}{}{}
\newcommand{\revisionst}[1]{}
\fi

\newif\ifsecondrevision
\secondrevisiontrue     
\secondrevisionfalse  

\ifsecondrevision
\newenvironment{secondrevision}{\color{red}}{}
\newcommand{\secondrevisionst}[1]{{\color{red}\st{#1}}}
\else
\newenvironment{secondrevision}{}{}
\newcommand{\secondrevisionst}[1]{}
\fi

\let\vec\bm
\let\ten\bm

\begin{frontmatter}

\title{Unsupervised full-field Bayesian inference of orthotropic hyperelasticity \\from a single biaxial test: a myocardial case study}

\author[inst1]{Rogier P. Krijnen}
\author[inst2]{Akshay Joshi}
\author[inst3]{Siddhant Kumar}
\author[inst1]{Mathias Peirlinck\corref{cor1}}
\affiliation[inst1]{organization={Dept. of BioMechanical Engineering, Delft University of Technology, Netherlands}}
\affiliation[inst2]{organization={Dept. of Mechanical Engineering, Indian Institute of Science, Bengaluru, India}}
\affiliation[inst3]{organization={Dept. of Materials Science and Engineering, Delft University of Technology, Netherlands}}

\begin{abstract}
Cardiac muscle tissue exhibits highly non-linear hyperelastic and orthotropic material behavior during passive deformation. 
Traditional constitutive identification protocols therefore combine multiple loading modes and typically require multiple specimens and substantial handling. 
In soft living tissues, such protocols are challenged by inter- and intra-sample variability and by manipulation-induced alterations of mechanical response, which can bias inverse calibration.
In this work we exploit spatially heterogeneous full-field kinematics as an information-rich alternative to multimodal testing.
We {\secondrevisionst{adapt }}{\secondrevision{recast}} EUCLID, an unsupervised method for the automated discovery of constitutive models, towards Bayesian parameter inference for highly nonlinear, orthotropic constitutive models.
Using synthetic myocardial tissue slabs, we demonstrate that a single heterogeneous biaxial experiment, combined with sparse reaction-force measurements, enables robust recovery of Holzapfel–Ogden parameters with quantified uncertainty, across multiple noise levels.
The inferred responses agree closely with ground-truth simulations and yield credible intervals that reflect the impact of measurement noise on orthotropic material model inference.
Our work supports single-shot, uncertainty-aware characterization of nonlinear orthotropic material models from a single biaxial test, reducing sample demand and experimental manipulation.
\end{abstract}
\begin{keyword}
material model inference \sep
full-field kinematics \sep
uncertainty quantification \sep
anisotropic hyperelasticity \sep
experimental tissue testing \sep
Bayesian inference
\end{keyword}

\cortext[cor1]{Correspondence:}
\ead{mplab-me@tudelft.nl}

\end{frontmatter}

\section{Introduction}
Soft biological tissues exhibit complex mechanical responses due to their fibrous, anisotropic microstructure. 
Accurately characterizing this behavior is essential for constructing predictive constitutive models that support simulation-driven design, diagnostics, and treatment planning \citep{Peirlinck2021,Niederer2019}.
However, unlike engineered materials, soft tissues are subject to considerable biological variability, both within a patient, across a patient cohort, and across experimental studies \citep{Reeps2012,Roccabianca2014,Budday2017,Matous2017,Famaey2026}. 
This intra- and inter-sample variability is even more pronounced in the myocardium, where strong spatial gradients in myofiber orientation and laminar architecture govern regional mechanical function \citep{Rohmer2007, Wilson2022}. Here, we \uline{investigate the impact that the underlying microstructural organization variability has on the mechanical behavior of a single sample.}

Traditional approaches to constitutive modeling of such tissues typically rely on multiple mechanical tests across different deformation modes including uniaxial tension or compression, biaxial tension, or shear testing \citep{Dokos2002,Sommer2015,Kakaletsis2021}. 
These protocols often require combining data from different specimens and, as such, ignore and smooth out intrinsic intra- and inter-sample variability into an average response \citep{Staber2017}.
Moreover, these protocols involve strong assumptions, such as spatially homogeneous stress and strain fields within the region of interest \citep{Holzapfel2009, Guan2018, Martonova2024}. 
{\revision{Such assumptions are problematic for living tissues, where spatial heterogeneity is the norm, and the mechanical response may depend on subtle differences in fiber orientation, geometry, inter-sample variable stiffness, or mounting.}}
In addition, invasive sample manipulations, such as cutting, clamping, and gluing, can alter the intrinsic mechanical behavior that these tests seek to measure \citep{Fehervary2018, Alloisio2024, Vervenne2026}.

These challenges have prompted the development of single-sample, single-experiment methodologies that aim to infer full constitutive behavior from a single loading protocol. 
This shift has been enabled by advances in full-field deformation measurements, such as digital image correlation and imaging-based 3D kinematics \citep{Wang2021,Jailin2024,Meng2025}. 
When applied to samples undergoing spatially heterogeneous deformation, such data enable the inference of multiple material parameters simultaneously, reducing the need for multiple loading modes. 
Indeed, unsupervised identification techniques, such as the virtual fields method (VFM) \citep{Grediac2006,Pierron2012}, finite element model updating (FEMU) \citep{Kavanagh1971,Peirlinck2018b,Elouneg2021}, variational systems identification \citep{Wang2021}, the equilibrium gap method (EGM) \citep{Claire2004,Peyraut2025}, and EUCLID \citep{Flaschel2021, Thakolkaran2022,Joshi2022,Thakolkaran2025}, demonstrate that heterogeneous strain states, once seen as a nuisance, can be leveraged as rich sources of information, obviating the need for homogenizing assumptions. In this study, we 
\uline{assess the extent to which a simple biaxial tensile testing protocol on a single myocardial tissue slab can support identification of the underlying orthotropic behavior}.

Yet, a key limitation persists in most current workflows: the lack of uncertainty quantification. 
Material parameters are frequently reported as point estimates, with little accounting for the inherent measurement noise or biological variability. 
In the context of myocardial modeling, where even adjacent samples can differ markedly in stiffness and fiber orientation \citep{Dokos2002, NiellesVallespin2017}, such deterministic inference may be misleading \citep{Hauseux2018}. 
Recent work has begun to address this gap through Bayesian approaches to inverse modeling, which offer not only best-fit estimates but also posterior distributions reflecting uncertainty in data and model form \citep{Ozturk2021,Joshi2022,Aggarwal2023,Peshave2024,Anton2024,Tac2024,Linka2025}.
Within the unsupervised setting, Bayesian-EUCLID forms an interesting contribution towards the inference of myocardial tissue behavior from full-field displacement fields and force measurements.
In prior work, Joshi et al. demonstrated the feasibility of using Gibbs sampling for full-field inference in two-dimensional problems \citep{Joshi2022}, where the material response could be inferred from a linear system of equations for the unknown material parameters in a rich constitutive feature library. 

In contrast, the setting of this study involves two distinct and compounding challenges. 
First, the underlying constitutive models are highly nonlinear \citep{Costa2001,Holzapfel2009,Martonova2024}, with strongly cross-correlated parameters \citep{Guan2018,Laita2025} that give rise to complex posterior landscapes. 
Second, we transition to three-dimensional full-field data.
While the latter provides richer information, it also substantially increases the computational cost of evaluating the likelihood function.
These factors independently and jointly limit the practical utility of Markov Chain Monte Carlo (MCMC) methods, including Gibbs sampling, due to the high cost of repeated likelihood evaluations and the poor mixing behavior in high-dimensional, correlated spaces. 
{\revision{These limitations }}\uline{{\revision{motivate}} adapting the Bayesian EUCLID framework with more scalable inference strategies, such as Stochastic Variational Inference (SVI)}, which approximates posterior distributions through optimization rather than sampling, enabling \uline{tractable inference in settings with both complex models and large data volumes}.

{\revision{Towards these goals, this study presents an extensive numerical study using an unsupervised, full-field, Bayesian framework for the inference of orthotropic hyperelastic material parameters from a single synthetic heterogeneous biaxial tensile test.}}
Our approach builds on the Bayesian-EUCLID framework \citep{Joshi2022}, extended via SVI to capture parametric uncertainty in a highly nonlinear, a priori chosen constitutive model form. 
The myocardium is modeled using the well-established Holzapfel--Ogden constitutive law \citep{Holzapfel2009, Avazmohammadi2019}, and heterogeneity in the deformation field is introduced either through microstructural variation (i.e., intrinsic fiber-sheet-interlaminar architectural heterogeneity), geometric tissue sample modification, or both. 
We demonstrate that the spatial richness of the {\revision{synthetic}} heterogeneous displacement field combined with reaction force data enables robust parameter inference with uncertainty quantification from a single test. 
We additionally validate the inferred material models using unseen invariant values and assess sensitivity to displacement noise.
{\revision{Finally, we restrict attention to a controlled numerical setting, which allows us to isolate the effects of noise, heterogeneity, and model nonlinearity on inference performance, factors that are typically confounded in experimental data.}}
\uline{Overall, this work highlights the potential of full-field, unsupervised, and uncertainty-aware approaches to redefine how we efficiently characterize complex biological tissues, taking into account minimal tissue availability and minimizing tissue manipulation.}

\section{Unsupervised Stochastic Variational constitutive parameter inference}

\begin{figure}[h]
    \centering
    \includegraphics[width=1.0\linewidth]{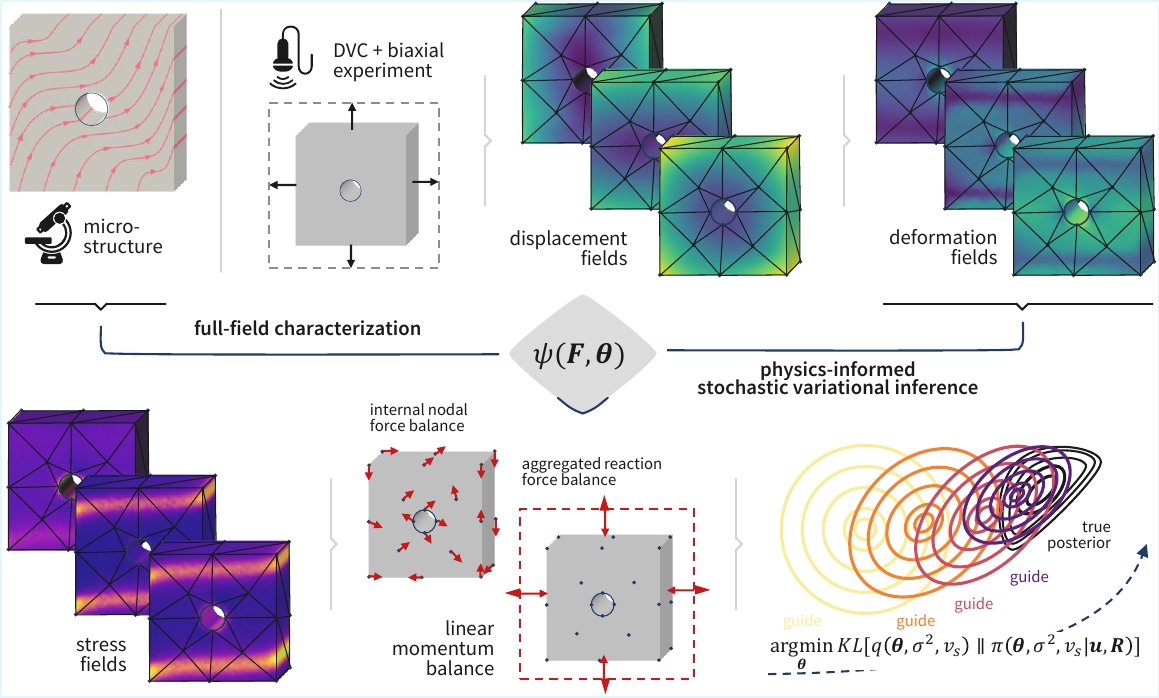}
    \caption{\textbf{Schematic overview of our unsupervised full-field stochastic variational inference framework for orthotropic hyperelastic tissue behavior.}
    Quantification of the tissue's microstructural organization, such as spatially varying fiber orientations, provides essential architectural information across the sample domain. In parallel, point-wise measurements of displacements $\vec{u}$ and reaction forces $\vec{R}$ are acquired under quasi-static biaxial loading using digital volume correlation. The sample geometry is discretized into a finite element mesh, enabling reconstruction of continuous deformation fields $\vec{F}$ from the measured data.
    A nonlinear orthotropic constitutive model $\psi$, parameterized by a set of cross correlated material parameters $\vec{\theta}$, serves as the mechanistic basis for inference. Given the reconstructed deformation and microstructural organization fields, the model predicts stress responses at the element level, which are used to compute internal and external nodal forces.
    Residuals are formulated through the weak form of the conservation of linear momentum, minimized pointwise for unconstrained degrees of freedom and in aggregate at boundaries with known reaction forces. The inverse problem is cast as a stochastic variational inference task, where a variational guide distribution $q(\vec{\theta})$ approximates the true posterior $\pi\left(\vec{\theta}, \sigma^2, v_s \mid \vec{u}, \vec{R}\right)$ by minimizing the Kullback-Leibler divergence $\mathrm{KL} \left(q \left( \vec{\theta} \right) \, || \, \pi\left(\vec{\theta}, \sigma^2, v_s \mid \vec{u}, \vec{R}\right) \right)$. Our physics-informed constitutive inference framework enables efficient and scalable inference from complex high dimensional full-field deformation datasets, accommodating nonlinear material behaviors and cross-correlated parameter effects.}
    \label{fig:overview}
\end{figure}

Conventional approaches for characterizing myocardial material behavior typically rely on spatially averaged measurements and simplifying assumptions of spatially homogeneous microstructural organization. 
{\revision{However, this intrinsic homogenization obscures the spatial heterogeneity present within a tissue sample.}}
To address this, we infer the constitutive parameter vector in an unsupervised setting from experimentally accessible data: 
(i) global reaction forces,
(ii) full-field deformation, and 
(iii) microstructural organization across the slab. 
In this work, we assume digital volume correlation (DVC) data derived from concomitant through-thickness imaging, such as ultrasound \citep{Navy2025}, Magnetic Resonance Imaging \citep{Estrada2020}, or X-Ray Imaging \citep{Davis2024} to obtain three-dimensional deformation fields. 
{\revision{In parallel, microstructural maps can be obtained via SHG or micro-CT \citep{Alberini2024,Kakaletsis2021,Maes2022}.}}
Among established inverse constitutive identification strategies such as FEMU, VFM, and EGM, we {\revision{adopt an FE-based weak-form residual formulation building on}} EUCLID \citep{Flaschel2021,Joshi2022}.
{\revision{In contrast to FEMU, which requires repeated finite element solves for each parameter update and load step, we evaluate the weak form of momentum balance directly from the measured displacement field (see Section~\ref{ssec:euclidprelim}), thereby avoiding repeated forward equilibrium solves and suggesting an expected computational advantage, particularly for high-dimensional or highly nonlinear problems.}} 
{\revision{The present formulation is related in spirit to both VFM and EGM as full-field alternatives to FEMU. 
It differs from classical VFM in that it uses the Bubnov--Galerkin finite-element test space induced by the reconstruction mesh rather than a reduced set of hand-crafted or parameter-specific virtual fields, and from EGM in that it relies on the weak rather than strong form of momentum balance, thereby reducing sensitivity to noise in the measured kinematics.}}
Building on this residual formulation, {\revision{we extend the method to a fully three-dimensional setting with nonlinear orthotropic materials and recast the classic deterministic parameter optimization into an unsupervised Bayesian inference problem.
Leveraging Stochastic Variational Inference (SVI)}}, 
we infer posterior distributions over the constitutive parameters while accounting for uncertainty in the equilibrium residuals induced by noisy displacement and reaction-force measurements. 
We identify orthotropic hyperelastic parameters from a single biaxial test; the overall workflow is summarized in Figure~\ref{fig:overview} and detailed in Section~\ref{ssec:stochvarinf}.

\subsection{EUCLID preliminaries}
\label{ssec:euclidprelim}
We first define the kinematic and force data available from a single experiment, and then derive the equilibrium residuals that underpin deterministic parameter identification.
Consider a myocardial tissue specimen undergoing quasi-static deformation in a three-dimensional reference domain $\Omega \subset \mathbb{R}^3$. 
Depending on the tissue slab's extraction orientation protocol (see Section~ \ref{sec:synthdatagen}), the underlying myocardial microstructure variation induces diverse and heterogeneous strain states.
In this work, this heterogeneity enters through spatially varying structural fields, while the constitutive parameter vector $\vec{\theta}$ is assumed to be spatially homogeneous and thus constant over $\Omega$.
This separation attributes spatial variability to the structural inputs while keeping the inverse problem focused on a single parameter vector.
Optionally, additional tissue manipulations, e.g., cutting a hole in the tissue slab or local cardiac ablation, can further promote a highly heterogeneous deformation state during biaxial tensile testing.

Boundary conditions are applied such that Dirichlet conditions are enforced on $\partial \Omega_u \subset \partial \Omega$ and Neumann conditions on the remainder, $\partial \Omega_t = \partial \Omega \setminus \partial \Omega_u$. 
For simplicity, our analysis focuses on displacement-controlled loading (i.e., Dirichlet boundary conditions), while noting that applied forces in load-controlled scenarios are equivalent to reaction forces under displacement control. The dataset comprises $n_t$ snapshots of displacement measurements,
\begin{equation}
\left\{ \vec{u}^{a,t} \in \mathbb{R}^3 : a=1,\dots,n_n;\ t=1,\dots,n_t \right\},
\end{equation}
recorded at $n_n$ reference points comprising the measurement node set:
\begin{equation}
\left\{ \vec{X}^a \in \Omega : a=1,\dots,n_n \right\}.
\end{equation}
Additionally, for each snapshot, $n_\beta$ reaction forces 
\begin{equation}
\left\{ R^{\beta,t} \in \mathbb{R} : \beta=1,\dots,n_\beta;\ t=1,\dots,n_t \right\}
\end{equation}
are measured at selected Dirichlet boundaries (e.g., using load cells). 
Since $n_\beta \ll n_n$, the reaction force data are sparse compared to the dense displacement field.
For brevity, the superscript $(\cdot)^t$ is omitted in the subsequent discussion, although the numerical procedure is applied independently to every snapshot.
Given $\left\{\vec{u}^{a} \right\}$ and $\left\{ R^{\beta} \right\}$, our goal is to infer the parameter set $\vec{\theta}$  of an a priori assumed constitutive model $\psi(\ten{F},\vec{\theta})$. 

To connect the measurements to equilibrium, we first reconstruct a continuous kinematic field and its deformation gradient.
The reference domain $\Omega$ is discretized using linear tetrahedral elements, evaluated with a single quadrature point at its barycenter, yielding 
\begin{equation}\label{disp-field-approx}
\vec{u}(\vec{X}) = \sum_{a=1}^{n_n}N^a(\vec{X})\ \vec{u}^{a}.
\end{equation}
Here, $N^a : \Omega \to \mathbb{R}$ represents the shape function associated with the measurement node $\vec{X}^a$. 
We subsequently approximate the corresponding deformation gradient field as
\begin{align}\label{eq:defgrad}
\ten{F}(\vec{X}) = \ten{I} + \sum_{a=1}^{n_n}\vec{u}^{a}\otimes \nabla N^a(\vec{X}),
\end{align}
where $\ten{I}$ is the identity matrix, and $\nabla$ represents the gradient operator with respect to the reference coordinates $\vec{X}$.

Assuming hyperelasticity, i.e., no dissipative energy losses within the material, the first Piola Kirchhoff stress follows as
\begin{equation}\label{eq:PderivF}
    \ten{P}(\ten{F}) = \frac{\partial \psi(\ten{F}, \vec{\theta})}{\partial \ten{F}}
\end{equation}
as the constitutive relation between the first Piola Kirchhoff stress $\ten{P}$ and the deformation gradient $\ten{F}$ defined through the free energy function $\psi$ and parameterized using the constitutive parameter set $\vec{\theta}$ \cite{Planck1897,Coleman1959,Peirlinck2024}.
In particular, $\ten{P}$ depends on the unknown parameters through $\psi(\ten{F}, \vec{\theta})$.

We leverage the conservation of linear momentum to guide the learning of the constitutive model parameter set, which eliminates the need for stress labels \cite{Flaschel2021,Thakolkaran2022}. 
Assuming quasi-static loading conditions and negligible body forces, the weak form of the linear momentum balance in our reference domain $\Omega$ is given by
\begin{equation}\label{eq:weak-form}
    \int_{\Omega} \ten{P}(\ten{F}, \vec{\theta}) : \nabla\vec{v}\, \mathrm{d}V - \int_{\partial\Omega_t} \bar{\ten{t}} \cdot \vec{v}\, \mathrm{d}S = 0 \quad \forall \ \text{admissible } \vec{v}\,,
\end{equation}
where $\bar{\ten{t}}$ denotes the prescribed traction acting on $\partial\Omega_t$ per unit reference area and $\vec{v}$ is a test function that is sufficiently regular and vanishes on the Dirichlet boundary $\partial\Omega_u$. 
We prefer the weak formulation over the strong form, as it avoids the need for second order derivatives with respect to the displacement field, which are sensitive to measurement noise. 
In our displacement-controlled setting, non-instrumented free surfaces are typically traction free (so $\bar{\ten{t}} = 0$ there).
Measured boundary forces enter through reactions on $\partial\Omega_u$, which we incorporate below.

Let $\mathcal{D} = \{(a,i): a=1,\dots,n_n;\ i=1,2,3\}$ denote all the displacement degrees of freedom in our domain $\Omega$, which we partition into $\mathcal{D}^\text{free}$, the unconstrained degrees of freedom, and $\mathcal{D}^\text{fix}_\beta$ (with $\beta=1,\dots,n_\beta$) denoting the degrees of freedom under Dirichlet constraints that contribute to the observed reaction force $R^\beta$.
Using a Bubnov–Galerkin discretization, we approximate the admissible test function $\vec{v}$ as
\begin{equation}
\vec{v}(\vec{X}) = \sum_{a=1}^{n_n} N^a(\vec{X})\,\vec{v}^a,\qquad \text{with} \quad  v_i^a = 0 \quad \forall\ (a,i)\in\bigcup_{\beta=1}^{n_\beta} \mathcal{D}^\text{fix}_\beta\,,
\end{equation}
and introducing our constitutive relation \eqref{eq:PderivF}, our weak form \eqref{eq:weak-form} reduces to
\begin{equation}\label{eq:reducedWeakForm}
    \sum_{a=1}^{n_n} v_i^a f_i^a = 0\,, \quad \text{with} \quad f_i^a = \underbrace{\int_{\Omega} P_{ij}\,\nabla_j N^a\,\mathrm{d}V}_{\text{internal force}} - \underbrace{\int_{\partial\Omega_t} \bar{t}_i N^a\,\mathrm{d}S}_{\text{external force}}\,.
\end{equation}
The integrals are evaluated by numerical quadrature on the same discretization as \eqref{disp-field-approx}.
Since the test functions are arbitrary on $\mathcal{D}^\text{free}$, the force residual must vanish at every unconstrained degree of freedom,
\begin{equation}\label{eq:free-constraints}
    f_i^a = 0 \quad \forall\ (a,i) \in \mathcal{D}^\text{free}\,.
\end{equation}
At the fixed degrees of freedom, the internal and external forces are counteracted by the reaction force imposed by the Dirichlet constraints. 
Because point-wise reactions are not accessible experimentally, we treat them as unavailable. 
Instead, we use the measured reaction forces integrated over instrumented boundary segments, which yields the global balance constraint
\begin{equation}\label{eq:fix-constraints}
\sum_{(a,i)\in \mathcal{D}^\text{fix}_\beta} f^a_i = R^\beta \qquad \forall \qquad \beta=1,\dots,n_\beta,
\end{equation}
where the summation is carried out over all point-wise forces associated with the degrees of freedom in the $\beta^\text{th}$ Dirichlet constraint, denoted as $\mathcal{D}^\text{fix}_\beta$.

Together, \eqref{eq:free-constraints} and \eqref{eq:fix-constraints} constitute a system of equations that defines the inverse problem of constitutive model inference \citep{Flaschel2021,Thakolkaran2022,Thakolkaran2025}.
We deterministically solve for $\vec{\theta}^{\star}$ by minimizing the equilibrium residuals over all snapshots:
\begin{equation}\label{eq:detloss}
    \vec{\theta}^{\star} \leftarrow \arg\min_{\vec{\theta}} \sum_{t=1}^{n_t} \Bigg[
    \underbrace{\sum_{(a,i)\ \in \ \mathcal{D}^\text{free}} \left(f_i^{a,t}\right)^2}_{\text{free degrees of freedom}}
    + \underbrace{\sum_{\beta=1}^{n_\beta} \Bigl(\lambda_{r} R^{\beta,t} - \sum_{(a,i)\ \in\ \mathcal{D}^\text{fix}_\beta} \lambda_{r} f_i^{a,t}\Bigr)^2}_{\text{fixed degrees of freedom}}
    \Bigg]\,,
\end{equation}
such that the displacement and reaction force data satisfy the physics-based constraints \eqref{eq:free-constraints} and \eqref{eq:fix-constraints}.

\subsection{Stochastic variational inference}
\label{ssec:stochvarinf}
In this study, we assume an a priori known constitutive law \eqref{eq:HO-model} and \secondrevision{aim to infer $\pi \left( \vec{\theta}, \sigma^2, v_s \, | \, \vec{u},\vec{R} \right)$, i.e. the posterior distribution of the constitutive parameter set $\vec{\theta}$ given the data $\{\vec{u}, \vec{R} \}$.} 
Leveraging Bayes' theorem, we combine prior beliefs about $\vec{\theta}$ with a likelihood model that accounts for the impact of measurement noise on the physics-based residual constraints conditioned on the displacement and reaction force data.

{\revision{
We recast the deterministic optimization problem in \eqref{eq:detloss} within a Bayesian inference framework by introducing the forward model $\vec{g}${\secondrevision{, a convenient and vectorized form of $f^{a}_{i}$ from \eqref{eq:reducedWeakForm}}},
and define the measured set of reaction forces, $\vec{R}$, as

\begin{equation}
\begin{aligned}
    \vec{g}^t(\vec{\theta}) =
    \begin{bmatrix}
        \left\{f_i^{a,t}\right\}_{(a,i) \in \mathcal{D}^\text{free}} \\
        \left\{\lambda_{r}\sum_{(a,i) \in \mathcal{D}^\text{fix}_\beta} f_i^{a,t}\right\}_{\beta = 1}^{n_\beta}
    \end{bmatrix}, & \quad 
    \vec{R}^t =
    \begin{bmatrix}
        \left\{0\right\} \\
        \left\{\lambda_{r}R^{\beta, t} \right\}_{\beta = 1}^{n_\beta}
    \end{bmatrix}, 
\end{aligned}
\end{equation}

for a given snapshot $t=1,\ldots,n_t$.
We stack all predicted snapshots and measured forces into two global vectors
\begin{equation}
\begin{aligned}
    \vec{g}(\vec{\theta})=\left[\begin{array}{c}
    \vec{g}^1(\vec{\theta}) \\
    \vdots \\
    \vec{g}^{n_t}(\vec{\theta})
\end{array}\right], \quad \vec{R}=\left[\begin{array}{c}
    \vec{R}^1 \\
    \vdots \\
    \vec{R}^{n_t}
\end{array}\right]
\end{aligned}
\end{equation}

This construction removes the need to carry superscript $(\cdot)^t$ in the subsequent probabilistic expressions while still retaining the multi-snapshot information.
Note that $\vec{g}(\vec{\theta})$ implicitly depends on observed displacements through the constitutive response leading to the predicted forces in $f_i^{a}$.

We assume that the difference between the predicted and measured forces follow a multivariate normal distribution with zero mean and isotropic variance, reflecting the assumption of additive Gaussian noise:

\begin{equation}
    \vec{R} = \vec{g}(\vec{\theta}) + \vec{\epsilon}, \quad \vec{\epsilon} \sim \mathcal{N}(\vec{0}, \sigma^2 \ten{I})
\end{equation}

which yields the likelihood
\begin{equation}\label{eq:likelihood}
    \pi \left(\vec{R} \mid \vec{\theta}, \sigma^2, \vec{u} \right) \, \propto \,
     \exp\left( -\frac{1}{2\sigma^2} \|\,\vec{g}(\vec{\theta}) - \vec{R}\,\|^2 \right)
\end{equation}
}}

In our case, both displacement field and reaction force measurements are affected by sensor noise, and we treat $\sigma^2$ as an effective residual variance that captures this combined uncertainty.
Given that displacement propagates in a nonlinear fashion to the force residuals, we expect the residual variance to be dominated by displacement noise.
{\secondrevision{We note here that the noise model adopted in the likelihood is an assumption, in accordance with related studies \citep{Joshi2022, Thorat2025}.
In practice, the true residual noise distribution could exhibit a more complex structure. }}

\begin{figure}[H]
    \centering
    \includegraphics[width=0.96\linewidth]{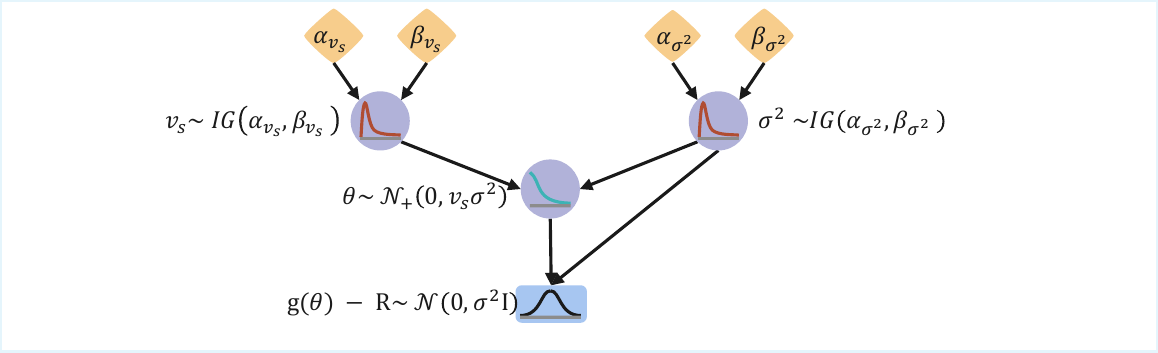}
    \caption{\textbf{Schematic of the hierarchical inference model.} 
    This hierarchical Bayesian model places priors on both the residual variance in the likelihood and the constitutive model parameters. The red circular distributions represent hyperpriors for $v_s$ and $\sigma^2$, the light-blue circular distribution represents the prior on the material model parameters, and the black rectangular distribution represents the likelihood induced by the linear momentum balance residuals. The black arrows denote hierarchical dependencies. The hyperpriors are inverse-gamma distributed, the constitutive parameters are distributed according to a truncated multivariate normal distribution, and the likelihood is assumed to be normally distributed.
    }
    \label{fig:hierarchy}
\end{figure}

Next, we place priors on the residual noise variance $\sigma^2$ and the parameter vector $\vec{\theta}$.
{\revision{Typically, such variance is interpreted as a measurement error, but it can also consist of solver tolerance or discretization error especially when no measurement errors are assumed.}}
Since the variance of the measurement noise $\sigma^2$ is unknown and must be inferred, we assign an inverse-gamma prior:
\begin{equation}\label{eq:prior01}
    \sigma^2 \sim \mathcal{IG} \, \left( \alpha_\sigma, \beta_\sigma \right)
\end{equation}
Concomitantly, we enforce physical constraints on the parameters by requiring them to be positive and therefore enforce a truncated normal distribution for the prior (truncated on the left at zero). Here, we introduce a scaling hyperprior $v_s$ to decouple the overall measurement residual noise variance $\sigma^2$ from the underlying parameter prior variance:    
\begin{equation}\label{eq:prior02}
    \vec{\theta} \sim \mathcal{N}_+ \, \left(\vec{0}, v_s \, \sigma^2 \ten{I} \right)
\end{equation}
We place an inverse-gamma hyperprior on $v_s$:
\begin{equation}\label{eq:prior03}
    v_s \sim \mathcal{IG} \, \left( \alpha_{v_s}, \beta_{v_s} \right)
\end{equation}
The hierarchical structure \citep{Joshi2022, Thorat2025, Peirlinck2019} of our prior distributions is showcased in Figure \ref{fig:hierarchy}.

Combining the prior distributions \eqref{eq:prior01}--\eqref{eq:prior03} with the likelihood \eqref{eq:likelihood} yields the posterior:
{\secondrevision{
\begin{equation}\label{eq:posterior}
    \pi \left( \vec{\theta}, \sigma^2, v_{s} \mid \vec{u} , \vec{R} \right) \, 
    = 
    \frac{\pi \left(\vec{u},  \vec{R}, \vec{\theta}, \sigma^2, v_s \right) }{\pi \left ( \vec{u}, \vec{R}  \right )}
    = 
    \, \frac{\pi \left( \vec{R} \mid \vec{\theta}, \sigma^2, \vec{u} \right) \, \pi \left( \vec{\theta} \mid \sigma^{2},v_{s} \right) \, \pi \left( \sigma^2 \right) \, \pi \left( v_s \right)}{\pi \left ( \vec{u}, \vec{R}  \right )},
\end{equation}

}}
This posterior can, in principle, be approximated using \textit{sampling-based} inference or \textit{variational inference}. 
In simpler inference problems, sampling-based methods such as MCMC can often be relied on to approximate the posterior distribution. For example, Gibbs sampling has been successfully applied to two-dimensional full-field problems, where the material response could be inferred by solving a linear system of equations on a library of constitutive features \citep{Joshi2022}. 
In those cases, the constitutive models were potentially nonlinear in input but remained linear in their parameters, which keeps the posterior geometry relatively tractable for MCMC. 
In our setting, however, the constitutive models are nonlinear in both inputs and parameters, which induces strong parameter interactions and complex (often multimodal) posteriors. 
Together with the cost of evaluating data-heavy 3D full-field likelihoods, this makes conventional MCMC-based posterior exploration computationally unattractive in practice.

We therefore adopt stochastic variational inference (SVI) to approximate the posterior efficiently using gradient-based optimization.
SVI has been shown to substantially accelerate inference relative to MCMC in related settings \citep{Rodrigues2022,Thorat2025}. 
{\secondrevisionst{In SVI, we approximate the posterior distribution $\pi$ with a tractable family of parametric guide distributions $q$.}} 
{\secondrevision{More specifically, in SVI we approximate the posterior $\pi(\vec{\theta},\sigma^2,v_s \mid \vec{u},\vec{R})$ using a parametric guide family $q_{\vec{\Phi}} \left ( \vec{\theta}, \sigma^2, v_s \right )$.
To determine the best approximation within this variational family, we minimize the Kullback--Leibler (KL) divergence \citep{Kullback1968,Cover2006} between 
both distributions,
i.e.,
\begin{equation}
    \vec{\Phi}^{\star} \leftarrow \arg \min_{\vec{\Phi}} \operatorname{KL}\!\left(
    q_{\vec{\Phi}}(\vec{\theta}, \sigma^2, v_s)
    \,\|\, 
    \pi(\vec{\theta}, \sigma^2, v_s \mid \vec{u}, \vec{R})
    \right).
\label{eq:argminKL}
\end{equation}

Direct minimization of \eqref{eq:argminKL} is intractable because the posterior density depends on the marginal likelihood $\pi\left(\vec{u}, \vec{R}\right)$, as shown in \eqref{eq:posterior}.
Applying the definition of the KL divergence gives
\begin{equation}
\operatorname{KL}\!\left(
q_{\vec{\Phi}}(\vec{\theta}, \sigma^2, v_s)
\,\|\,
\pi(\vec{\theta}, \sigma^2, v_s \mid \vec{u}, \vec{R})
\right)
=
\mathbb{E}_{q_{\vec{\Phi}}}\!\left[
\log \frac{
q_{\vec{\Phi}}(\vec{\theta}, \sigma^2, v_s)
}{
\pi(\vec{\theta}, \sigma^2, v_s \mid \vec{u}, \vec{R})
}
\right].
\end{equation}
Substituting the posterior from \eqref{eq:posterior} and separating numerator and denominator inside the logarithm yields
\begin{equation}
\operatorname{KL}\!\left(
q_{\vec{\Phi}}(\vec{\theta}, \sigma^2, v_s)
\,\|\,
\pi(\vec{\theta}, \sigma^2, v_s \mid \vec{u}, \vec{R})
\right)
=
-\mathcal{L}(\vec{\Phi})
+
\log \pi(\vec{u}, \vec{R}),
\end{equation}
where
\begin{equation}
\mathcal{L}(\vec{\Phi})
:=
\mathbb{E}_{q_{\vec{\Phi}}}\!\left[
\log \pi(\vec{u}, \vec{R}, \vec{\theta}, v_s, \sigma^2)
-
\log q_{\vec{\Phi}}(\vec{\theta}, \sigma^2, v_s)
\right].
\end{equation}
is the evidence lower bound (ELBO). 
This reformulation is useful because the marginal likelihood $\pi\left(\vec{u}, \vec{R}\right)$, i.e. the evidence, no longer appears inside the optimization objective: it is constant with respect to $\vec{\Phi}$. As a result, minimizing the KL divergence in \eqref{eq:argminKL} is equivalent to maximizing the ELBO,
\begin{equation}
\vec{\Phi}^{\star}
\leftarrow
\arg \max_{\vec{\Phi}} \mathcal{L}(\vec{\Phi}).
\end{equation}
Equivalently, and using that the KL divergence is non-negative,
\begin{equation}
\log \pi(\vec{u}, \vec{R})
=
\mathcal{L}(\vec{\Phi})
+
\operatorname{KL}\!\left(
q_{\vec{\Phi}}(\vec{\theta}, \sigma^2, v_s)
\,\|\,
\pi(\vec{\theta}, \sigma^2, v_s \mid \vec{u}, \vec{R})
\right),
\end{equation}
which shows that $\mathcal{L}(\vec{\Phi})$ is indeed a lower bound on the log-evidence. 
In practice, this converts an intractable posterior-divergence minimization into a tractable variational optimization problem, since $\mathcal{L}(\vec{\Phi})$ depends only on the joint density $\pi\left(\vec{u}, \vec{R}, \vec{\theta}, v_s, \sigma^2\right)$ and the chosen guide distribution $q_{\vec{\Phi}}\left(\vec{\theta}, \sigma^2, v_s\right)$.}}


{\secondrevision{What remains is to choose a suitable guide distribution to approximate the posterior.}}
We employ a mean-field variational family $q_{\vec{\Phi}}$ \citep{Parisi1988,Consonni2007}, which assumes statistical independence between the quantities of interest,
\begin{equation}\label{eq:meanfieldapprox}
    q_{\vec{\Phi}} \left( \vec{\theta}, \, \sigma^2, \, v_s \right) = q \left( \vec{\theta} \right) \, q \left( \sigma^2 \right) \, q \left( v_s \right)
\end{equation}
with factors:
\begin{alignat}{3}
    &q \left( \vec{\theta} \right) \quad & = &\quad \mathcal{N}_+ \left(\vec{\mu}^q, \vec{\Sigma}^q \circ \ten{I}\right) \\
    &q \left( \sigma^2  \right) \quad     & = &\quad \mathcal{IG} \left(\alpha_{\sigma^2}^q, \beta_{\sigma^2}^q\right) \\
    &q \left( v_s \right) \quad          & = &\quad \mathcal{IG} \left(\alpha_{v_s}^q, \beta_{v_s}^q\right),
\end{alignat}
where $\circ$ denotes the element-wise Hadamard product. 
The full set of variational parameters is denoted by $\vec{\Phi} = \left\{\vec{\mu}^q,\vec{\Sigma}^q,\alpha_{\sigma^2}^q,\beta_{\sigma^2}^q,\alpha_{v_s}^q,\beta_{v_s}^q\right\}$, and hyperparameters are reported in Table \ref{tab:parameters}. 
We specifically highlight $\vec{\mu}^q$, and $\vec{\Sigma}^q$ to be the location and scale parameters of the truncated normal distribution.
These correspond to the mean and standard deviation of the underlying and untruncated normal distribution used in the definition of $\mathcal{N}_{+}$.
The entire inference pipeline is implemented in \texttt{NumPyro} \citep{Phan2019}, enabling automatic differentiation and stochastic optimization with ADAM{\revisionst{ using the reparameterization trick}} \citep{Kingma2017, Kingma2013}. 
Doing so, we alleviate the need to explicitly define the minimization procedure as was done in Thorat et al. \citep{Thorat2025}.
We initialize the optimizer with a learning rate of $r_{0}=0.01$, and apply an exponential learning rate decay $\gamma$, i.e. $r_{i} = r_{0} \gamma^{i/n_{T}}$ at epoch i with $n_{T}$ denoting the total number of epochs.
{\revision{The parameter $\lambda_{r}$ was heuristically set to $10$ after preliminary checks indicated stable parameter fitting behavior.}}
{\revision{A sensitivity study indicated that increasing the number of snapshots $n_t$ beyond two did not improve validation performance (see \ref{ap:loadstep-sensitivity}).}}

\begin{table}[H]
\centering
\caption{\textbf{Hyperparameters and simulation settings used in the stochastic variational inference framework and synthetic tissue slab data generation.} 
The upper rows list optimization settings and hierarchical Bayesian prior parameters used for the SVI guide distributions and model priors. 
The bottom rows summarize the geometry and loading conditions for the training and validation slab specimens, respectively.} 
\label{tab:parameters}
\begin{tabular}{lcl}
\hline
\multicolumn{1}{l}{\textbf{Parameter}} & \textbf{Notation} & \textbf{Value} \\ \hline
\textit{SVI hyperparameters:}\\
$\quad$ weight for reaction force balance & $\lambda_{r}$    & $10$\\
$\quad$ Optimizer  & $-$    & Adam  \\
$\quad$ Number of epochs  & $n_{T}$    & $20{,}000$  \\
$\quad$ Learning rate scheduler  & $-$    & exponential  \\
$\quad$ Base learning rate  & $r_{0}$    & $0.01$ \\
$\quad$ Learning decay rate  & $\gamma$    & $0.1$ \\

\textit{SVI guide hyperparameters:}\\
$\quad$ Initial mean of $\mathcal{N}_{+}$ distribution & $\mu_{\theta}$    & $10 + \epsilon^2 \times \{1,\dots, 1\}, \quad \epsilon \sim \mathcal{N}(0, 1)$ \\ 
$\quad$ Initial variance of $\mathcal{N}_{+}$ distribution & $\sigma^2_{\theta}$   & $10^{-18} \times \{1,\dots, 1\}$ \\ 
$\quad$ Initial scale parameter for $v_s$  & $\alpha_{v_s}$    & $10^{4}$ \\ 
$\quad$ Initial rate parameter for $v_s$  & $\beta_{v_s}$    & $10^{4}$ \\ 
$\quad$ Initial scale parameter for $\sigma^2$  & $a_{\sigma^2}$    & $10^{4}$ \\ 
$\quad$ Initial rate parameter for $\sigma^2$  & $b_{\sigma^2}$    & $10^{4}$ \\ 

\textit{SVI model prior hyperparameters:}\\
$\quad$ prior mean of $\mathcal{N}_{+}$ distribution & $\mu_{\theta}$    & $10^{-10} \times \{1,\dots, 1\}$ \\ 
$\quad$ prior variance of $\mathcal{N}_{+}$ distribution & $\sigma^2_{\theta}$  & $\{1,\dots, 1\}$ \\ 
$\quad$ prior scale parameter for $v_s$  & $\alpha_{v_s}$    & $1$ \\ 
$\quad$ prior rate parameter for $v_s$  & $\beta_{v_s}$    & $1$ \\ 
$\quad$ prior scale parameter for $\sigma^2$  & $\alpha_{\sigma^2}$    & $1$ \\ 
$\quad$ prior rate parameter for $\sigma^2$  & $\beta_{\sigma^2}$    & $1$ \\ 
\hline

\textit{Training slab specimen:}\\
$\quad$ Number of nodes in mesh for FEM-based data generation  & - & $16,856$  \\
$\quad$ Number of reaction force constraints & $n_{\beta}$   & $5$   \\ 
$\quad$ Number of data snapshots    & $n_{t}$     & 2 \\

$\quad$ Loading parameter & $\delta$  & $\{0.5 \times t: t=2,3\}$ \\
$\quad$ Loading ratios & $-$  & $\{(1:2), (1:1), (2:1)\}$\\
\hline

\textit{Ground-truth material model parameters $\vec{\theta}_{gt}$:}\\
$\quad$ Linear scaling parameters $a_{\star}$ {\revision{[MPa]}} & $a_{}, a_{f}, a_{n}, a_{fs}$ & {\revision{$\{0.809, 1.911,  0.227, 0.547\} \times 10^{-3}$}}\\ 
$\quad$ Exponential scaling parameters $b_{\star}$ [-] & $b_{}, b_{f}, b_{n}, b_{fs}$ & $\{7.474, 22.063, 34.802, 5.691\}$\\
$\quad$  Bulk modulus [MPa] & $K$ & $0.1$\\ 
\hline
\end{tabular}
\end{table}

\newpage
\section{Synthetic data generation}
\label{sec:synthdatagen}
To assess our Bayesian inference framework under controlled conditions, we generate synthetic biaxial tensile testing data from high-resolution simulations of varying microstructurally and geometrically heterogeneous myocardial tissue slabs.

\subsection{Myofiber architecture and tissue slab slicing}
Informed by prior histological characterizations of myocardial tissue \citep{Rohmer2007,Wilson2022,Holzapfel2009}, we model our studied specimens as a layered orthotropic material in which cardiomyocytes align along main myofiber directions, packed by endomysial collagen into sheets, and connected via perimysial collagen. 
This organization defines three locally orthogonal directions: the fiber direction $\vec{f}$, the sheet direction $\vec{s}$, and the sheet-normal direction $\vec{n}$. 
To investigate the role of heterogeneity in unsupervised constitutive parameter identifiability, we simulate myocardial tissue slabs that exhibit both \textit{microstructure}-induced and \textit{geometry}-induced heterogeneity.

\begin{figure}[h]
    \centering
    \includegraphics[width=1.0\linewidth]{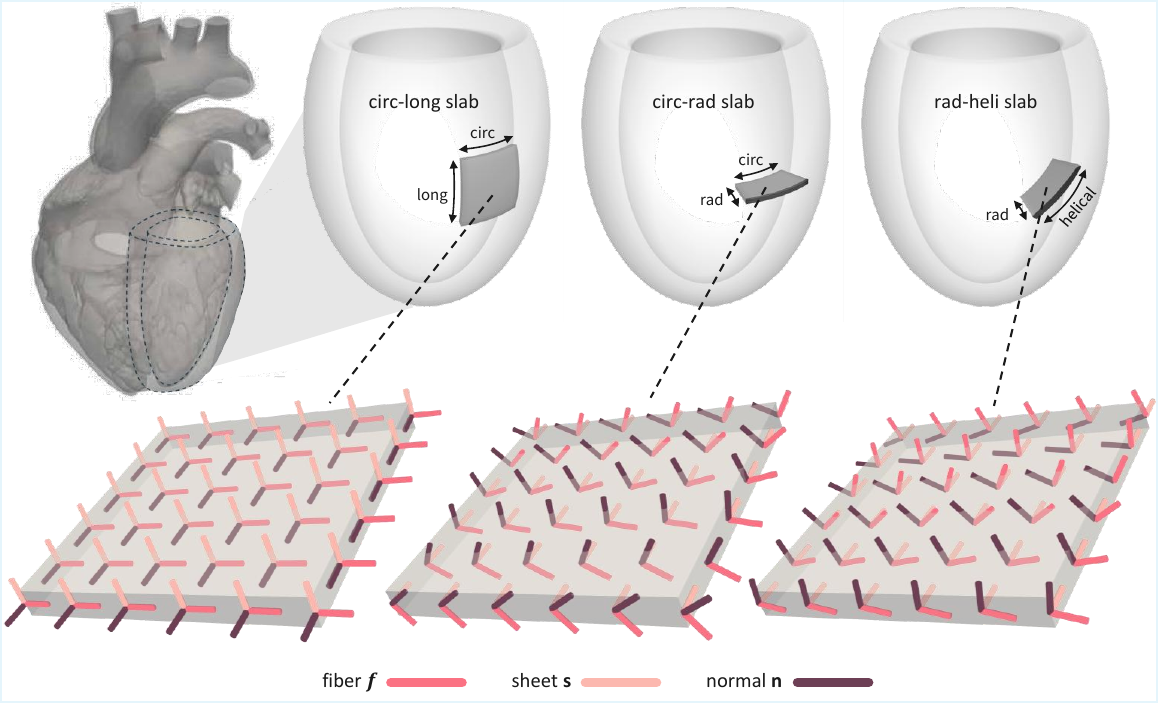}
    \caption{\textbf{Specimen slicing and microstructural organization.} Top: Schematic of the left ventricle and three considered slicing orientations to extract myocardial tissue slabs: circumferential–longitudinal (circ-long), circumferential–radial (circ-rad), and a rotated variant of the latter (rad-heli). Bottom: Each slicing orientation leads to a distinct microstructural configuration within the specimen. In circ-long slabs, the fiber direction lies in-plane and the sheet direction spans the through-thickness. Circ-rad slabs display a transmural fiber rotation from epicardium to endocardium, with the sheet direction aligned in the radial plane. The rad-heli orientation rotates the slab around the radial axis, aligning all three microstructural directions approximately within the plane of the specimen. Fiber directions (\(\vec{f}\)) are shown in red, sheet directions (\(\vec{s}\)) in pink, and normal directions (\(\vec{n}\)) in brown. 
    }
    \label{fig:myocardslicingsamples}
\end{figure} 

Leveraging the intrinsic spatially varying myofiber architecture in the human heart \citep{Lombaert2012}, we extract three microstructurally distinct synthetic tissue slabs from the left ventricular free wall. 
All samples are assumed to be $10 \times 10 \times 1$ mm$^3$ in size, representative of slabs from a healthy human left ventricular wall.
{\revision{These slab dimensions should be interpreted as an idealized benchmark geometry for controlled 3D full-field constitutive inference.}}
We further detail the sample geometry and applied boundary conditions in \ref{ap:appendix_tissueslab}.

Figure \ref{fig:myocardslicingsamples} showcases the microstructural variability across the slab domain, with directional fiber \(\vec{f}\), sheet \(\vec{s}\) and normal \(\vec{n}\) vectors discretized element-wise, and their spatial distribution depends on the chosen slicing orientation:
\begin{itemize}
    \item \textbf{circ-long:} sections aligned with the ventricular circumferential–longitudinal plane. The in-plane $\vec{f}$ orientation and through-thickness $\vec{s}$ direction enable near-planar fiber alignment. We consider both aligned and off-axis variants depending on their angular deviation from the square edge. This slab specimen type aligns with the biaxial tension testing slabs studied by \cite{Sommer2015}.
    
    \item \textbf{circ-rad:} sections cut in the circumferential–radial plane. These samples exhibit a transmural fiber rotation from approximately $-60^\circ$ at the epicardium to $+60^\circ$ at the endocardium, while the sheet direction $\vec{s}$ remains in-plane and radially aligned.

    \item \textbf{rad-heli:} samples based on a radial-helical cut. Given the natural transmural variation of the myofiber angle $\vec{f}$ from $-60^\circ$ in the sub-epicardium to $+60^\circ$ in the sub-endocardium \citep{Lombaert2012}, this configuration places all three microstructural directions ($\vec{f}, \vec{s}, \vec{n}$) approximately within the sample plane by rotating the sample an additional $45^\circ$ along the radial axis. The resultant fiber orientation ranges from $-15^\circ$ up to $105^\circ$, aligning the $\vec{f}$ fiber in the sample plane near the bottom edge of the sample and aligning the $\vec{n}$ fiber in the sample plane near the top edge.
\end{itemize}

To investigate the influence of geometric heterogeneity, we introduce a second set of specimens with a central circular occlusion with a radius of $1$ mm, see \ref{ap:appendix_tissueslab}. 
These samples are expected to introduce additional deformation heterogeneity that we hypothesize may improve constitutive parameter identifiability under limited data conditions, e.g., when inferring orthotropic behavior from a single biaxial tensile test.

\subsection{Biaxial loading emulation}
We simulate a digital volume correlation experiment by modeling a biaxial loading protocol on the microstructurally and geometrically homogeneous and heterogeneous tissue slabs described above using the finite element method (FEM) implemented in ABAQUS/Standard \citep{Abaqus}, see for example \citep{Peirlinck2018b}. 
Our virtual tissue slabs are subjected to displacement-controlled symmetric biaxial loading, with loading ratios $\lambda_1 : \lambda_2 = (1:2), (1:1), (2:1)$ defined in terms of the prescribed edge stretches along the two in-plane loading directions. 
The maximum stretch applied to the samples is $15\%$ of the initial edge length, consistent with maximal values used in experiments \citep{Sommer2015}.
We use linear tetrahedral elements to discretize the domain and record nodal displacements along with reaction forces for a sequence of $n_t$ load steps (see Figure \ref{fig:geometry_and_bcs})

To define the constitutive response, we use the local volume change $J = \det \left( \ten{F} \right)$ and multiplicatively decompose the deformation gradient $\ten{F}$ \eqref{eq:defgrad} into its volumetric $\ten{F}^{\rm{vol}}$ and isochoric $\bar{\ten{F}}$ parts \citep{Flory1961},
\beq
\ten{F} 
= \ten{F}^{\rm{vol}} \bar{\ten{F}}
\quad \mbox{with} \quad
\ten{F}^{\rm{vol}} = J^{\frac{1}{3}}\ten{I}
\quad \mbox{and} \quad
\bar{\ten{F}} = J^{-\frac{1}{3}}\ten{F}.
\label{eq:volsplit}
\eeq
We then introduce the following deformation (pseudo-)invariants \citep{Spencer1984, Menzel2004}:
\beq
\begin{aligned}
    &\bar{I}_1 & = \quad &[\bar{\ten{F}}^{\scas{T}}  \bar{\ten{F}} ] : \ten{I} \\
    &\bar{I}_{\rm{4ff}} & = \quad &[\bar{\ten{F}}^{\scas{T}} \bar{\ten{F}} ] : [\vec{f}^0 \otimes \vec{f}^0] \\
    &\bar{I}_{\rm{4nn}} & = \quad &[\bar{\ten{F}}^{\scas{T}} \bar{\ten{F}} ] : [\vec{n}^0 \otimes \vec{n}^0] \\
    &\bar{I}_{\rm{4fs}} & = \quad &[\bar{\ten{F}}^{\scas{T}} \bar{\ten{F}} ] : [\vec{f}^0 \otimes \vec{s}^0]
\end{aligned}
\label{eq:invs}
\eeq
where $\vec{f}^0$, $\vec{s}^0$, $\vec{n}^0$ denote the local microstructure fiber, sheet, and normal unit vector orientations in the undeformed reference configuration.
We adopt the orthotropic, compressible Holzapfel-Ogden (HO) material model \citep{Holzapfel2009,Guan2018} using the following strain energy density function:
\begin{equation}
    \begin{aligned}
        \psi(\ten{F}, \vec{\theta}) &= \frac{a}{2b} \left[ \exp\left(b(\bar{I}_1 - 3)\right) - 1 \right]
        + \sum_{i \, = \, f,n} \frac{a_i}{2b_i} \left[ \exp\left(b_i\left<\bar{I}_{4ii} - 1\right>^2\right) - 1 \right] \\
        &+ \frac{a_{fs}}{2b_{fs}} \left[ \exp\left(b_{fs} \bar{I}_{4fs}^2\right) - 1 \right]
        + \frac{K}{2} \left( \frac{J^2 - 1}{2} - \ln J \right)
    \end{aligned}
    \label{eq:HO-model}
\end{equation}
to compute the First Piola Kirchhoff stress tensor $\ten{P}$ following \eqref{eq:PderivF}.
{\revision{Note that this variation from Guan et al. relies on the invariant $I_{4nn}$, whereas the original model used the invariant $I_{4ss}$.}}
This variation of the HO-model was implemented using the recently proposed universal user material subroutine in Abaqus \citep{Peirlinck2024}. 

In \eqref{eq:HO-model}, $\vec{\theta} = \{a, b, a_f, b_f, a_n, b_n, a_{fs}, b_{fs}, K\}$ (see Table \ref{tab:parameters}) denotes the to-be-inferred ground-truth material parameters. 
The Macaulay brackets $\langle x \rangle$ ensure that fiber and sheet term strain energy contributions activate only under tension \citep{Wittrick1965,Wriggers2016,Peirlinck2024}.
Lastly, a bulk modulus is chosen based on the distribution of $J$ in the circ-long-hom samples and common compressibility considerations of the myocardium \citep{McEvoy2018, Yin1996} to limit volumetric strains while retaining numerical stability.

\begin{figure}[!h]
\centering
\includegraphics[width=1.0\textwidth]{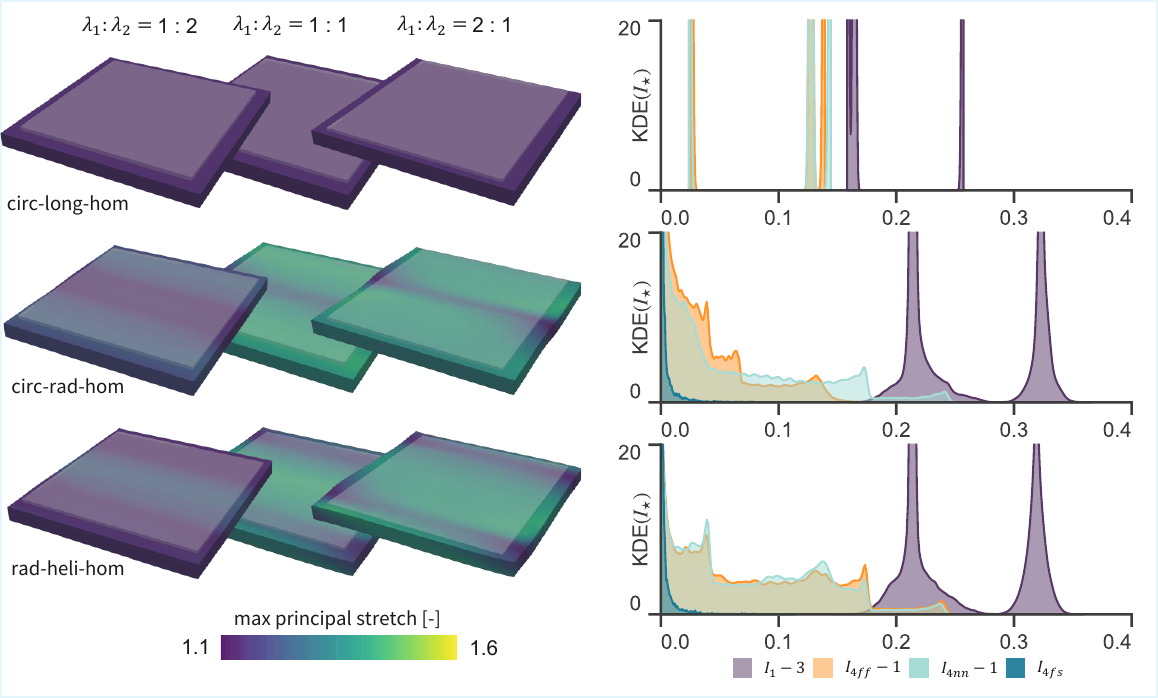}
\caption[]{\textbf{Deformation profiles of geometrically homogeneous tissue slabs with varying microstructural heterogeneity.}
Maximum principal stretches are shown for circ-long-hom (top row), circ-rad-hom (middle row), and rad-heli-hom (bottom row) specimens subjected to biaxial loading protocols ($\lambda_1 : \lambda_2 = (1:2), (1:1), (2:1)$) up to $\lambda^{\text{max}} = 1.15$.
Rightmost columns display distributions of invariants $I_1-3$, $I_{4ff}-1$, $I_{4nn}-1$, and $I_{4fs}$ for each specimen type respectively using approximated kernel density estimations denoted as $\mathrm{KDE}(I_*)$. 
The kernel density estimation is truncated at an arbitrary level of 20 to show the spread of the invariants more clearly.
}
\label{fig:synth_train_deform_hom}
\end{figure}

\begin{figure}[!h]
    \centering
    \includegraphics[width=1.0\textwidth]{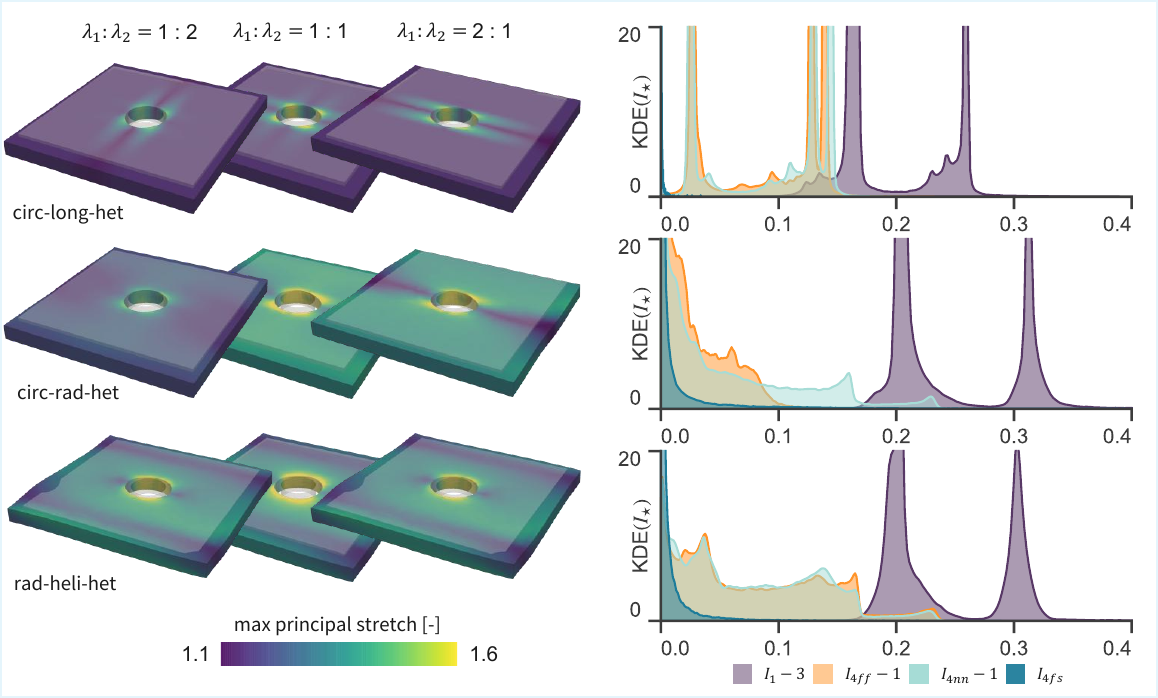}
    \caption[]{\textbf{Deformation profiles of geometrically and microstructurally heterogeneous tissue slabs.}
    Maximum principal stretches circ-long-het (top row), circ-rad-het (middle row), and rad-heli-het (bottom row) specimens under biaxial loading ($\lambda_1 : \lambda_2 = (1:2), (1:1), (2:1)$). 
    Invariant distributions of $I_1-3$, $\bar{I}_{4ff}-1$, $\bar{I}_{4nn}-1$, and $I_{4fs}$ are shown in the rightmost panels using approximated kernel density estimations denoted as $\mathrm{KDE}(I_*)$.
    The kernel density estimation is truncated at an arbitrary level of 20 to show the spread of the invariants more clearly.}
    \label{fig:synth_train_deform_het}
\end{figure}

To quantify how microstructural and geometric heterogeneity enrich the deformation space relevant for inference, Figures~\ref{fig:synth_train_deform_hom}--\ref{fig:synth_train_deform_het} report the resulting distributions of the key invariants activated under biaxial loading, which motivates our single-test identification setup.

Figure \ref{fig:synth_train_deform_hom} presents the deformation response of geometrically homogeneous slabs with varying microstructural architectures.
The qualitative results of the maximum principal stretch are shown for all loading conditions and homogeneous tissue samples.
On the right side of the figure we show kernel density estimations of the biaxially activated invariants, truncated at $20$ to visualize the spread of these invariants.
The \textit{circ-long-hom} configuration represents a classic homogenized myocardial sample, as excised by \citep{Sommer2015}, with assumed uniform fiber-sheet-normal orientations.
Under biaxial loading, this results in a uniform deformation pattern with isotropic {invariants $I_1$ ranging from $3.158$ to $3.257$} across the three loading protocols. 
Anisotropic contributions are moderate, {with $I_{4ff}$ and $I_{4nn}$ spanning ranges from $1.025$ to $1.139$ and $1.024$ to $1.144$ respectively}. 
Fiber-sheet shear contributions remain negligible, with $I_{4fs}$ limited to $0.000$ up to machine precision.
Introducing transmural fiber variation in the \textit{circ-rad-hom} slab increases local deformation heterogeneity. 
The isotropic deformation {invariant $I_1$ expands to a broader range of $3.160$ to $3.367$}, {while $I_{4ff}$ and $I_{4nn}$ reach $1.194$ and $1.252$, respectively}.
Notably, fiber-sheet shear contributions increase, with $I_{4fs}$ reaching up to $0.099$.
In the \textit{rad-heli-hom} specimen, which introduces a more asymmetric helical fiber distribution, deformation heterogeneity is further amplified.
The isotropic {invariant $I_1$ ranges from $3.147$ to $3.359$, while $I_{4ff}$ increases significantly, spanning up to $1.248$}.
{The $I_{4nn}$ component reaches $1.251$}, and fiber-sheet shear contributions grow to $0.114$.
The distributions of invariants shown in the right column of Figure \ref{fig:synth_train_deform_hom} emphasize these trends.
The circ-long-hom specimen exhibits single homogeneous singular values for $I_{4ff}$ and $I_{4nn}$, while $I_{4fs}$ remains zero.
In contrast, circ-rad-hom and rad-heli-hom configurations display broader distributions across all invariants, particularly for fiber-sheet shear deformations.
Consistent with these observations, we note tissue bulging effects in the thickness direction.
Circ-rad configurations exhibit bulging centrally, whereas rad-heli configurations deform near the specimen edges.
In particular, the rad-heli-hom configuration shows noticeable thickness bulging coinciding with bands of reduced strain, where fibers align in-plane.
Overall, microstructure-induced heterogeneity alone enriches the activated invariant space under a single biaxial protocol, which motivates using these slabs as training configurations in this study.

Figure \ref{fig:synth_train_deform_het} illustrates the deformation response of geometrically heterogeneous slabs, where a central occlusion is combined with varying microstructural organization.
In the \textit{circ-long-het} specimen, geometric heterogeneity alone induces significant deformation diversity. 
The isotropic {invariant $I_1$ spans from $3.070$ to $3.390$, with $I_{4ff}$ and $I_{4nn}$ reaching up to $1.232$ and $1.209$, respectively}. 
Fiber-sheet shear contributions, represented by $I_{4fs}$, remain limited, peaking at $0.034$.
Adding microstructural heterogeneity in the \textit{circ-rad-het} configuration further amplifies deformation heterogeneity. 
{Here, $I_1$ varies between $3.141$ and $3.5477$, while $I_{4ff}$ and $I_{4nn}$ attain maximum principal stretches of $1.218$ and $1.243$ respectively}. 
Notably, fiber-sheet shear contributions increase substantially, with $I_{4fs}$ rising to $0.4342$.
In the \textit{rad-heli-het} specimen, characterized by asymmetric helical fiber variation and geometric heterogeneity, {isotropic $I_1$ deformation spans $3.136$ to $3.518$. 
Anisotropic contributions reach $1.242$ for $I_{4ff}$, $1.241$ for $I_{4nn}$}, and $1.289$ for the fiber-sheet shear term $I_{4fs}$, which confirms this configuration as the most mechanically diverse.
The distributions of invariants reflect these differences. 
The circ-long-het slab shows moderate spreading compared to its homogeneous counterpart, but fiber-sheet shearing remains limited. 
In contrast, the circ-rad-het and rad-heli-het configurations exhibit pronounced broadening in all invariants, particularly in $I_{4fs}$, underscoring the synergistic effect of combined geometrical and microstructural heterogeneities.
Similar to the homogeneous cases, bulging effects are evident. 
Circ-rad configurations exhibit central bulging, while rad-heli configurations deform near specimen edges. 
The rad-heli-het specimen shows the most pronounced bulging, exceeding initial thickness. 
Localized deformation bands with reduced strain are again visible, while regions surrounding the central occlusion display elevated strains, highlighting the amplification of local deformation due to geometric discontinuities.
These observations reinforce the rationale for using such heterogeneous configurations as training specimens, enhancing parameter identifiability through enriched mechanical diversity.

\section{Inference results and robustness to measurement noise}
\label{sec:inference_results}
{\revision{In this section, we assess the training and fitting performance of the method. 
We track the evolution of the fit over optimization epochs and compute error metrics from the inferred posterior. 
The convergence of the posterior standard deviations of the material parameters was verified, confirming that the uncertainty estimates stabilize prior to the final optimization epoch and for the results reported here. 
We then assess robustness under both spatially uncorrelated and spatially correlated displacement perturbations.
}}
\subsection{Noise-free inference: identifiability gains from microstructural and geometrical heterogeneity}
\label{ssec:results_nonoise}
\begin{figure}[!h]
     \includegraphics[width=1.0\textwidth]{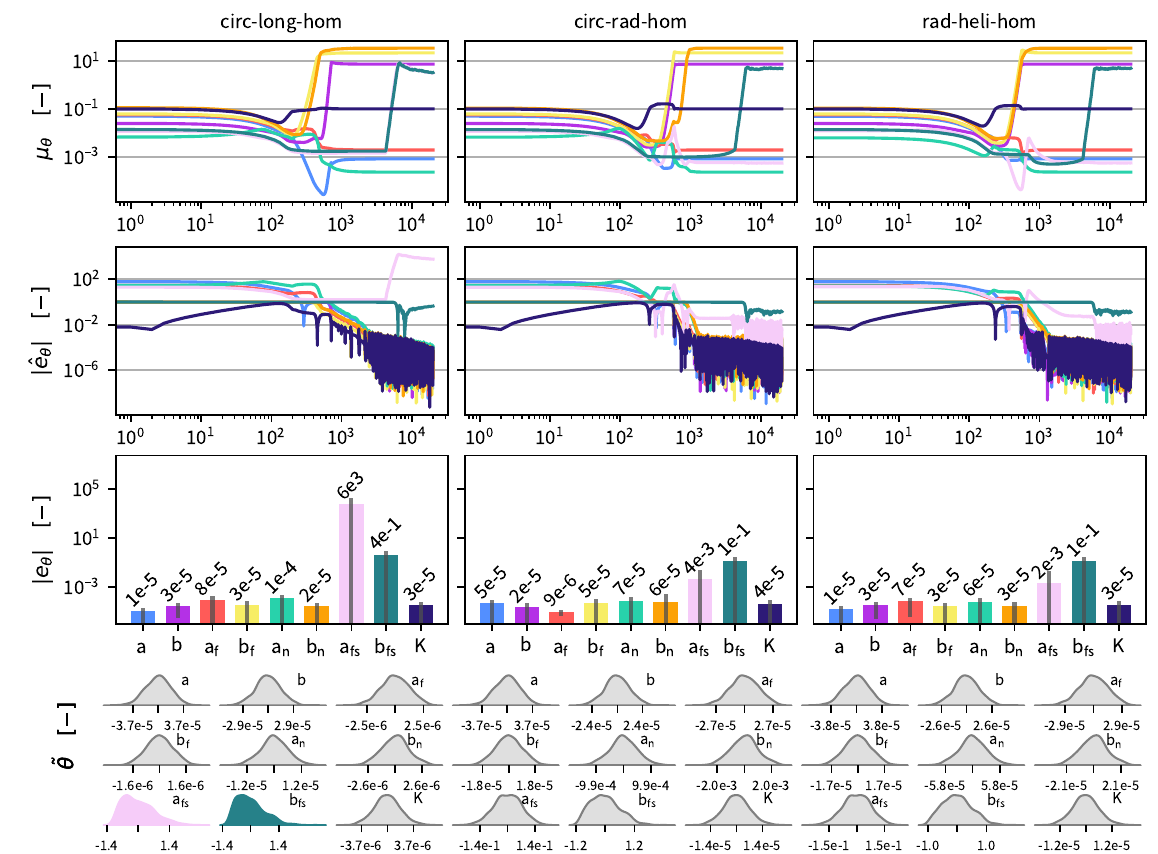}
    \caption[]{\textbf{Noise-free parameter inference on geometrically homogeneous and microstructurally heterogeneous tissue slabs.}
    From left to right: circ-long-hom, circ-rad-hom, and rad-heli-hom specimens.
    From top to bottom: (i) evolution of inferred material parameters $\theta$ over optimization epochs,
    (ii) evolution of absolute relative location parameter errors $\vert \hat{e}_{\theta} \vert$,
    (iii) absolute relative errors $\vert e_{\theta} \vert$ at the final epoch with corresponding 95\% credibility interval, and
    (iv) rescaled posterior distributions $\tilde{\theta}$ (drawn from the variational guide).
    The plots show rapid convergence and small final errors for most parameters across all three specimens, while the shear-coupling parameters ($a_{fs}$, $b_{fs}$) display slower convergence and broader posteriors in several cases.
    Distributions are shown in grayscale when the absolute unbiased skewness satisfies $|s_{\tilde{\theta}}|<0.5$.
    }
     \label{fig:paraminf_hom}
\end{figure}

\begin{figure}[!h]
     \includegraphics[width=1.0\textwidth]{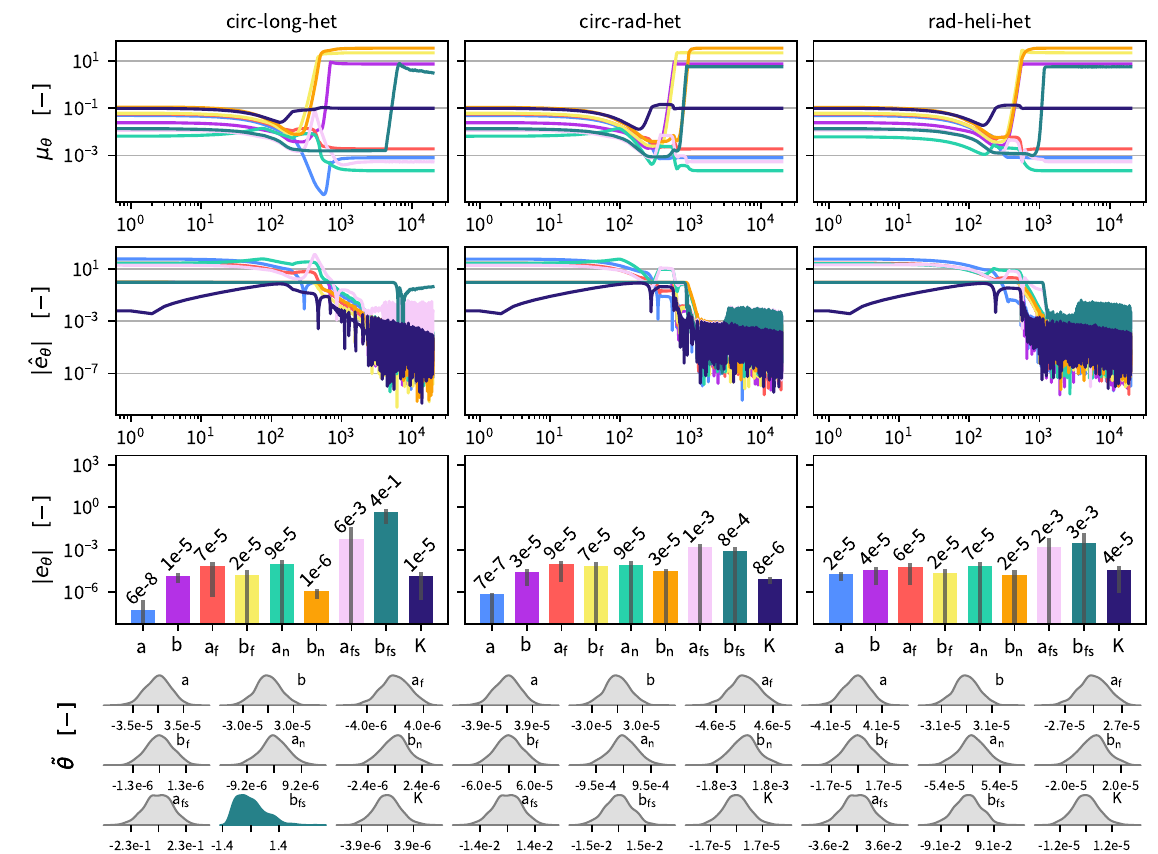}
    \caption[]{\textbf{Noise-free parameter inference on geometrically and microstructurally heterogeneous tissue slabs.}
    From left to right: circ-long-het, circ-rad-het, and rad-heli-het specimens.
    From top to bottom: (i) evolution of inferred material parameters $\theta$ over optimization epochs,
    (ii) evolution of absolute relative location parameter errors $\vert \hat{e}_{\theta} \vert$,
    (iii) absolute relative errors $\vert e_{\theta} \vert$ at the final epoch with corresponding 95\% credibility interval, and
    (iv) rescaled posterior distributions $\tilde{\theta}$ (drawn from the variational guide).
    The plots show stable convergence for all parameter groups across the three specimens, and the posterior samples indicate narrower spreads for the shear-coupling parameters ($a_{fs}$, $b_{fs}$) compared with the homogeneous counterparts.
    Distributions are shown in grayscale when the absolute unbiased skewness satisfies $|s_{\tilde{\theta}}|<0.5$.
    }
     \label{fig:paraminf_het}
\end{figure}

Figures \ref{fig:paraminf_hom} and \ref{fig:paraminf_het} summarize the inference performance of our framework in recovering the ground-truth HO parameter set $\vec{\theta}_{\text{gt}}$ (Table \ref{tab:parameters}) based on combined full-field deformation and reaction force data. Each figure shows, from top to bottom: 
(i) the evolution of the inferred parameters' guide location $\mu_{\theta}$ over optimization epochs, 
(ii) the corresponding absolute relative errors $e_{\mu_{\theta}}$,
(iii) final relative errors after convergence, and 
(iv) posterior distributions $\tilde{\theta}$, scaled to highlight skewness and relative spread. 
Here, we highlight posterior distributions in color when the unbiased skewness value $|s_{\tilde{\theta}}|$ is strictly greater than $0.5$. 
We define the following four metrics: the relative point error metric $\hat{e}_{\theta}$ between the mean of the underlying and untruncated normal distribution and the ground truth, and the relative error $e_{\theta}$ between samples from the posterior and the ground truth:
\begin{equation}
\begin{aligned}
    \hat{e}_{\theta} &= \frac{\mu_{\theta} - \theta_{\text{gt}}}{\theta_{\text{gt}}}
    \qquad
    & e_{\theta} &= \frac{\theta - \theta_{\text{gt}}}{\theta_{\text{gt}}}
\end{aligned}
\end{equation}
Here, we define $\mu_{\theta}=\mu^{q}$ as the mean of the underlying untruncated normal distribution used in the guide distribution.
Furthermore, we define the unbiased skewness $s_{\theta}$:
\begin{equation}
\begin{aligned}
    & s_{\tilde{\theta}} &= \frac{\sqrt{N(N-1)}}{N-2} \cdot \frac{\frac{1}{N} \sum_{n=1}^{N} (\tilde{\theta}_n - \mu_{\tilde{\theta}})^{3}}{\left( \frac{1}{N} \sum_{n=1}^{N} (\tilde{\theta}_n - \mu_{\tilde{\theta}})^{2} \right)^{3/2}}
    \qquad
    \text{with}
    \qquad
    & \tilde{\theta} &= \frac{\theta - \mu_{\theta}}{\mu_{\theta}}
\end{aligned}
\end{equation}
with $\tilde{\theta}$ denoting the relative error between samples taken from the guide distribution and the scaled posterior,
$\mu_{\tilde{\theta}}$ denoting the mean of the distribution $\tilde{\theta}$,
and $N$ equaling the number of samples drawn from the posterior distribution.
Unless stated otherwise, we draw $N=1{,}000$ samples from the inferred guide distribution
shown in the bottom rows of Figures \ref{fig:paraminf_hom} and \ref{fig:paraminf_het}, respectively, to visualize the shape, mean, and spread of the approximated posterior distribution. 
For completeness, the full set of converged parameter estimates, posterior scales (the standard deviation of the underlying and untruncated normal distribution), and relative errors for all six training specimens is reported in \ref{ap:noise-free-parameter-results} (Tables~\ref{tab:noise-free-parameter-estimates}--\ref{tab:noise-free-relative-error}). In the main text, we therefore focus on the parameter groups that drive the qualitative differences between Figures~\ref{fig:paraminf_hom} and \ref{fig:paraminf_het}, and summarize the remaining terms compactly.

Focusing on the \textit{geometrically homogeneous} specimens (Figure~\ref{fig:paraminf_hom}), the plots show rapid convergence and narrow posteriors for the isotropic $(a,b,K)$ and primary anisotropic $(a_f,b_f,a_n,b_n)$ parameter groups across all three microstructural settings.
Quantitatively, these parameters reach small final deviations, with the largest magnitudes remaining below \CRMsigmabn \ in this homogeneous-geometry group (Table~\ref{tab:noise-free-relative-error}).
In contrast, the fiber--sheet shear-coupling terms $(a_{fs},\,b_{fs})$ are the dominant outliers.
In the fully homogeneous \textit{circ-long-hom} specimen, the shear-coupling terms remain poorly identified, with large final errors ($\hat{e}_{a_{fs}}=$\CLMEafs \ and $\hat{e}_{b_{fs}}=$\CLMEbfs; Table~\ref{tab:noise-free-relative-error}). 

Introducing transmural and helical fiber variation improves identifiability of $a_{fs}$ substantially (from $\mathcal{O}(10^{3})$ to $\mathcal{O}(10^{-3}\!-\!10^{-2})$; Table~\ref{tab:noise-free-relative-error}), while $b_{fs}$ remains the least accurate inferred parameter in the homogeneous-geometry group (e.g., $\hat{e}_{b_{fs}}=$\CRMEbfs \ for \textit{circ-rad-hom} and \RHMEbfs \ for \textit{rad-heli-hom}).
Notably, $b_{fs}$ also converges more slowly than the remaining parameters: it stabilizes only after $\sim$\CRMCSbfs\ epochs in \textit{circ-rad-hom}\revisionst{, and it shows no clear plateau by \RHMCSbfs epochs in \textit{rad-heli-hom}}, whereas most other parameters stabilize within $\sim$ \CRMCSbn\  epochs (rows (i)--(ii)).
These trends are consistent with the broader and more skewed posteriors for $(a_{fs},\,b_{fs})$ in Figure~\ref{fig:paraminf_hom} (bottom row).

For the \textit{geometrically heterogeneous} specimens (Figure~\ref{fig:paraminf_het}), the plots show stable convergence across parameter groups and generally narrower posterior spreads than in Figure~\ref{fig:paraminf_hom}. 
For the isotropic and primary anisotropic parameters, final deviations remain small overall, with the largest magnitudes staying below \CLTEan \ across the heterogeneous-geometry group (Table~\ref{tab:noise-free-relative-error}).
The most pronounced quantitative change occurs in the inferred shear-coupling parameters.
With combined geometric heterogeneity and fiber rotation, both $a_{fs}$ and $b_{fs}$ converge to small final deviations: in \textit{circ-rad-het}, $\hat{e}_{a_{fs}}=$\CRTEafs \ and $\hat{e}_{b_{fs}}=$\CRTEbfs, and in \textit{rad-heli-het}, $\hat{e}_{a_{fs}}=$\RHTEafs \ and $\hat{e}_{b_{fs}}=$\RHTEbfs \ (Table~\ref{tab:noise-free-relative-error}). 
In contrast, without fiber rotation (\textit{circ-long-het}), $b_{fs}$ remains comparatively inaccurate ($\hat{e}_{b_{fs}}=$\CLTEbfs), consistent with the broader posterior in the corresponding column of Figure~\ref{fig:paraminf_het}.
Compared with the homogeneous-geometry cases in Figure~\ref{fig:paraminf_hom}, the rotated-fiber heterogeneous specimens (\textit{circ-rad-het} and \textit{rad-heli-het}) \revisionst{ reach a stable $b_{fs}$ trajectory with small final deviations within $\mathcal{O}(10^{4})$ epochs ($\sim 5\times10^{3}$ and $\sim 10^{4}$ epochs, respectively), rather than requiring $\sim 3\times10^{4}$ to $>10^{5}$ epochs} reach a stable $b_{fs}$ trajectory with small final deviations within \CRTCSbfs\ epochs, rather than requiring $\sim$ \CRMCSbfs\ epochs, consistent with the tighter posterior spreads in row (iv).

Across both Figures~\ref{fig:paraminf_hom} and \ref{fig:paraminf_het}, several parameters exhibit non-monotonic transients before converging.
In particular, multiple traces (including $b$, $b_f$, $a_n$, and $a_{fs}$) show a decrease after roughly $10^{2}$ epochs, followed by recovery towards their converged values by $\mathcal{O}(10^{3})$ epochs.
Consistent excursions appear in the error plots over the same epoch range, reflecting epochs where the inferred trajectories pass through the ground-truth value (i.e., where the signed relative deviation changes sign) without immediately stabilizing.
This behavior is consistent with strong parameter coupling and a nonconvex loss landscape in nonlinear orthotropic inference, where the optimizer can traverse multiple basins before stabilizing.
\clearpage
{\revision{In Table \ref{tab:noise-free-hyperparameter-estimates}, we report the inferred hyperparameters of the inverse-gamma distributions associated with $v_s$ and $\sigma^2$ for the noise-free experiments.
Despite the absence of added displacement noise, the posterior mean of $\sigma^2$ is on the order of $4.0\times10^{-6}$, which we attribute to residual numerical uncertainty in the forward and inference procedures.
In contrast, the posterior mean of $v_s$ is several orders of magnitude larger (approximately $4.1\times10^{6}$), indicating that, in this regime, uncertainty is predominantly governed by variability in the inferred material parameters rather than observational noise or numerical uncertainty.}}

\subsection{{\revision{Robustness to displacement noise: Gaussian white}}}
\label{ssec:results_noise}
We next assess robustness to displacement measurement noise by perturbing the measured full-field kinematics with additive Gaussian white noise at two levels, namely low ($\sigma_u = 1 \times 10^{-4}\,\mathrm{mm}$) and high ($\sigma_u = 1 \times 10^{-3}\,\mathrm{mm}$) \citep{Flaschel2021, Thakolkaran2022, Joshi2022, rahmaniNewApproachInverse2013}.
{\revision{No smoothing, stabilization, or displacement-field assimilation is applied prior to inference, such that the imposed perturbations act as a conservative stress test of the framework.}}
We restrict attention to the geometrically heterogeneous training specimens, since these provided the most reliable baseline inference performance in the noise-free setting (Section~\ref{ssec:results_nonoise}).

\begin{figure}[!h]
    \includegraphics[width=1.0\textwidth]{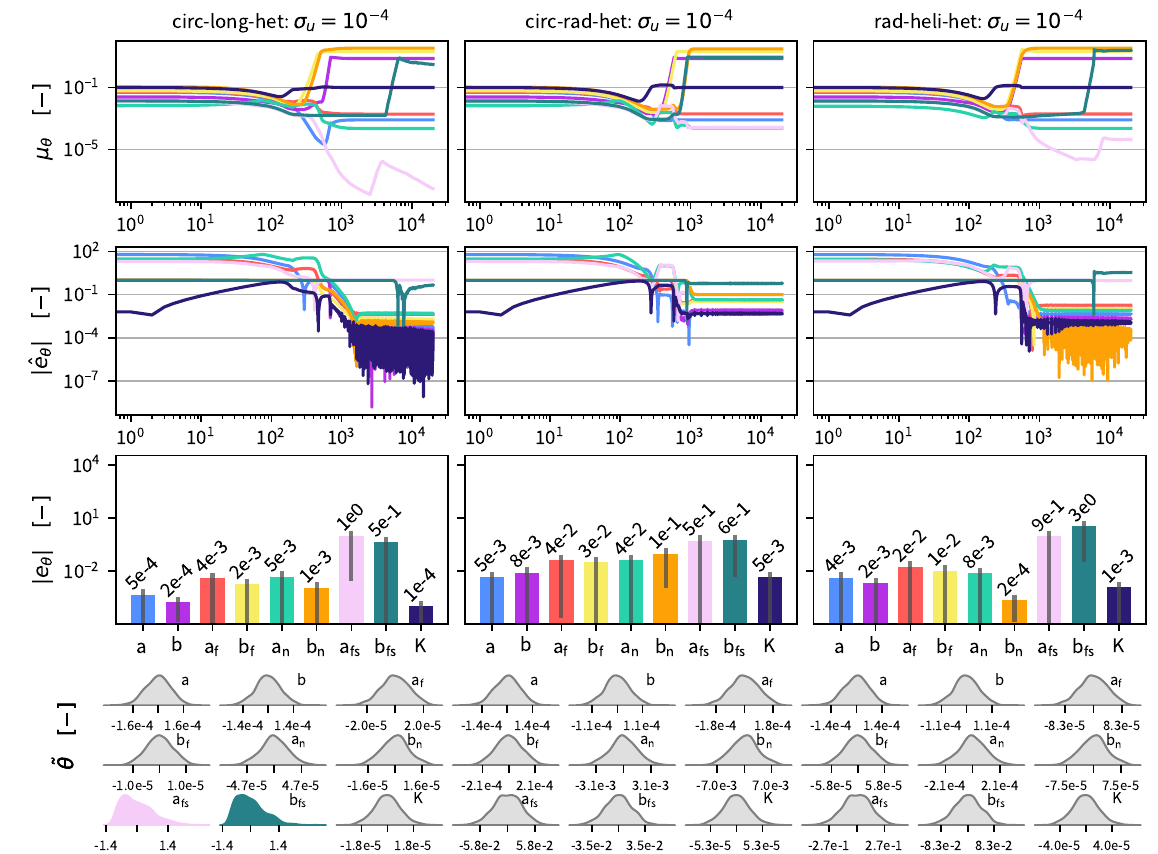}
    \caption[]{\textbf{Parameter inference on geometrically and microstructurally heterogeneous tissue slabs under low displacement measurement noise.}
    From left to right: circ-long-het, circ-rad-het, and rad-heli-het specimens.
    From top to bottom: (i) evolution of inferred material parameters $\theta$ over optimization epochs,
    (ii) evolution of absolute relative location parameter errors $\vert \hat{e}_{\theta}\vert$,
    (iii) absolute relative errors $\vert e_{\theta} \vert$ at the final epoch with corresponding 95\% credibility interval, and
    (iv) rescaled posterior distributions $\tilde{\theta}$ (drawn from the variational guide).
    The plots show that most parameters maintain stable convergence under low noise, while $a_{fs}$ and $b_{fs}$ exhibit the largest variability across specimens, reflected by slower convergence and broader posteriors.
    Here, uncorrelated Gaussian white displacement noise with standard deviation $\sigma_u=10^{-4}\,\mathrm{mm}$ is added to the full-field kinematics.
    Distributions are shown in grayscale when the absolute unbiased skewness satisfies $|s_{\tilde{\theta}}|<0.5$.
    }
    \label{fig:paraminf_het_low_noise}
\end{figure}

Figure~\ref{fig:paraminf_het_low_noise} summarizes inference under low displacement noise for the three geometrically heterogeneous training specimens.
Across all three microstructural settings, the trajectories in rows (i) and (ii) indicate that the low-noise setting largely preserves the convergence patterns of the noise-free heterogeneous baseline in Figure~\ref{fig:paraminf_het}, and differences emerge most clearly in the shear-coupling terms.
For \textit{circ-long-het}, final relative errors for isotropic and primary anisotropic parameters stay within a narrow band, with representative values between $10^{-4}$ and $5\times10^{-3}$ (Table~\ref{tab:noisy-relative-error}).
In contrast, the parameter trajectory for $a_{fs}$ does not settle to a clear plateau in row (i), and $b_{fs}$ exhibits increased stochasticity together with a mild downward drift.
The corresponding relative errors for the shear-coupling terms are of order one, with $|\hat{e}_{a_{fs}}|\approx 1$ and $|\hat{e}_{b_{fs}}|\approx 0.5$.
This behavior is mirrored in the posterior samples in row (iv), which show pronounced skewness and visible truncation for $(a_{fs},b_{fs})$, consistent with elevated uncertainty and weak practical identifiability under this noise level.
Whereas in the noise-free heterogeneous setting (Figure~\ref{fig:paraminf_het}) the shear parameter $a_{fs}$ could still be identified up to a relative error on the order of $10^{-2}$, the low-noise case now produces errors for both $a_{fs}$ and $b_{fs}$ of the same order of magnitude as their true values.
In the helically distributed fiber specimens \textit{circ-rad-het} and \textit{rad-heli-het}, the posterior samples in row (iv) do not exhibit comparable skewness or truncation for the inferred parameters.
Compared with the noise-free case, the final relative errors for isotropic and primary anisotropic parameters are slightly higher, remaining mostly between $5\times10^{-3}$ and $5\times10^{-2}$ for \textit{circ-rad-het} and between $3\times10^{-4}$ and $5\times10^{-2}$ for \textit{rad-heli-het} (excluding $a_{fs}$ and $b_{fs}$), while the majority of parameters still converge to clearly defined plateaus in rows (i) and (ii).
For the shear-coupling terms, the posterior shapes of $(a_{fs},b_{fs})$ become more regular in row (iv), yet their error traces in row (ii) plateau at comparatively elevated levels.
In the noise-free case (Figure~\ref{fig:paraminf_het}), we achieved rotated-fiber shear-coupling errors $|\hat{e}_{a_{fs}}|$ of order $10^{-3}$ and $|\hat{e}_{b_{fs}}|$ of at most $10^{-3}$ in the \textit{circ-rad-het} and \textit{rad-heli-het} specimens (Table~\ref{tab:noise-free-relative-error}).
Under low noise here, these errors increase to magnitudes between roughly $5\times10^{-1}$ and order one, with the largest outlier observed for $b_{fs}$ in \textit{rad-heli-het} (Table~\ref{tab:noisy-relative-error}).
Strikingly, the shear-coupling parameters stabilize in the \textit{circ-rad-het} and \textit{rad-heli-het} configurations after roughly \CRTLNCSafs\ and \RHTLNCSafs\ epochs respectively, even though their relative errors remain high.
Combined with the lack of skewness and truncation in their posteriors, this pattern is consistent with the parameters having moved away from the prior and having settled in a suboptimal local minimum under low-noise conditions.

\begin{figure}[!h]
    \includegraphics[width=1.0\textwidth]{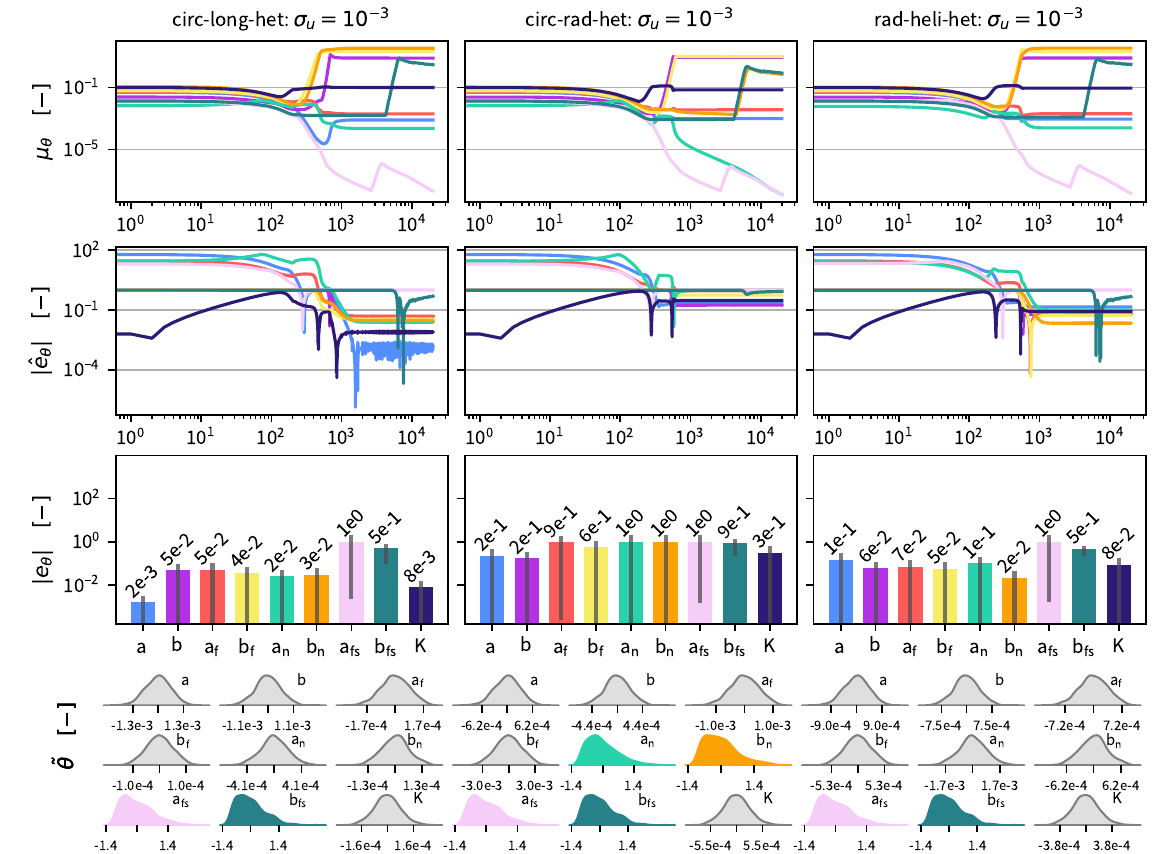}
    \caption[]{\textbf{Parameter inference on geometrically and microstructurally heterogeneous tissue slabs under high displacement measurement noise.}
    From left to right: circ-long-het, circ-rad-het, and rad-heli-het specimens.
    From top to bottom: (i) evolution of inferred material parameters $\theta$ over optimization epochs,
    (ii) evolution of absolute relative location parameter errors $\vert \hat{e}_{\theta} \vert$,
    (iii) absolute relative errors $\vert e_{\theta} \vert$ at the final epoch with corresponding 95\% credibility interval, and
    (iv) rescaled posterior distributions $\tilde{\theta}$ (drawn from the variational guide).
    The plots show increased variability and slower stabilization for several parameters compared with the low-noise case, with the shear-coupling parameters ($a_{fs}$, $b_{fs}$) exhibiting the largest spread in the posterior samples across specimens.
    Here, displacement noise with standard deviation $\sigma=10^{-3}\,\mathrm{mm}$ is added to the full-field kinematics.
    Distributions are shown in grayscale when the absolute unbiased skewness satisfies $|s_{\tilde{\theta}}|<0.5$.
    }
    \label{fig:paraminf_het_high_noise}
\end{figure}

Figure~\ref{fig:paraminf_het_high_noise} assesses parameter inference under high displacement measurement noise ($\sigma_u=10^{-3}\,\mathrm{mm}$) for the three geometrically heterogeneous specimens, using the same layout as in Figure~\ref{fig:paraminf_het_low_noise}. 
Relative to the low-noise case, the trajectories in rows (i) and (ii) show a marked loss of stability and accuracy, most prominently in the shear-coupling parameters $(a_{fs},b_{fs})$. 
Across all three specimens, these shear terms no longer attain reliable plateaus, their estimates for $a_{fs}$ collapse towards numerical near-zero values, and the corresponding relative errors remain of order one (Table~\ref{tab:noisy-relative-error}). 
In addition, the primary anisotropic sheet-direction parameters $(a_n,b_n)$ in the \textit{circ-rad-het} specimen fail to settle to stable values and reach relative errors close to unity, indicating a breakdown of identifiability in this parameter group under high noise. 
By contrast, the isotropic and primary anisotropic parameters remain comparatively more robust in \textit{circ-long-het}, where final relative errors for these parameters mostly lie between $10^{-3}$ and $3\times10^{-2}$. 
In the \textit{rad-heli-het} specimen, these relative parameter errors fall between $2\times10^{-2}$ and $10^{-1}$ (Table~\ref{tab:noisy-relative-error}). 
The \textit{circ-rad-het} configuration clearly performs worst overall, with many isotropic and primary anisotropic parameters exhibiting relative errors in the range $2\times10^{-1}$ to $1$, representing a sharp degradation compared with both the noise-free and low-noise cases. 
{\revision{Similar to the low-noise setting, parameters that do stabilize under high noise typically do so around \CLTHNCSan\ epochs for \textit{circ-long-het}, \textit{circ-rad-het}, and the \textit{rad-heli-het}.}} 
\ref{ap:noisy-parameter-results} provides the corresponding noisy hyperparameter and constitutive parameter estimates, together with their relative errors and posterior scale parameters (Tables~\ref{tab:noisy-hyperparameter-estimates}--\ref{tab:noisy-relative-error}), which quantitatively underpin the trends observed in Figures~\ref{fig:paraminf_het}--\ref{fig:paraminf_het_high_noise}.

Comparing Figures~\ref{fig:paraminf_het}, \ref{fig:paraminf_het_low_noise}, and \ref{fig:paraminf_het_high_noise} show a clear, parameter-selective degradation of identifiability as displacement noise increases from zero to low to high levels.
In the noise-free heterogeneous baseline (Figure~\ref{fig:paraminf_het}), isotropic and primary anisotropic parameters achieve small final deviations, with $|\hat{e}_{\theta}|$ typically below $5\times10^{-4}$, while the shear-coupling terms $(a_{fs},b_{fs})$ are already more sensitive in \textit{circ-long-het} but remain well identified in \textit{circ-rad-het} and \textit{rad-heli-het} (Table~\ref{tab:noise-free-relative-error}).
Under low noise (Figure~\ref{fig:paraminf_het_low_noise}), these isotropic and primary anisotropic errors increase by roughly one to two orders of magnitude yet mostly remain below $10^{-1}$, whereas the shear-coupling errors in \textit{circ-long-het}, \textit{circ-rad-het}, and \textit{rad-heli-het} rise to order one or larger, consistent with the broadened and, in some cases, truncated posteriors for $(a_{fs},b_{fs})$. 
At high noise (Figure~\ref{fig:paraminf_het_high_noise}), this trend intensifies: the shear-coupling parameters lose practical identifiability in all three specimens, with $|\hat{e}_{a_{fs}}|$ and $|\hat{e}_{b_{fs}}|$ remaining of order one and the corresponding posterior samples strongly skewed, while selected primary anisotropic parameters in \textit{circ-rad-het} (notably $a_f$, $a_n$, and $b_n$) also exhibit near-unity relative errors and non-stabilizing trajectories. 
By contrast, most isotropic parameters and the bulk modulus $K$ retain comparatively moderate deviations even at the highest noise level, particularly in \textit{circ-long-het} and \textit{rad-heli-het}, indicating a degree of robustness in the volumetric and baseline stiffness sector despite the pronounced degradation in shear-related and directional anisotropic terms.

{\revision{Table \ref{tab:noisy-hyperparameter-estimates} summarizes the inferred hyperparameters for experiments with additive Gaussian white displacement noise.
For low noise levels ($\sigma_u = 10^{-4}$), the posterior mean of $\sigma^2$ increases to approximately $1.3 \times 10^{-5}$, while for higher noise levels ($\sigma_u = 10^{-3}$), it increases by roughly an order of magnitude.
At the same time, the posterior mean of $v_s$ decreases from approximately $1.2 \times 10^{6}$ (low noise) to $1.1 \times 10^{5}$ (high noise).
This trend indicates a progressive shift in the dominant source of uncertainty: as displacement noise increases, uncertainty associated with the likelihood (data noise) becomes more prominent relative to uncertainty attributed to the material model parameters.
}}

{\revision{
\subsection{Robustness to displacement noise: spatially correlated}
\label{ssec:results_spat_noise}
We next consider an additional robustness study with spatially correlated displacement noise to further reflect the structured nature of experimental full-field measurement errors.
Implementation details of the noise model are provided in \ref{ap:spatial-noise}.

\begin{figure}[!h]
    \includegraphics[width=1.0\textwidth]{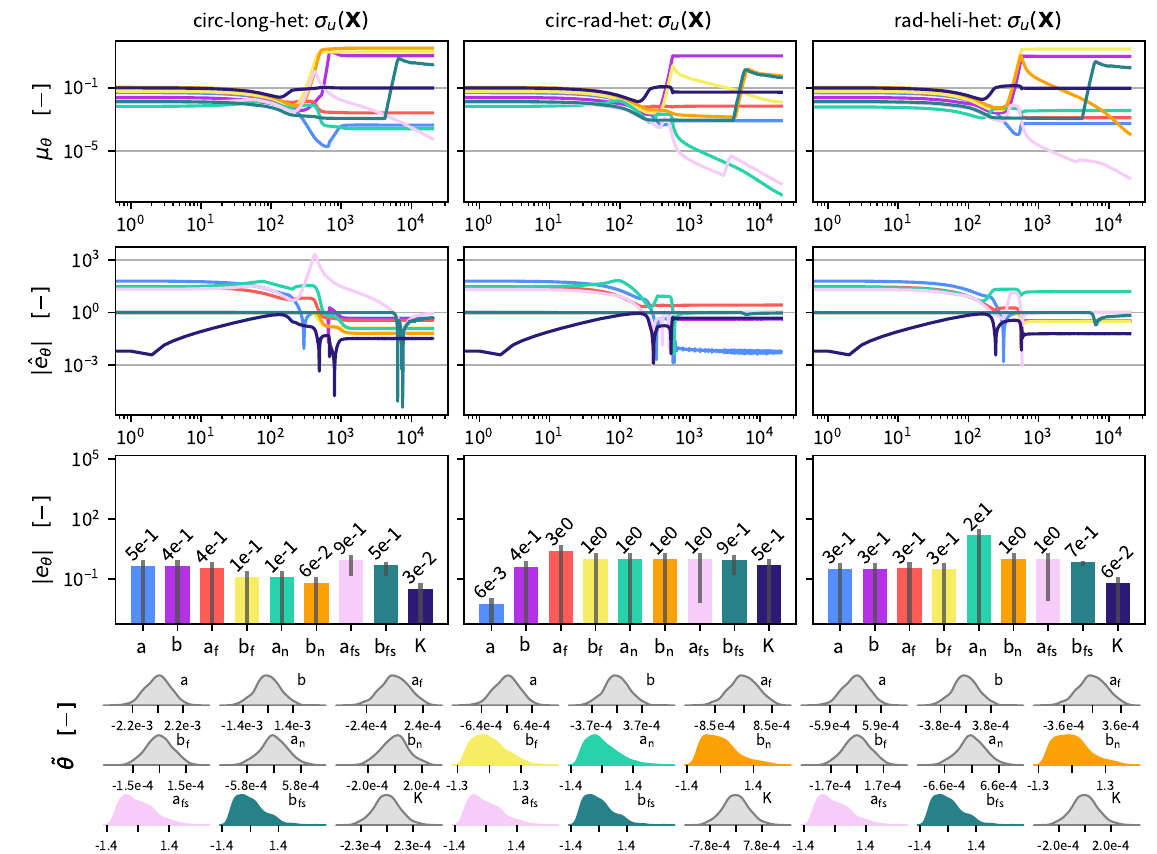}
    \caption[]{\textbf{Parameter inference on geometrically and microstructurally heterogeneous tissue slabs under spatially correlated displacement noise.}
    From left to right: \textit{circ-long-het}, \textit{circ-rad-het}, and \textit{rad-heli-het} specimens.
    From top to bottom: (i) evolution of inferred material parameters $\theta$ over optimization epochs,
    (ii) evolution of absolute relative location parameter errors $\vert \hat{e}_{\theta} \vert$,
    (iii) absolute relative errors $\vert e_{\theta} \vert$ at the final epoch with corresponding 95\% credibility interval, and
    (iv) rescaled posterior distributions $\tilde{\theta}$ (drawn from the variational guide).
    Here, a spatially varying and spatially correlated displacement noise field $\sigma_{u}(\vec{X})$ is added to the full-field kinematics, as defined in \ref{ap:spatial-noise}.
    The plots show increased variability and slower stabilization for several parameters, with the shear-coupling parameters ($a_{fs}$, $b_{fs}$) again exhibiting the strongest loss of identifiability across all specimens. Notably, the structured nature of the noise induces additional degradation in primary anisotropic parameters (such as $b_{f}$, $a_{n}$, and $b_{n}$) in the \textit{circ-rad-het} and \textit{rad-heli-het} specimens, while the \textit{circ-long-het} specimen remains comparatively more robust.
    Distributions are shown in grayscale when the absolute unbiased skewness satisfies $\vert s_{\tilde{\theta}}\vert < 0.5$.}
    \label{fig:paraminf_het_spat_noise}
\end{figure}

In Figure~10, we assess parameter inference under a spatially correlated displacement noise field $\sigma_u(\mathbf{X})$ for the three geometrically heterogeneous specimens.
The resulting degradation pattern is broadly similar to that observed under high Gaussian white noise, with the shear-coupling parameters $(a_{fs}, b_{fs})$ again showing the strongest loss of identifiability across all specimens.
Most clearly, in the \textit{circ-rad-het} specimen, the parameters $(b_f, a_n, b_n)$ also degrade substantially, with relative errors of order of magnitude $\mathcal{O}(1)$.
In the \textit{rad-heli-het} specimen, $b_n$ deteriorates similarly, whereas $a_n$ still stabilizes but at a larger relative error of $-1.516 \times 10^{1}$.
In contrast, the \textit{circ-long-het} specimen does not show additional parameter inference deterioration relative to the high Gaussian white-noise case, indicating that the impact of spatially correlated noise remains strongly specimen- and parameter-dependent.
The posterior distributions in row (iv) reflect the same trend.
Across all specimens, $(a_{fs}, b_{fs})$ retain the broadest uncertainty and, in several cases, also exhibit pronounced skewness, consistent with asymmetric and weak practical constraint under the structured noise perturbation.
The \textit{circ-rad-het} specimen additionally shows increased uncertainty in $b_f$, $a_n$, and $b_n$, while the \textit{rad-heli-het} specimen shows a further broadening in $b_n$ relative to the spatially uncorrelated Gaussian high-noise case.
Overall, isotropic and primary anisotropic parameters that remain practically identifiable still tend to stabilize after roughly $10^3$ epochs.
In contrast, $a_{fs}$ shows persistent drift toward numerically near-zero values, whereas $b_{fs}$ converges only much later, around $1.7 \times 10^4$ epochs.
}}

{\revision{In Table \ref{tab:spat-hyperparameter-estimates}, we present the corresponding results for spatially correlated displacement noise.
In this case, the posterior mean of $\sigma^2$ increases further to approximately $1.0 \times 10^{-4}$, representing an additional order-of-magnitude increase compared to the high Gaussian white noise setting.
Concurrently, the posterior mean of $v_s$ decreases to $\mathcal{O}(10^{4})$.
Consistent with the previous observations, these results indicate a continued shift in the uncertainty structure, with the likelihood term increasingly dominating the overall uncertainty as both the magnitude and spatial complexity of the noise increase.}}

\section{Validation beyond training}
We now assess how well parameter sets inferred from single-shot heterogeneous biaxial tests generalize beyond the specific training protocol.
To this end, we first compare Cauchy stress responses of {\revision{$1,000$ parameter samples drawn from the inferred posterior under classical homogeneous uni-modal deformation modes against the ground-truth parameterized Holzapfel--Ogden law.}}
Subsequently, we probe more complex mixed-modal deformation states, sampled from a physiologically constrained invariant space, and compare the resulting strain energy {\revision{between $1,000$ parameter samples drawn from the inferred posterior and ground-truth parameters to assess generalization under heterogeneous load paths.}}
{\revision{Finally, in the following subsections we exclusively evaluate the isochoric parameter misfit of the strain-energy model (equation \ref{eq:HO-model}) in order to ascertain internal consistency.}}

\subsection{Uni-modal deformation mode validation}
\label{ssec:unimodalvalidation}
We reproduce homogeneous, uni-modal loading protocols, i.e.\ classical incompressible biaxial tension
and triaxial shear tests, following the experimental protocols of \cite{Dokos2002,Sommer2015}. 
We consider incompressible biaxial tension, where $\lambda_{\rm f}$ and $\lambda_{\rm n}$ are prescribed and $\lambda_{\rm s}$ follows from incompressibility, and triaxial shear tests, where the principal stretches are fixed to unity and a single shear component is activated per test. 
The resulting stress expressions are summarized in \ref{ap:homogeneous_validation_protocols}.
Constitutive parameters inferred from our unsupervised single-shot heterogeneous biaxial protocol are then used to assess, in a controlled fashion, the predictive capability of the proposed framework relative to established multi-modal calibration pipelines for myocardial tissue 
\citep{Kakaletsis2021,Guan2018,Martonova2024, Peirlinck2018b}.

\begin{figure}[!h]
    \centering
    \includegraphics[width=0.80\linewidth]{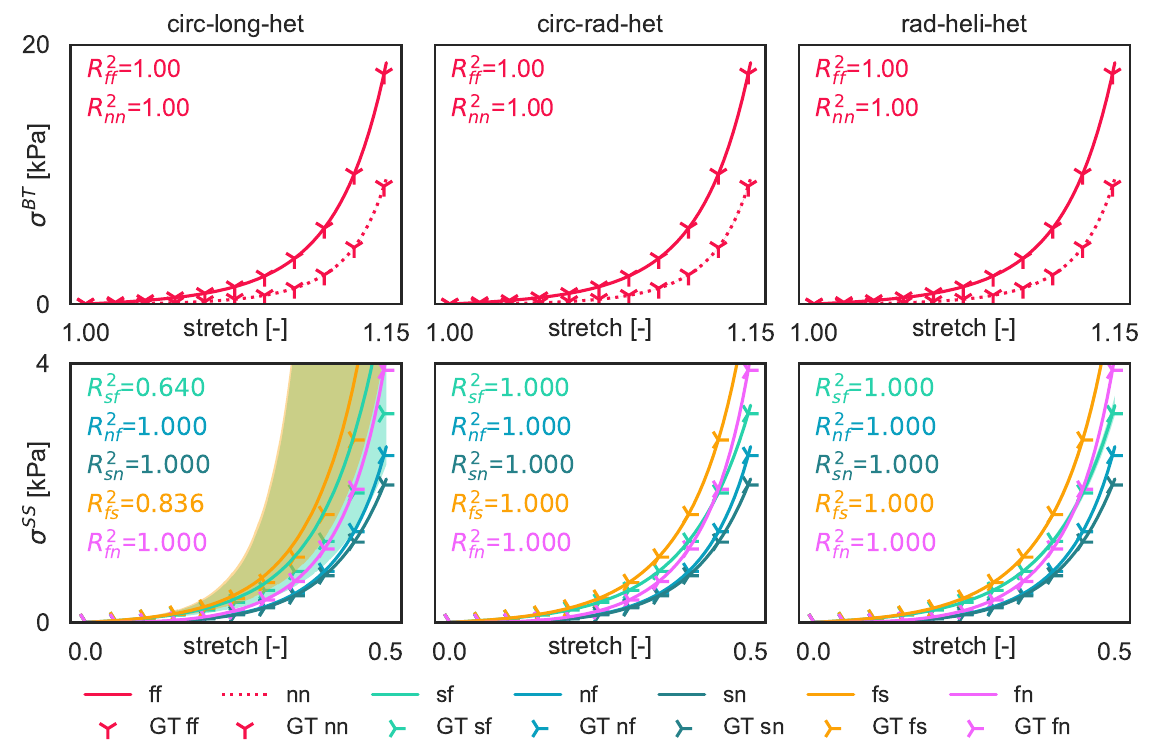}
    \caption[]{\textbf{Noise-free single-shot inference validation under homogeneous uni-modal loading.} Cauchy stresses predicted by the ground-truth HO parameter set (tri markers) and by the inferred noise-free single-shot full-field posterior (shaded $95\%$ credible intervals) for three heterogeneous configurations: circ--long--het (left), circ--rad--het (middle), and rad--heli--het (right).
    Rows correspond to classical incompressible biaxial tension and triaxial shear modes.
    Panel-wise $R^{2}$ values quantify agreement between posterior mean predictions and ground-truth stresses.
    Noise-free inference generalizes perfectly across all modes for circ-rad-het and rad-heli-het ($R^{2}\approx 1$ throughout), while circ-long-het shows the largest deviations in cross-fiber shear ($\sigma_{\mathrm{fs}}$, $\sigma_{\mathrm{sf}}$).}
    \label{fig:nonoise-stressvalid}
\end{figure}
Figure~\ref{fig:nonoise-stressvalid} reports noise-free validation across the three geometrically heterogeneous configurations: circ–long–het (left), circ–rad–het (middle), and rad–heli–het (right).
Each column compares Cauchy stresses from the ground-truth parameter set (tri markers) with predictions based on the noise-free single-shot full-field inference parameter sets (shaded $95\%$ credible intervals) under classical biaxial tension and triaxial shear tests. 
Leveraging parameters inferred from the noise-free \textit{circ-long--het} slab, the inferred biaxial stress-stretch curves match the ground-truth mechanical response of the tissue with \(R^{2}=1.00\) for both biaxial tension cases. 
The inferred \textit{nf}-, \textit{sn}-, and \textit{ns}-shear responses are also indistinguishable from the ground-truth response, whereas the inferred cross-fiber shear behavior shows lower agreement with respect to the ground-truth model, with \(R^{2}=0.89\) and \(R^{2}=0.77\) for \(\sigma_{\mathrm{fs}}\) and \(\sigma_{\mathrm{sf}}\) respectively.
Aggregated over all tests in this configuration, the overall coefficient of determination amounts to \(R^{2}=0.96\). 
These deviations are consistent with the limited deformational excitation of the $I_{4fs}$ shear-coupling invariant in the circ-long-het specimen (Figure~\ref{fig:synth_train_deform_het}) and with the weaker identifiability of $(a_{fs},b_{fs})$ observed in the previous section.
Leveraging the \textit{circ-rad-het} slab, the inferred constitutive parameter sets perfectly reproduce the ground-truth stresses across all biaxial tension and shear tests with \(R^{2}=1.00\) in every panel. 
Posterior confidence intervals remain tight, indicating precise recovery of the underlying constitutive parameters for this orientation, including the shear-coupling response.
Focusing on the \textit{rad-heli-het} specimen, the inferred stress curves again overlay the ground-truth for all biaxial tensile and shear tests with \(R^{2}=1.00\).
{\revision{Overall, the noise-free within-model-class validation shows that single-shot full-field inference generalizes reliably across all three heterogeneous configurations, with slight deviations confined to cross-fiber shear in the circ-long-het case, which is in agreement with the parameter-level identifiability analysis in Figure \ref{fig:paraminf_het}.}}

\begin{figure}[!h]
    \centering
    \includegraphics[width=0.80\linewidth]{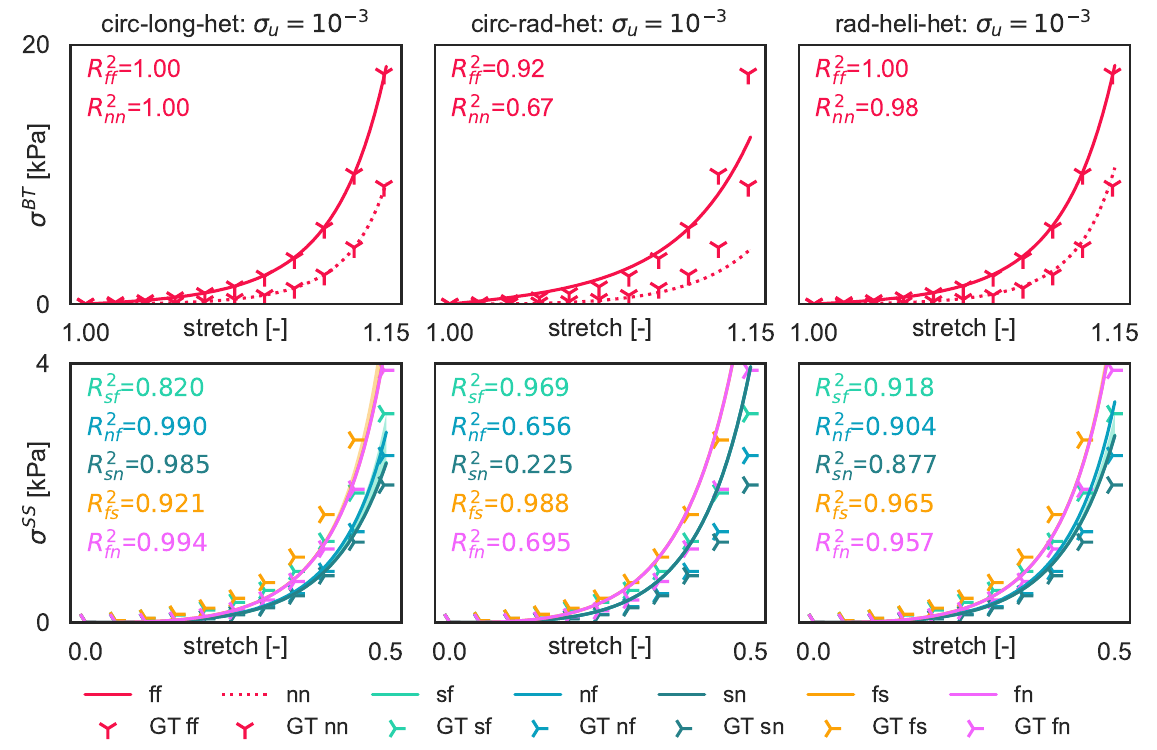}
    \caption[]{\textbf{High Gaussian white noise single-shot inference validation under homogeneous uni-modal loading.} Cauchy stresses predicted by the ground-truth HO parameter set (tri markers) and by the inferred single-shot full-field posterior under Gaussian white displacement noise $\sigma_u = 10^{-3}\,\mathrm{mm}$ (shaded $95\%$ credible intervals) for circ--long--het (left), circ--rad--het (middle), and rad--heli--het (right).
    Rows correspond to the same biaxial tension and triaxial shear protocols as in Figure~\ref{fig:nonoise-stressvalid}.
    Panel-wise $R^{2}$ values quantify agreement between posterior mean predictions and ground-truth stresses.
    High noise primarily degrades cross-fiber shear predictivity and widens credible intervals, most notably for circ-rad-het, while biaxial tension and non-cross-fiber shear modes remain well captured, particularly for circ-long-het and rad-heli-het.}
    \label{fig:highnoise-stressvalid}
\end{figure}
Figure~\ref{fig:highnoise-stressvalid} reports high-noise ($\sigma_u = 10^{-3}$) validation for the same three heterogeneous single-shot slab specimens.  
Inferring from displacement and reaction force data in the \textit{circ--long--het} specimen, the inferred biaxial tension behavior remains well aligned with the ground-truth, with panel-wise \(R^2\) values close to one. 
Similarly, the \textit{nf}-, \textit{sn}-, and \textit{ns}-shear responses retain correct magnitude and slope. 
Noise primarily affects the inferred cross-fiber shear stresses \(\sigma_{\mathrm{fs}}\) and \(\sigma_{\mathrm{sf}}\), where credible intervals widen and small systematic deviations between ground-truth and inferred curves appear. 
Aggregated over all tests in this configuration, the overall coefficient of determination remains high at \(R^{2}=0.96\).
Leveraging the \textit{circ–rad–het} slab, stresses computed from the high-noise inferred parameter sets no longer match the uni-modal ground-truth responses perfectly. 
Mean trends are preserved across biaxial and shear protocols, but panel-wise \(R^{2}\) values are lower than in the \textit{circ-long-het} and \textit{rad-heli-het} configurations, particularly in cross-fiber and sheet-direction shear. 
Nevertheless, the aggregated coefficient of determination still reaches \(R^{2}=0.73\).
Lastly, we note the best uni-modal predictive nature of the inferred parameter set from the high-noise \textit{rad-heli-het} specimen, with {\revisionst{with }}all panel-wise $R^{2}$-values greater than $0.87$ and an aggregated coefficient of determination of \(R^{2}=0.94\). 
{\revision{Overall, at high displacement noise $(\sigma_{u}=10^{-3})$ recovery of all HO-model parameters is no longer reliable as also reflected in the results in Section~\ref{ssec:results_noise}.
Simultaneously, we observe that certain posterior-predictive quantities - specifically stress responses in selected loading modes - remain in reasonable agreement with the ground truth.
The most notable of which are biaxial tension and non-cross-fiber triaxial shear. 
This indicates that, even when parameter identification breaks down, the inferred model can still provide informative predictions for specific deformation modes.}}

\begin{figure}[!h]
    \centering
    \includegraphics[width=0.80\linewidth]{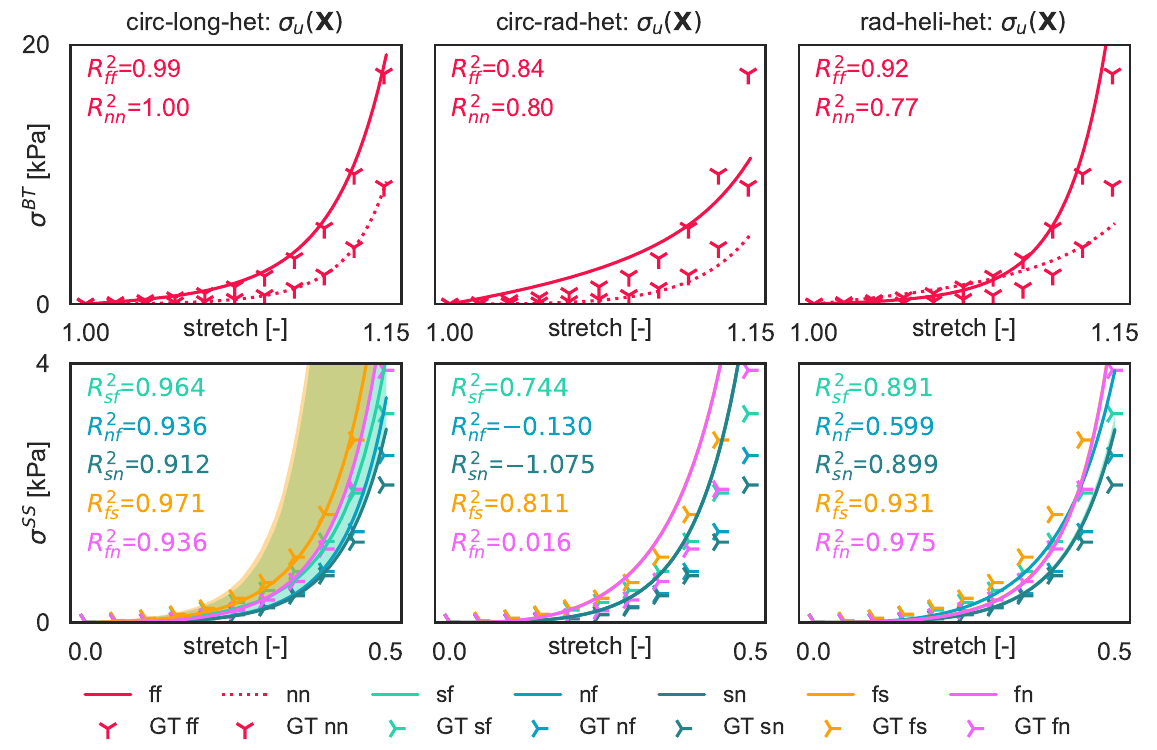}
    \caption[]{\textbf{Spatially correlated noise single-shot inference validation under homogeneous uni-modal loading.} Cauchy stresses predicted by the ground-truth HO parameter set (tri markers) and by the inferred single-shot full-field posterior under displacement noise $\sigma_u(\vec{X})$ (shaded $95\%$ credible intervals) for circ--long--het (left), circ--rad--het (middle), and rad--heli--het (right).
    Rows correspond to the same biaxial tension and triaxial shear protocols as in Figure~\ref{fig:nonoise-stressvalid}.
    Panel-wise $R^{2}$ values quantify agreement between posterior mean predictions and ground-truth stresses.
    Spatially correlated noise mainly introduces larger posterior stress uncertainty in the circ-long-het sample. It further introduces a mismatch in n-directional stress in biaxial tension for the circ-rad-het and rad-heli-het samples.}
    \label{fig:spat-noise-stressvalid}
\end{figure}

{\revision{Lastly, Figure \ref{fig:spat-noise-stressvalid} 
presents the uni-modal stress validation for the specimen-specific parameter sets inferred under spatially correlated displacement noise.
For the \textit{circ-long-het} specimen, the inferred biaxial tensile response remains well aligned with the ground truth, with panel-wise $R^2$ values close to one.
We observe increased uncertainty in the $\sigma_{fs}$ and $\sigma_{sf}$ stresses, together with a slight mismatch in shear behavior.
For the \textit{circ-rad-het} specimen, the predictive behavior degrades compared to the high Gaussian white-noise case.
The panel-wise $R^2$ values are lower than for \textit{circ-long-het}, and now yield an aggregated coefficient of determination of $R^2 = 0.28$, mainly due to the mismatch in the $(nf, sn, fn)$ shear directions. 
For the \textit{rad-heli-het} specimen, performance also decreases, most clearly in the $\sigma_{nn}$ and $\sigma_{nf}$ directions, but it still attains an aggregated coefficient of determination of $R^2 = 0.86$.}}

\subsection{Mixed-modal deformation validation}
\label{ssec:mixmodalvalidation}
To probe generalization under mixed deformation modes, we sample deformation states in invariant spaces between the homogeneous loading paths.
We draw $(I_1,I_{4ff},I_{4nn},I_{4fs})$ uniformly within the minimum and maximum values induced by the homogeneous protocols in the uni-modal deformation, see also \eqref{eq:hom_defgrad_app}, and retain only physiologically admissible states.
Specifically, we accept samples that satisfy near-incompressibility $0.9999 \le J \le 1.0001$ and whose maximum principal Cauchy stress does not exceed $20.3~\mathrm{kPa}$, consistent with reported peak stresses for left-ventricular biaxial tests on human myocardial tissue \citep{Sommer2015}.
The invariant-to-kinematics reconstruction and stress evaluation used for filtering are detailed in \ref{ap:mixed_modal_sampling}.
We repeat this procedure until $5{,}000$ admissible mixed-modal states are obtained.
\begin{figure}[!h]
    \centering
    \includegraphics[width=0.80\linewidth]{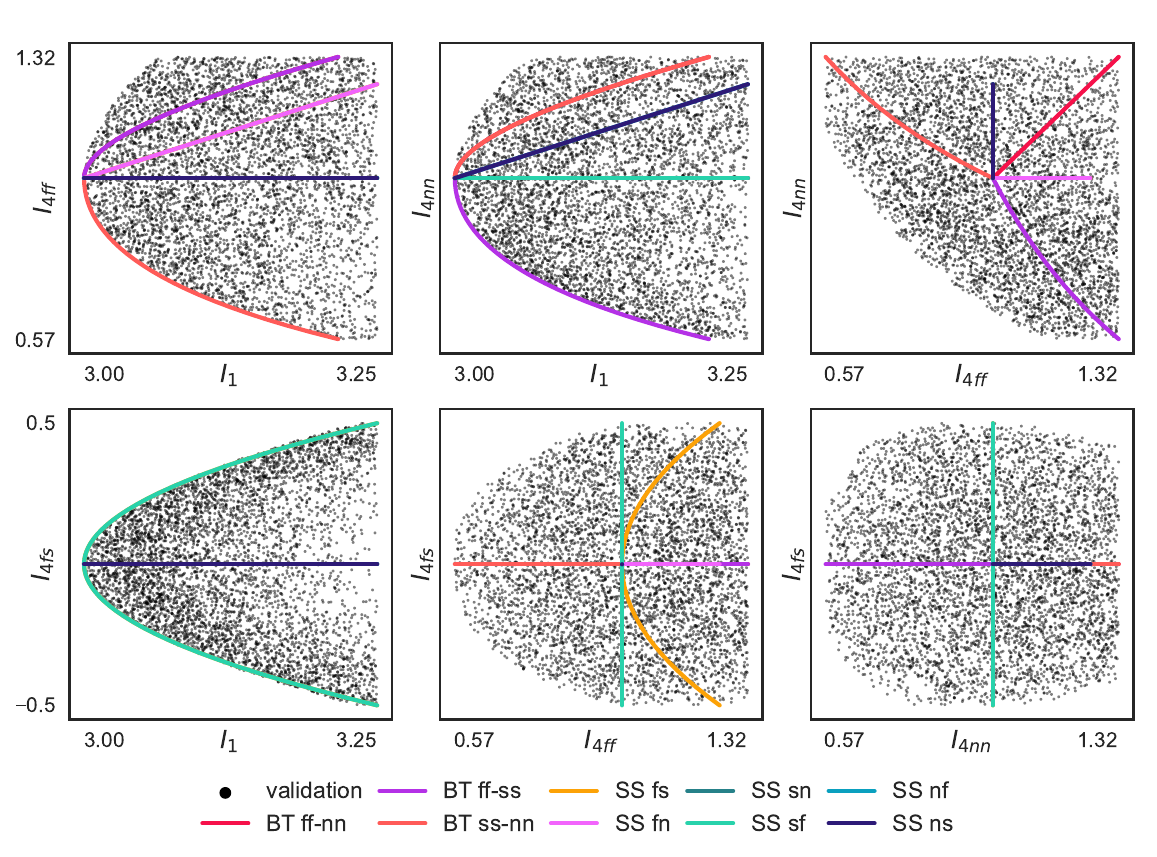}
    \caption[]{\textbf{Mixed-modal validation set construction in invariant space.}
    Invariant trajectories induced by uni-modal biaxial and single-shear protocols (colored curves) compared with randomly sampled invariant combinations that satisfy (quasi-)incompressibility and a physiological bound on maximum principal stress detailed in \ref{ap:mixed_modal_sampling}.
    The uni-modal protocols excite low-dimensional paths, whereas the accepted multi-modal deformation modes populate the space between these uni-modal paths, yielding mixed-modal deformation states for further and richer out-of-training validation.}
    \label{fig:invariant-correlation}
\end{figure}
Figure~\ref{fig:invariant-correlation} illustrates the resulting invariant coverage.
The thin sweeping colored arcs represent the uni-modal loading conditions from the previous section, highlighting that these tests excite the invariants along low-dimensional paths.
In contrast, our randomly sampled deformation gradients populate the space between these arcs and thus correspond to more complex mixed-modal testing protocols.
The accepted samples lie between $3$ and $3.25$ in $I_1$, between $0.57$ and $1.32$ for $I_{4ff}$ and $I_{4nn}$, and between $-0.5$ and $0.5$ for $I_{4fs}$, and they are spread out approximately uniformly within this constrained region.
This construction supports the view that the resulting invariant set spans a physiologically relevant range of deformation states that is well suited for out-of-training validation of the inferred versus ground-truth constitutive parameter set.

\begin{figure}
    \centering
        \includegraphics[width=0.80\linewidth]{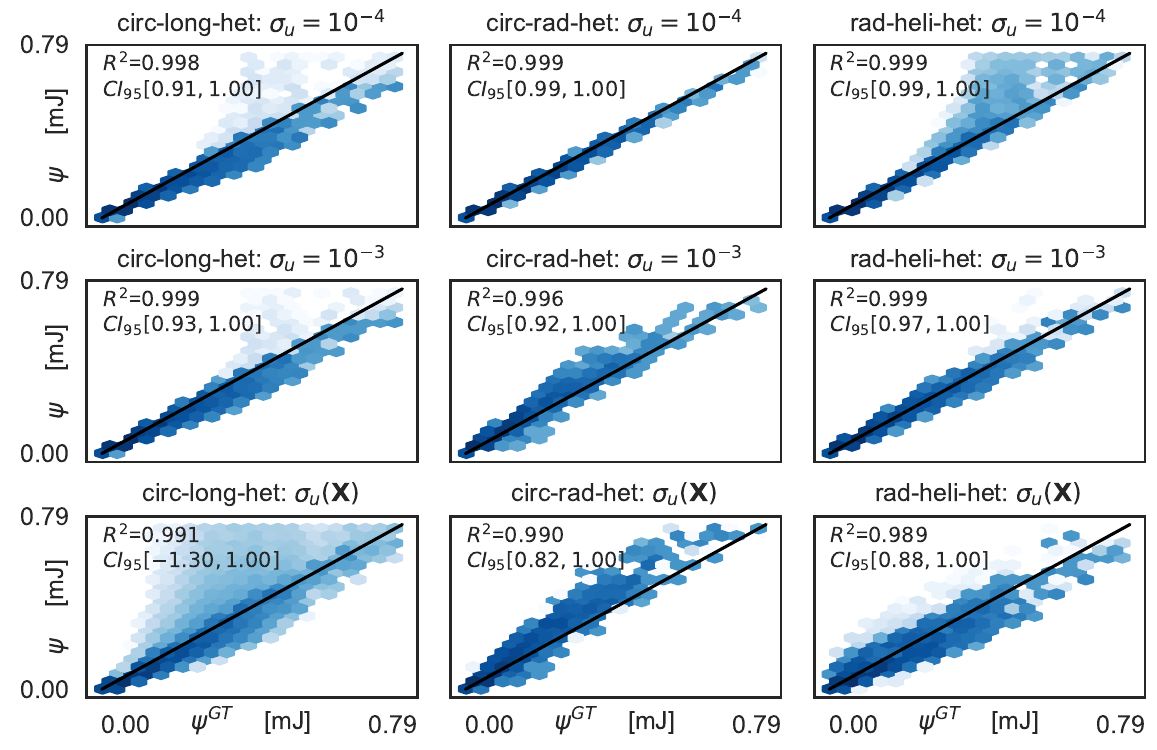}
    \caption[]{\textbf{Strain-energy generalization under mixed-modal deformations.}
    Correlation between strain energy predicted by inferred parameter posteriors, $\Psi$, and the ground-truth strain energy, $\Psi^{\mathrm{GT}}$, over $5{,}000$ accepted mixed-modal deformation states sampled in invariant space (Figure~\ref{fig:invariant-correlation}).
    Results are shown for geometrically heterogeneous training specimens under low and high displacement measurement noise, i.e. $\sigma_u = 10^{-4}$ and $10^{-3}\,\mathrm{mm}$ respectively, as well as for spatially correlated noise.
    Hexbin density emphasizes the dominant response region, and the black diagonal indicates perfect agreement.
    Despite noise-driven degradation of select parameter groups, predicted energies remain strongly correlated with the ground-truth across mixed-modal states (all $R^{2}\geq 0.9$). The 95\% confidence interval shows degrading scores as the noise level increases and as spatially correlated noise is added. Note that $R^2$ denotes the median score for all samples from the posterior.}
    \label{fig:strain_energy_correlation}
\end{figure}
More specifically, we compute strain energies using both the ground-truth parameter set and the inferred parameter sets obtained from both low and high displacement measurement noise-affected heterogeneous training experiments.
The correlation between the strain energy predicted by the inferred parameter sets $\Psi$ and the ground truth $\Psi^{\mathrm{GT}}$ is shown as a hexbin plot in Figure~\ref{fig:strain_energy_correlation}.
The black diagonal centerline indicates optimal behavior, i.e. the inferred mechanical response matches the expected ground-truth response perfectly.
In line with our prior results for the weakest identifiability of the noise-affected \textit{circ-long--het} specimen in the parameter inference and uni-modal stress inference analysis, this specimen also yields the lowest strain-energy agreement across the multi-modal deformation modes with $R^2=0.91$ in the low-noise setting and $R^2=0.931$ in the high-noise setting.
Across the three heterogeneous slabs tested, we observe larger deviations away from the diagonal, indicating that a subset of randomly sampled invariant combinations, together with corresponding posterior samples, lead to strain energies that differ more strongly from the ground-truth.
The \textit{circ-rad-het} specimen shows a very high coefficient of determination in the low-noise case ($R^2=0.989$), and its performance slightly degrades to $R^2=0.915$ in the high-noise case.
Despite this drop, \textit{circ-rad-het} exhibits the tightest clustering of hexbin intensities around the diagonal, suggesting that most mixed-modal states are still well captured, with a relatively small number of outliers contributing to the $R^2$ reduction.
Lastly, the most consistent performing sample is the \textit{rad-heli-het} with an $R^2$ score of $0.962$ in the low noise setting and $0.968$ in the high noise setting.

Overall, the mixed-modal strain-energy validation indicates that, even though high displacement noise can substantially perturb individual parameter estimates---particularly the shear-coupling and some directional anisotropic terms---the inferred models retain strong predictive capability at the level of strain energy across a broad set of physiologically admissible deformation states (all $R^{2}\geq 0.9$).
This finding complements the stress-based validation and suggests that the proposed single-shot full-field identification framework can {\revision{retain strong predictive accuracy beyond the specific training protocol}}, provided that geometric and microstructural heterogeneity are sufficiently rich and measurement noise remains within a moderate range.

\clearpage
\section{Discussion}
\label{sec:discussion}
The ability to infer complex orthotropic hyperelastic models from a single full-field biaxial tensile testing experiment represents an interesting avenue in biomechanical tissue characterization.
Conventional constitutive identification workflows for complex orthotropic tissues {\revision{such as the myocardium}}, typically rely on assumed microstructural homogeneity across the tissue specimen and leverage multiple specialized uni-modal loading protocols to identify constitutive parameter sets \citep{Dokos2002, Sommer2015, avazmohammadi_integrated_2018}.
This reflects a broader challenge in calibrating orthotropic HO laws for the myocardium: parameter uniqueness is generally difficult to guarantee under limited excitation, motivating multi-modal protocols, sensitivity-informed reparameterizations, or model reduction \citep{Guan2018,Gao2015,Palit2018,Lazarus2022}
This work demonstrates that a single heterogeneous biaxial experiment, when combined with full-field kinematics and sparse reaction-force measurements, can support uncertainty-aware inference of a nonlinear orthotropic Holzapfel--Ogden (HO) parameter set using stochastic variational inference. 
Beyond the identifiability perspective, this single-shot strategy reduces experimental tissue manipulation and is particularly attractive for heterogeneous biological tissues, where sample availability or repeat testing is limited.
Across the considered slab configurations, two consistent findings emerge: (i) identifiability is primarily governed by the degree to which the experiment excites the relevant invariant subspace, 
and (ii) even when individual HO parameters are not recovered accurately under varying levels of displacement measurement noise, the inferred models can remain highly predictive for stresses and strain energy over both uni-modal and mixed-modal deformation states. 
These observations clarify when single-shot full-field inference can be expected to recover the full parameter set, and when it should instead be interpreted as delivering a reliable posterior predictive constitutive response with quantified uncertainty.

\textbf{Heterogeneity as an experimental design principle for orthotropic constitutive parameter inference.}
Dominant gains in constitutive parameter identifiability arise from deliberately enriching the deformation space within a single test. 
Microstructural heterogeneity, introduced through slab orientation and the associated transmural fiber rotation, and geometric heterogeneity, introduced through a central occlusion, both broaden the activated distributions of $I_1$, $I_{4ff}$, $I_{4nn}$, and $I_{4fs}$ (Figures~\ref{fig:synth_train_deform_hom}--\ref{fig:synth_train_deform_het}). 
These experimental protocol choices increase the sensitivity of the equilibrium residuals to the full parameter vector $\vec{\theta}$. 
In the noise-free setting, this explains the contrast between homogeneous and heterogeneous specimens in Figures \ref{fig:paraminf_hom}-\ref{fig:paraminf_het}. 
Most isotropic and primary anisotropic parameters are inferred robustly across configurations, while the shear-coupling parameters $(a_{fs}, b_{fs})$ become practically identifiable only when the single slab biaxial tension experiment produces appreciable $I_{4fs}$ excitation. 
The \textit{circ-rad-het} and \textit{rad-heli-het} slab configurations provide the most consistent recovery of these terms, reflecting the combined effect of geometry-induced strain gradients and orientation-driven multiaxial activation of the microstructure.
In this sense, our results translate the multi-modal calibration message from prior myocardium studies into a single-shot setting: rather than combining separate stretch and shear protocols, we aim to excite the same (otherwise weakly constrained) coupling modes within one heterogeneous experiment \citep{Dokos2002, Sommer2015, Guan2018}.
These results further suggest that future protocols could deliberately combine geometrical and microstructural tailoring to maximize parameter observability within a single specimen.

\textbf{Experimental displacement measurement noise selectively impacts constitutive parameter identifiability.}
Displacement noise degrades constitutive parameter identifiability selectively rather than uniformly. 
For low noise ($\sigma_u = 10^{-4}$\,mm), Figure ~\ref{fig:paraminf_het_low_noise} shows stable convergence and comparatively narrow posteriors for the isotropic and primary anisotropic parameters. 
In contrast, $(a_{fs}, b_{fs})$ show the largest posterior broadening and, in some cases, pronounced skewness or truncation of the inferred posterior parameter distribution. 
{\revision{For high noise ($\sigma_u = 10^{-3}$\,mm), Figure ~\ref{fig:paraminf_het_high_noise} showcases how the shear-coupling terms lose practical identifiability across specimens, and additional degradation can appear in directional anisotropic parameters, most clearly in \textit{circ-rad-het}. 
These results are consistent with global sensitivity and inverse uncertainty quantification studies of the underlying Holzapfel-Ogden constitutive model \citep{Laita2025,Lazarus2022,Balaban2016}, which show that parameter sensitivity and practical identifiability depend strongly on the deformation modes and invariant support represented in the available data.}}
Two implications follow.
Full-field single-shot inference strongly benefits from experimental designs that amplify invariants associated with weaker energy modes. 
{\revision{In addition, posterior-shape diagnostics such as truncation and skewness provide useful indicators of noise-driven practical non-identifiability and should be interpreted alongside point estimates.}}

\textbf{Identifiability versus predictive capability.}
An important result is that imperfect recovery of ground-truth HO parameters does not necessarily imply poor constitutive prediction, as also noted in prior HO calibration and in vivo estimation studies \citep{Gao2015, Peirlinck2018b}. 
Figures~\ref{fig:nonoise-stressvalid}-\ref{fig:highnoise-stressvalid} show that posterior predictive stresses under classical homogeneous uni-modal protocols remain accurate in aggregate, even at high noise. 
Figure~\ref{fig:strain_energy_correlation} shows strong strain-energy correlation across 5,000 physiologically admissible mixed-modal deformation states, with all $R^2 \geq 0.9$. 
When viewed together with the parameter-level degradation in Figures~\ref{fig:paraminf_het_low_noise}-\ref{fig:paraminf_het_high_noise}, these results indicate that distinct parameter combinations can yield near-indistinguishable macroscopic responses over the rich explored invariant deformation space \citep{Guan2018, Lazarus2022}. 
In such cases, the posterior predictive response, together with its credible interval, becomes more relevant for downstream simulation than any single best-fit parameter vector and is consistent with inference perspectives that emphasize parameter distributions rather than unique point estimates \citep{Hauseux2018,Joshi2022,Linka2025,Ge2025}. 
This perspective is particularly pertinent for soft tissues, where measurement noise, limited loading diversity, and intrinsic variability constrain practical identifiability.

\textbf{Some heterogeneous configurations remain more predictive under noise.}
Our validation results also reveal slab-specific trends. 
The \textit{circ-rad-het} configuration can be more sensitive to displacement noise in the uni-modal and mixed-modal validations, despite exhibiting a rich heterogeneous deformation profile in Figure~\ref{fig:synth_train_deform_het}. 
This observation indicates that heterogeneity alone is not sufficient. 
Identifiability also depends on how displacement measurement noise perturbs the invariants that control weaker energy modes and how these perturbations propagate through the equilibrium residuals, which is consistent with the broader observation that the subset of HO terms supported by data depends on the deformation modes represented in the calibration dataset \cite{Guan2018}. 
In contrast, \textit{rad-heli-het} exhibits the most consistent predictive performance across stress and energy validations in Figures~\ref{fig:nonoise-stressvalid}-\ref{fig:strain_energy_correlation}, making it a promising slicing orientation when a single-shot protocol is prioritized and robust in-plane excitation of multiple microstructural constituent contributions is feasible.

\textbf{Outlook.}
{\revision{The present findings should be interpreted as a controlled identifiability and robustness study rather than as a full experimental validation of the workflow.
By construction, the synthetic setting assumes known boundary conditions, known microstructural fields, thin slices, and an \textit{a priori} Holzapfel--Ogden model form, which allows us to isolate how heterogeneity, model nonlinearity, and displacement noise affect inference performance.
From an inverse-identification perspective, the chosen slab thickness is conservative, since thicker specimens would in principle provide more internal kinematic information if measured reliably.
Our noise studies should likewise be interpreted as deliberately conservative robustness tests. 
We intentionally did not apply smoothing, stabilization, or displacement-field assimilation prior to inference. 
We instead first considered spatially uncorrelated Gaussian white noise as a coarse worst-case perturbation of the full-field kinematics, with noise magnitudes chosen in line with prior inverse and Bayesian identification studies \citep{Flaschel2021,Thakolkaran2022,Joshi2022,rahmaniNewApproachInverse2013}.
At the same time, realistic full-field displacement measurements are typically more structured, with spatial correlations and degradation near boundaries and interfaces. 
The added spatially correlated noise study therefore represents a first step towards a more experimentally realistic setting. 
Nevertheless, it remains an idealized robustness test rather than a complete model of DIC/DVC measurement error \citep{Peyraut2025,jafari_bayesian_2025}.
From an experimental perspective, these results suggest that successful application will depend not only on sufficiently rich invariant excitation within a single specimen, but also on careful treatment of boundary-condition uncertainty, missing near-boundary data, uncertainty or misregistration in the measured microstructural fields, and model form discrepancy.}}
Automated model discovery frameworks provide an interesting avenue to address both practical non-identifiability and model form discrepancy in settings with complex loading patterns and varying amounts of measurement noise \cite{Martonova2024,Flaschel2021,Joshi2022,Thakolkaran2025,Thorat2025,Linka2023}.
Complementary directions include explicit probabilistic models for displacement measurement error, experimental designs that target weakly activated invariants, and alternative variational weighting strategies to improve robustness under noisy full-field data \citep{Peyraut2025,jafari_bayesian_2025,Marek2019}.
Variational families that capture posterior dependencies more accurately, combined with posterior predictive checks tailored to equilibrium residuals, would further strengthen uncertainty quantification.
Finally, the present study focused on spatially homogeneous material parameters with spatially varying structure fields.
Extending the framework to spatially varying parameter fields with appropriate regularization could be valuable for truly heterogeneous tissues, while raising additional identifiability and experimental design challenges \citep{Ghouli2025,Chaurasiya2025}.

Overall, our results support single-shot, uncertainty-aware identification of nonlinear orthotropic constitutive behavior from a single heterogeneous biaxial test. 
Our main experimental design message is that the relevant invariant subspace must be sufficiently activated, with particular emphasis on the weak shear-coupling mode. 
At the same time, the validation results show that posterior predictive accuracy can remain high even when individual parameters are poorly identifiable due to noise.

\section{Conclusion}

We presented a physics-informed stochastic variational framework that recovers the full constitutive parameter set of a nonlinear orthotropic hyperelastic model from a single biaxial stretch experiment.
By exploiting intrinsic microstructural heterogeneity, or introducing simple geometric heterogeneity when needed, the method identifies not only the dominant tensile parameters but also shear-coupling terms that are typically weakly excited in biaxial tension.
Our results demonstrate that an effective balance between the underlying microstructure and biaxial loading improves identifiability, and that this balance can be set through slicing and tissue harvesting.
Circumferential–radial slabs and rotated cuts that place both fiber and sheet directions in the loading plane enable the recovery of the full parameter set under noise-free conditions.
{\revision{Under realistic yet spatially stationary, displacement measurement noise levels, our framework yields stress responses and strain energies in close agreement with the ground-truth mechanical response and provides posterior uncertainty estimates for the underlying constitutive parameters.}}
{\revision{Isotropic and primary anisotropic parameters are generally robust, whereas shear-coupling parameters and selected sheet-direction terms are more susceptible to degradation under displacement noise, particularly for spatially varying and boundary amplified perturbation.}}
Adding geometric heterogeneity, such as a central occlusion, systematically improves convergence and tightens posterior distributions, especially for shear-coupling parameters.
These results indicate that slicing protocols and experimental design can be used deliberately to enhance observability in full-field inverse characterization.
More broadly, the framework enables localized characterization of orthotropic nonlinear biological tissues from a single heterogeneous full-field experiment.
{\revision{As such, this work provides a controlled but important first step toward translating single-shot, uncertainty-aware constitutive inference to real full-field tissue experiments.}}
This reduces experimental tissue manipulation, improves reproducibility, and supports studies where available tissue is limited.

\appendix
\setcounter{figure}{0}

\section*{Acknowledgements}
M.P. acknowledges support through the NWO Veni Talent Award 20058. 

\clearpage
\bibliographystyle{elsarticle-num}
\bibliography{library}

@Article{Peirlinck2021,
  author    = {Peirlinck, M. and Costabal, F. Sahli and Yao, J. and Guccione, J. M. and Tripathy, S. and Wang, Y. and Ozturk, D. and Segars, P. and Morrison, T. M. and Levine, S. and Kuhl, E.},
  journal   = {Biomechanics and Modeling in Mechanobiology},
  title     = {Precision medicine in human heart modeling: Perspectives, challenges, and opportunities},
  year      = {2021},
  issn      = {1617-7940},
  month     = feb,
  number    = {3},
  pages     = {803--831},
  volume    = {20},
  doi       = {10.1007/s10237-021-01421-z},
  publisher = {Springer Science and Business Media LLC},
}

@Article{Peirlinck2018b,
  author    = {Peirlinck, Mathias and Sack, Kevin L. and De Backer, Pieter and Morais, Pedro and Segers, Patrick and Franz, Thomas and De Beule, Matthieu},
  journal   = {International Journal for Numerical Methods in Biomedical Engineering},
  title     = {Kinematic boundary conditions substantially impact in silico ventricular function},
  year      = {2018},
  issn      = {2040-7947},
  month     = oct,
  number    = {1},
  volume    = {35},
  doi       = {10.1002/cnm.3151},
  publisher = {Wiley},
}

@Article{Peirlinck2019,
  author    = {Peirlinck, M. and Sahli Costabal, F. and Sack, K. L. and Choy, J. S. and Kassab, G. S. and Guccione, J. M. and De Beule, M. and Segers, P. and Kuhl, E.},
  journal   = {Biomechanics and Modeling in Mechanobiology},
  title     = {Using machine learning to characterize heart failure across the scales},
  year      = {2019},
  issn      = {1617-7940},
  month     = jun,
  number    = {6},
  pages     = {1987--2001},
  volume    = {18},
  doi       = {10.1007/s10237-019-01190-w},
  publisher = {Springer Science and Business Media LLC},
}

@Misc{Thakolkaran2025,
  author    = {Thakolkaran, Prakash and Guo, Yaqi and Saini, Shivam and Peirlinck, Mathias and Alheit, Benjamin and Kumar, Siddhant},
  title     = {Can KAN CANs? Input-convex Kolmogorov-Arnold Networks (KANs) as hyperelastic constitutive artificial neural networks (CANs)},
  year      = {2025},
  copyright = {arXiv.org perpetual, non-exclusive license},
  doi       = {10.48550/ARXIV.2503.05617},
  publisher = {arXiv},
}

@Article{Flaschel2021,
  author   = {Flaschel, Moritz and Kumar, Siddhant and De Lorenzis, Laura},
  journal  = {Computer Methods in Applied Mechanics and Engineering},
  title    = {Unsupervised discovery of interpretable hyperelastic constitutive laws},
  year     = {2021},
  issn     = {0045-7825},
  pages    = {113852},
  volume   = {381},
  doi      = {10.1016/j.cma.2021.113852},
}

@Article{Rohmer2007,
  author   = {Rohmer, Damien and Sitek, Arkadiusz and Gullberg, Grant T.},
  journal  = {Investigative Radiology},
  title    = {Reconstruction and {Visualization} of {Fiber} and {Laminar} {Structure} in the {Normal} {Human} {Heart} from {Ex} {Vivo} {Diffusion} {Tensor} {Magnetic} {Resonance} {Imaging} ({DTMRI}) {Data}},
  year     = {2007},
  issn     = {0020-9996},
  month    = nov,
  number   = {11},
  pages    = {777},
  volume   = {42},
  doi      = {10.1097/RLI.0b013e3181238330},
  language = {en-US},
  urldate  = {2023-01-24},
}

@Article{NiellesVallespin2017,
  author   = {Nielles-Vallespin, Sonia and Khalique, Zohya and Ferreira, Pedro F. and de Silva, Ranil and Scott, Andrew D. and Kilner, Philip and McGill, Laura-Ann and Giannakidis, Archontis and Gatehouse, Peter D. and Ennis, Daniel and Aliotta, Eric and Al-Khalil, Majid and Kellman, Peter and Mazilu, Dumitru and Balaban, Robert S. and Firmin, David N. and Arai, Andrew E. and Pennell, Dudley J.},
  journal  = {Journal of the American College of Cardiology},
  title    = {Assessment of {Myocardial} {Microstructural} {Dynamics} by {In}{~}{Vivo} {Diffusion} {Tensor} {Cardiac} {Magnetic} {Resonance}},
  year     = {2017},
  issn     = {0735-1097},
  month    = feb,
  number   = {6},
  pages    = {661--676},
  volume   = {69},
  doi      = {10.1016/j.jacc.2016.11.051},
  urldate  = {2023-11-22},
}

@Article{Sommer2015,
  author   = {Sommer, Gerhard and Schriefl, Andreas J. and Andr{\"{a}}, Michaela and Sacherer, Michael and Viertler, Christian and Wolinski, Heimo and Holzapfel, Gerhard A.},
  journal  = {Acta Biomaterialia},
  title    = {Biomechanical properties and microstructure of human ventricular myocardium},
  year     = {2015},
  issn     = {1742-7061},
  month    = sep,
  pages    = {172--192},
  volume   = {24},
  doi      = {10.1016/j.actbio.2015.06.031},
  language = {en},
  urldate  = {2022-12-02},
}

@PhdThesis{Fehervary2018,
  author     = {Fehervary, Heleen and Smoljki{\'{c}}, Marija and Vander Sloten, Jos and Famaey, Nele},
  school     = {KU Leuven},
  title      = {Planar biaxial testing of soft biological tissues},
  year       = {2018},
  language   = {eng},
  pmid       = {26854936},
  shorttitle = {Planar biaxial testing of soft biological tissue using rakes},
}

@Article{Avazmohammadi2019,
  author   = {Avazmohammadi, Reza and Soares, Jo{\~{a}}o S. and Li, David S. and Raut, Samarth S. and Gorman, Robert C. and Sacks, Michael S.},
  journal  = {Annual Review of Biomedical Engineering},
  title    = {A {Contemporary} {Look} at {Biomechanical} {Models} of {Myocardium}},
  year     = {2019},
  issn     = {1545-4274},
  month    = jun,
  pages    = {417--442},
  volume   = {21},
  doi      = {10.1146/annurev-bioeng-062117-121129},
  language = {eng},
  pmcid    = {PMC6626320},
  pmid     = {31167105},
}

@Article{Yin1996,
  author   = {Yin, F. C. and Chan, C. C. and Judd, R. M.},
  journal  = {American Journal of Physiology-Heart and Circulatory Physiology},
  title    = {Compressibility of perfused passive myocardium},
  year     = {1996},
  issn     = {0363-6135},
  month    = nov,
  number   = {5},
  pages    = {H1864--H1870},
  volume   = {271},
  doi      = {10.1152/ajpheart.1996.271.5.H1864},
  urldate  = {2023-01-18},
}

@Article{McEvoy2018,
  author     = {McEvoy, Eoin and Holzapfel, Gerhard A. and McGarry, Patrick},
  journal    = {Journal of Biomechanical Engineering},
  title      = {Compressibility and {Anisotropy} of the {Ventricular} {Myocardium}: {Experimental} {Analysis} and {Microstructural} {Modeling}},
  year       = {2018},
  issn       = {0148-0731},
  month      = may,
  number     = {8},
  volume     = {140},
  doi        = {10.1115/1.4039947},
  shorttitle = {Compressibility and {Anisotropy} of the {Ventricular} {Myocardium}},
  urldate    = {2023-01-25},
}

@Misc{Wilson2022,
  author   = {Alexander J. Wilson, Daniel B. Ennis},
  month    = mar,
  title    = {Myocardial mesostructure and mesofunction},
  year     = {2022},
  doi      = {10.1152/ajpheart.00059.2022},
  language = {en},
  urldate  = {2024-11-05},
}

@Book{Planck1897,
  author    = {Planck, Max},
  publisher = {Veit \& Comp.},
  title     = {Vorlesungen {\"{u}}ber {Thermodynamik}, von dr. {Max} {Planck}},
  year      = {1897},
  language  = {de},
}

@Article{Coleman1959,
  author   = {Coleman, Bernard D. and Noll, Walter},
  journal  = {Archive for Rational Mechanics and Analysis},
  title    = {On the thermostatics of continuous media},
  year     = {1959},
  issn     = {1432-0673},
  month    = jan,
  number   = {1},
  pages    = {97--128},
  volume   = {4},
  doi      = {10.1007/BF00281381},
  language = {en},
  urldate  = {2024-11-22},
}

@Article{Estrada2020,
  author     = {Estrada, J. B. and Luetkemeyer, C. M. and Scheven, U. M. and Arruda, E. M.},
  journal    = {Experimental Mechanics},
  title      = {{MR}-u: {Material} {Characterization} {Using} {3D} {Displacement}-{Encoded} {Magnetic} {Resonance} and the {Virtual} {Fields} {Method}},
  year       = {2020},
  issn       = {1741-2765},
  month      = sep,
  number     = {7},
  pages      = {907--924},
  volume     = {60},
  doi        = {10.1007/s11340-020-00595-4},
  language   = {en},
  shorttitle = {{MR}-u},
  urldate    = {2024-02-08},
}

@Misc{Kingma2017,
  author     = {Kingma, Diederik P. and Ba, Jimmy},
  month      = jan,
  title      = {Adam: {A} {Method} for {Stochastic} {Optimization}},
  year       = {2017},
  language   = {en},
  publisher  = {arXiv},
  shorttitle = {Adam},
  urldate    = {2024-11-22},
}

@Misc{Phan2019,
  author    = {Phan, Du and Pradhan, Neeraj and Jankowiak, Martin},
  month     = dec,
  title     = {Composable {Effects} for {Flexible} and {Accelerated} {Probabilistic} {Programming} in {NumPyro}},
  year      = {2019},
  doi       = {10.48550/arXiv.1912.11554},
  language  = {en},
  publisher = {arXiv},
  urldate   = {2024-11-25},
}

@Article{Peshave2024,
  author   = {Peshave, A. and Pierron, F. and Lava, P. and Moens, D. and Vandepitte, D.},
  journal  = {Experimental Techniques},
  title    = {Practical {Uncertainty} {Quantification} {Guidelines} for {DIC}-{Based} {Numerical} {Model} {Validation}},
  year     = {2024},
  issn     = {1747-1567},
  month    = oct,
  doi      = {10.1007/s40799-024-00758-1},
  language = {en},
  urldate  = {2024-11-04},
}

@Article{Thakolkaran2022,
  author   = {Thakolkaran, Prakash and Joshi, Akshay and Zheng, Yiwen and Flaschel, Moritz and De Lorenzis, Laura and Kumar, Siddhant},
  journal  = {Journal of the Mechanics and Physics of Solids},
  title    = {{NN}-{EUCLID}: {Deep}-learning hyperelasticity without stress data},
  year     = {2022},
  issn     = {0022-5096},
  number   = {July},
  pages    = {105076},
  volume   = {169},
  doi      = {10.1016/j.jmps.2022.105076},
}

@Article{Wriggers2016,
  author   = {Wriggers, P. and Schr{\"{o}}der, J. and Auricchio, F.},
  journal  = {Advanced Modeling and Simulation in Engineering Sciences},
  title    = {Finite element formulations for large strain anisotropic material with inextensible fibers},
  year     = {2016},
  issn     = {2213-7467},
  month    = aug,
  number   = {1},
  pages    = {25},
  volume   = {3},
  doi      = {10.1186/s40323-016-0079-3},
  language = {en},
  urldate  = {2025-04-10},
}

@Article{Guan2018,
  author    = {Guan, Debao and Ahmad, Faizan and Theobald, Peter and Soe, Shwe and Luo, Xiaoyu and Gao, Hao},
  journal   = {Biomechanics and Modeling in Mechanobiology},
  title     = {On the AIC-based model reduction for the general Holzapfel--Ogden myocardial constitutive law},
  year      = {2018},
  issn      = {1617-7940},
  month     = apr,
  number    = {4},
  pages     = {1213--1232},
  volume    = {18},
  doi       = {10.1007/s10237-019-01140-6},
  publisher = {Springer Science and Business Media LLC},
}

@Article{Joshi2022,
  author    = {Joshi, Akshay and Thakolkaran, Prakash and Zheng, Yiwen and Escande, Maxime and Flaschel, Moritz and De Lorenzis, Laura and Kumar, Siddhant},
  journal   = {Computer Methods in Applied Mechanics and Engineering},
  title     = {Bayesian-EUCLID: Discovering hyperelastic material laws with uncertainties},
  year      = {2022},
  issn      = {0045-7825},
  month     = aug,
  pages     = {115225},
  volume    = {398},
  doi       = {10.1016/j.cma.2022.115225},
  publisher = {Elsevier BV},
}

@Article{Kakaletsis2021,
  author    = {Kakaletsis, Sotirios and Meador, William D. and Mathur, Mrudang and Sugerman, Gabriella P. and Jazwiec, Tomasz and Malinowski, Marcin and Lejeune, Emma and Timek, Tomasz A. and Rausch, Manuel K.},
  journal   = {Acta Biomaterialia},
  title     = {Right ventricular myocardial mechanics: Multi-modal deformation, microstructure, modeling, and comparison to the left ventricle},
  year      = {2021},
  issn      = {1742-7061},
  month     = mar,
  pages     = {154--166},
  volume    = {123},
  doi       = {10.1016/j.actbio.2020.12.006},
  publisher = {Elsevier BV},
}

@Article{Peirlinck2024,
  author   = {Peirlinck, Mathias and Hurtado, Juan A. and Rausch, Manuel K. and Tepole, Adri{\'{a}}n Buganza and Kuhl, Ellen},
  journal  = {Engineering with Computers},
  title    = {A universal material model subroutine for soft matter systems},
  year     = {2024},
  issn     = {1435-5663},
  month    = sep,
  doi      = {10.1007/s00366-024-02031-w},
  language = {en},
  urldate  = {2024-11-22},
}

@Article{Martonova2024,
  author     = {Martonov{\'{a}}, Denisa and Peirlinck, Mathias and Linka, Kevin and Holzapfel, Gerhard A. and Leyendecker, Sigrid and Kuhl, Ellen},
  journal    = {Computer Methods in Applied Mechanics and Engineering},
  title      = {Automated model discovery for human cardiac tissue: {Discovering} the best model and parameters},
  year       = {2024},
  issn       = {0045-7825},
  month      = aug,
  pages      = {117078},
  volume     = {428},
  doi        = {10.1016/j.cma.2024.117078},
  shorttitle = {Automated model discovery for human cardiac tissue},
  urldate    = {2024-10-22},
}

@Article{Holzapfel2009,
  author     = {Holzapfel, Gerhard A. and Ogden, Ray W.},
  journal    = {Philosophical Transactions of the Royal Society A: Mathematical, Physical and Engineering Sciences},
  title      = {Constitutive modelling of passive myocardium: a structurally based framework for material characterization},
  year       = {2009},
  issn       = {1364-503X, 1471-2962},
  month      = sep,
  number     = {1902},
  pages      = {3445--3475},
  volume     = {367},
  doi        = {10.1098/rsta.2009.0091},
  language   = {en},
  shorttitle = {Constitutive modelling of passive myocardium},
  urldate    = {2022-11-18},
}

@Article{Dokos2002,
  author    = {Dokos, Socrates and Smaill, Bruce H. and Young, Alistair A. and LeGrice, Ian J.},
  journal   = {American Journal of Physiology-Heart and Circulatory Physiology},
  title     = {Shear properties of passive ventricular myocardium},
  year      = {2002},
  issn      = {1522-1539},
  month     = dec,
  number    = {6},
  pages     = {H2650--H2659},
  volume    = {283},
  doi       = {10.1152/ajpheart.00111.2002},
  publisher = {American Physiological Society},
}

@Article{Grediac2006,
  author    = {Gr{\'{e}}diac, M. and Pierron, F. and Avril, S. and Toussaint, E.},
  journal   = {Strain},
  title     = {The Virtual Fields Method for Extracting Constitutive Parameters From Full-Field Measurements: a Review},
  year      = {2006},
  issn      = {1475-1305},
  month     = nov,
  number    = {4},
  pages     = {233--253},
  volume    = {42},
  doi       = {10.1111/j.1475-1305.2006.tb01504.x},
  publisher = {Wiley},
}

@Misc{Kingma2013,
  author    = {Kingma, Diederik P. and Welling, Max},
  title     = {Auto-Encoding Variational Bayes},
  year      = {2013},
  copyright = {arXiv.org perpetual, non-exclusive license},
  doi       = {10.48550/ARXIV.1312.6114},
  publisher = {arXiv},
}

@Article{Wittrick1965,
  author    = {Wittrick, W. H.},
  journal   = {AIAA Journal},
  title     = {A generalization of macaulay{\textquoteright}s method with applications in structural mechanics},
  year      = {1965},
  issn      = {1533-385X},
  month     = feb,
  number    = {2},
  pages     = {326--330},
  volume    = {3},
  doi       = {10.2514/3.2849},
  publisher = {American Institute of Aeronautics and Astronautics (AIAA)},
}

@Misc{Anton2024,
  author    = {Anton, David and Tr{\"{o}}ger, Jendrik-Alexander and Wessels, Henning and R{\"{o}}mer, Ulrich and Henkes, Alexander and Hartmann, Stefan},
  title     = {Deterministic and statistical calibration of constitutive models from full-field data with parametric physics-informed neural networks},
  year      = {2024},
  copyright = {Creative Commons Attribution 4.0 International},
  doi       = {10.48550/ARXIV.2405.18311},
  publisher = {arXiv},
}

@Article{Ozturk2021,
  author    = {Ozturk, Deniz and Kotha, Shravan and Ghosh, Somnath},
  journal   = {Journal of the Mechanics and Physics of Solids},
  title     = {An uncertainty quantification framework for multiscale parametrically homogenized constitutive models (PHCMs) of polycrystalline Ti alloys},
  year      = {2021},
  issn      = {0022-5096},
  month     = mar,
  pages     = {104294},
  volume    = {148},
  doi       = {10.1016/j.jmps.2021.104294},
  publisher = {Elsevier BV},
}

@Article{Wang2021,
  author    = {Wang, Z. and Estrada, J. B. and Arruda, E. M. and Garikipati, K.},
  journal   = {Journal of the Mechanics and Physics of Solids},
  title     = {Inference of deformation mechanisms and constitutive response of soft material surrogates of biological tissue by full-field characterization and data-driven variational system identification},
  year      = {2021},
  issn      = {0022-5096},
  month     = aug,
  pages     = {104474},
  volume    = {153},
  doi       = {10.1016/j.jmps.2021.104474},
  publisher = {Elsevier BV},
}

@Article{Alloisio2024,
  author    = {Alloisio, Marta and Wolffs, Joey J. M. and Gasser, T. Christian},
  journal   = {Journal of the Mechanical Behavior of Biomedical Materials},
  title     = {Specimen width affects vascular tissue integrity for in-vitro characterisation},
  year      = {2024},
  issn      = {1751-6161},
  pages     = {106520},
  volume    = {154},
  doi       = {10.1016/j.jmbbm.2024.106520},
  publisher = {Elsevier BV},
}

@Book{Pierron2012,
  author    = {Pierron, Fabrice and Gr{\'{e}}diac, Michel},
  publisher = {Springer New York},
  title     = {The Virtual Fields Method: Extracting Constitutive Mechanical Parameters from Full-field Deformation Measurements},
  year      = {2012},
  isbn      = {9781461418245},
  doi       = {10.1007/978-1-4614-1824-5},
}

@Article{Kavanagh1971,
  author    = {Kavanagh, Kenneth T. and Clough, Ray W.},
  journal   = {International Journal of Solids and Structures},
  title     = {Finite element applications in the characterization of elastic solids},
  year      = {1971},
  issn      = {0020-7683},
  month     = jan,
  number    = {1},
  pages     = {11--23},
  volume    = {7},
  doi       = {10.1016/0020-7683(71)90015-1},
  publisher = {Elsevier BV},
}

@Article{Tac2024,
  author    = {Ta{\c{c}}, Vahidullah and Rausch, Manuel K. and Bilionis, Ilias and Sahli Costabal, Francisco and Tepole, Adrian Buganza},
  journal   = {Engineering with Computers},
  title     = {Generative hyperelasticity with physics-informed probabilistic diffusion fields},
  year      = {2024},
  issn      = {1435-5663},
  month     = may,
  number    = {1},
  pages     = {51--69},
  volume    = {41},
  doi       = {10.1007/s00366-024-01984-2},
  publisher = {Springer Science and Business Media LLC},
}

@Article{Matous2017,
  author    = {Matou{\v{s}}, Karel and Geers, Marc G. D. and Kouznetsova, Varvara G. and Gillman, Andrew},
  journal   = {Journal of Computational Physics},
  title     = {A review of predictive nonlinear theories for multiscale modeling of heterogeneous materials},
  year      = {2017},
  issn      = {0021-9991},
  month     = feb,
  pages     = {192--220},
  volume    = {330},
  doi       = {10.1016/j.jcp.2016.10.070},
  publisher = {Elsevier BV},
}

@Article{Linka2025,
  author    = {Linka, Kevin and Holzapfel, Gerhard A. and Kuhl, Ellen},
  journal   = {Computer Methods in Applied Mechanics and Engineering},
  title     = {Discovering uncertainty: Bayesian constitutive artificial neural networks},
  year      = {2025},
  issn      = {0045-7825},
  month     = jan,
  pages     = {117517},
  volume    = {433},
  doi       = {10.1016/j.cma.2024.117517},
  publisher = {Elsevier BV},
}

@Article{Hauseux2018,
  author    = {Hauseux, Paul and Hale, Jack S. and Cotin, St{\'{e}}phane and Bordas, St{\'{e}}phane P. A.},
  journal   = {Applied Mathematical Modelling},
  title     = {Quantifying the uncertainty in a hyperelastic soft tissue model with stochastic parameters},
  year      = {2018},
  issn      = {0307-904X},
  month     = oct,
  pages     = {86--102},
  volume    = {62},
  doi       = {10.1016/j.apm.2018.04.021},
  publisher = {Elsevier BV},
}

@Article{Staber2017,
  author    = {Staber, B. and Guilleminot, J.},
  journal   = {Journal of the Mechanical Behavior of Biomedical Materials},
  title     = {Stochastic hyperelastic constitutive laws and identification procedure for soft biological tissues with intrinsic variability},
  year      = {2017},
  issn      = {1751-6161},
  month     = jan,
  pages     = {743--752},
  volume    = {65},
  doi       = {10.1016/j.jmbbm.2016.09.022},
  publisher = {Elsevier BV},
}

@Article{Elouneg2021,
  author    = {Elouneg, A. and Sutula, D. and Chambert, J. and Lejeune, A. and Bordas, S. P. A. and Jacquet, E.},
  journal   = {Computers \& Structures},
  title     = {An open-source FEniCS-based framework for hyperelastic parameter estimation from noisy full-field data: Application to heterogeneous soft tissues},
  year      = {2021},
  issn      = {0045-7949},
  month     = oct,
  pages     = {106620},
  volume    = {255},
  doi       = {10.1016/j.compstruc.2021.106620},
  publisher = {Elsevier BV},
}

@Article{Aggarwal2023,
  author    = {Aggarwal, Ankush and Jensen, Bj{\o}rn Sand and Pant, Sanjay and Lee, Chung-Hao},
  journal   = {Computer Methods in Applied Mechanics and Engineering},
  title     = {Strain energy density as a Gaussian process and its utilization in stochastic finite element analysis: Application to planar soft tissues},
  year      = {2023},
  issn      = {0045-7825},
  month     = feb,
  pages     = {115812},
  volume    = {404},
  doi       = {10.1016/j.cma.2022.115812},
  publisher = {Elsevier BV},
}

@Article{Peyraut2025,
  author    = {Peyraut, Alice and Genet, Martin},
  journal   = {Comptes Rendus. M{\'{e}}canique},
  title     = {Finite strain formulation of the discrete equilibrium gap principle: application to direct parameter estimation from large full-fields measurements},
  year      = {2025},
  issn      = {1873-7234},
  month     = jan,
  number    = {G1},
  pages     = {259--309},
  volume    = {353},
  doi       = {10.5802/crmeca.279},
  publisher = {Cellule MathDoc/Centre Mersenne},
}

@Article{Budday2017,
  author    = {Budday, S. and Sommer, G. and Birkl, C. and Langkammer, C. and Haybaeck, J. and Kohnert, J. and Bauer, M. and Paulsen, F. and Steinmann, P. and Kuhl, E. and Holzapfel, G. A.},
  journal   = {Acta Biomaterialia},
  title     = {Mechanical characterization of human brain tissue},
  year      = {2017},
  issn      = {1742-7061},
  month     = jan,
  pages     = {319--340},
  volume    = {48},
  doi       = {10.1016/j.actbio.2016.10.036},
  publisher = {Elsevier BV},
}

@Article{Reeps2012,
  author    = {Reeps, C. and Maier, A. and Pelisek, J. and H{\"{a}}rtl, F. and Grabher-Meier, V. and Wall, W. A. and Essler, M. and Eckstein, H.-H. and Gee, M. W.},
  journal   = {Biomechanics and Modeling in Mechanobiology},
  title     = {Measuring and modeling patient-specific distributions of material properties in abdominal aortic aneurysm wall},
  year      = {2012},
  issn      = {1617-7940},
  month     = sep,
  number    = {4},
  pages     = {717--733},
  volume    = {12},
  doi       = {10.1007/s10237-012-0436-1},
  publisher = {Springer Science and Business Media LLC},
}

@Article{Roccabianca2014,
  author    = {Roccabianca, S. and Figueroa, C. A. and Tellides, G. and Humphrey, J. D.},
  journal   = {Journal of the Mechanical Behavior of Biomedical Materials},
  title     = {Quantification of regional differences in aortic stiffness in the aging human},
  year      = {2014},
  issn      = {1751-6161},
  month     = jan,
  pages     = {618--634},
  volume    = {29},
  doi       = {10.1016/j.jmbbm.2013.01.026},
  publisher = {Elsevier BV},
}

@Article{Claire2004,
  author    = {Claire, D. and Hild, F. and Roux, S.},
  journal   = {International Journal for Numerical Methods in Engineering},
  title     = {A finite element formulation to identify damage fields: the equilibrium gap method},
  year      = {2004},
  issn      = {1097-0207},
  month     = jul,
  number    = {2},
  pages     = {189--208},
  volume    = {61},
  doi       = {10.1002/nme.1057},
  publisher = {Wiley},
}

@Article{Meng2025,
  author    = {Meng, Shuangshuang and Yousefi, Ali Akbar Karkhaneh and Avril, St{\'{e}}phane},
  journal   = {Computer Methods in Applied Mechanics and Engineering},
  title     = {Machine-learning-based virtual fields method: Application to anisotropic hyperelasticity},
  year      = {2025},
  issn      = {0045-7825},
  month     = feb,
  pages     = {117580},
  volume    = {434},
  doi       = {10.1016/j.cma.2024.117580},
  publisher = {Elsevier BV},
}

@Article{Jailin2024,
  author    = {Jailin, C. and Benady, A. and Legroux, R. and Baranger, E.},
  journal   = {Experimental Mechanics},
  title     = {Experimental Learning of a Hyperelastic Behavior with a Physics-Augmented Neural Network},
  year      = {2024},
  issn      = {1741-2765},
  month     = sep,
  number    = {9},
  pages     = {1465--1481},
  volume    = {64},
  doi       = {10.1007/s11340-024-01106-5},
  publisher = {Springer Science and Business Media LLC},
}

@Article{Costa2001,
  author    = {Costa, K. D. and Holmes, J. W. and Mcculloch, A. D.},
  journal   = {Philosophical Transactions of the Royal Society of London. Series A: Mathematical, Physical and Engineering Sciences},
  title     = {Modelling cardiac mechanical properties in three dimensions},
  year      = {2001},
  issn      = {1471-2962},
  month     = jun,
  number    = {1783},
  pages     = {1233--1250},
  volume    = {359},
  doi       = {10.1098/rsta.2001.0828},
  editor    = {Kohl, P. and Noble, D. and Hunter, P. J.},
  publisher = {The Royal Society},
}

@Article{Laita2025,
  author    = {Laita, Nicol{\'{a}}s and Mart{\'{i}}nez, Miguel {{\'{A}}}ngel and Doblar{\'{e}}, Manuel and Pe{\~{n}}a, Estefan{\'{i}}a},
  journal   = {Meccanica},
  title     = {On the myocardium modeling under multimodal deformations: a comparison between costa{\textquoteright}s, Holzapfel and Ogden{\textquoteright}s formulations},
  year      = {2025},
  issn      = {1572-9648},
  month     = apr,
  doi       = {10.1007/s11012-025-01959-7},
  publisher = {Springer Science and Business Media LLC},
}

@Article{Rodrigues2022,
  author    = {Rodrigues, F.},
  journal   = {Transportation Research Part B: Methodological},
  title     = {Scaling Bayesian inference of mixed multinomial logit models to large datasets},
  year      = {2022},
  issn      = {0191-2615},
  month     = apr,
  volume    = {158},
  doi       = {10.1016/j.trb.2022.01.005},
  publisher = {Transportation Research Part B: Methodological},
}

@Article{Consonni2007,
  author    = {Consonni, G. and Marin, J.},
  journal   = {Computational Statistics \& Data Analysis},
  title     = {Mean-field variational approximate Bayesian inference for latent variable models},
  year      = {2007},
  issn      = {0167-9473},
  month     = oct,
  volume    = {52},
  doi       = {10.1016/j.csda.2006.10.028},
  publisher = {Computational Statistics \& Data Analysis},
}

@Book{Parisi1988,
  author    = {Parisi, G.},
  publisher = {Redwood City, CA. : Addison-Wesley Pub. Co.},
  title     = {Statistical field theory},
  year      = {1988},
}

@Book{Kullback1968,
  author    = {Kullback, S.},
  publisher = {New York : Dover Publications},
  title     = {Information theory and statistics},
  year      = {1968},
}

@Book{Cover2006,
  author    = {Cover T., Joy A.},
  publisher = {Wiley-Interscience},
  title     = {Information theory and statistics},
  year      = {1968},
}

@Article{Lombaert2012,
  author    = {Lombaert, H. and Peyrat, J. and Croisille, P. and Rapacchi, S. and Fanton, L. and Cheriet, F. and Clarysse, P. and Magnin, I. and Delingette, H. and Ayache, N.},
  journal   = {IEEE Transactions on Medical Imaging},
  title     = {Human Atlas of the Cardiac Fiber Architecture: Study on a Healthy Population},
  year      = {2012},
  issn      = {1558-254X},
  month     = jul,
  number    = {7},
  pages     = {1436--1447},
  volume    = {31},
  doi       = {10.1109/tmi.2012.2192743},
  publisher = {Institute of Electrical and Electronics Engineers (IEEE)},
}

@Article{Flory1961,
  author    = {Flory, P. J.},
  journal   = {Transactions of the Faraday Society},
  title     = {Thermodynamic relations for high elastic materials},
  year      = {1961},
  issn      = {0014-7672},
  pages     = {829},
  volume    = {57},
  doi       = {10.1039/tf9615700829},
  publisher = {Royal Society of Chemistry (RSC)},
}

@InBook{Spencer1984,
  author    = {Spencer, A. J. M.},
  pages     = {1--32},
  publisher = {Springer Vienna},
  title     = {Constitutive Theory for Strongly Anisotropic Solids},
  year      = {1984},
  isbn      = {9783709143360},
  booktitle = {Continuum Theory of the Mechanics of Fibre-Reinforced Composites},
  doi       = {10.1007/978-3-7091-4336-0_1},
}

@Article{Menzel2004,
  author    = {Menzel, A.},
  journal   = {Biomechanics and Modeling in Mechanobiology},
  title     = {Modelling of anisotropic growth in biological tissues: A new approach and computational aspects},
  year      = {2004},
  issn      = {1617-7940},
  month     = oct,
  number    = {3},
  pages     = {147--171},
  volume    = {3},
  doi       = {10.1007/s10237-004-0047-6},
  publisher = {Springer Science and Business Media LLC},
}

@Article{Alberini2024,
  author    = {Alberini, Riccardo and Spagnoli, Andrea and Sadeghinia, Mohammad Javad and Skallerud, Bjorn and Terzano, Michele and Holzapfel, Gerhard A.},
  journal   = {Acta Biomaterialia},
  title     = {Second harmonic generation microscopy, biaxial mechanical tests and fiber dispersion models in human skin biomechanics},
  year      = {2024},
  issn      = {1742-7061},
  month     = sep,
  pages     = {266--280},
  volume    = {185},
  doi       = {10.1016/j.actbio.2024.07.026},
  publisher = {Elsevier BV},
}

@Article{Maes2022,
  author    = {Maes, Arne and Pestiaux, Camille and Marino, Alice and Balcaen, Tim and Leyssens, Lisa and Vangrunderbeeck, Sarah and Pyka, Grzegorz and De Borggraeve, Wim M. and Bertrand, Luc and Beauloye, Christophe and Horman, Sandrine and Wevers, Martine and Kerckhofs, Greet},
  journal   = {Nature Communications},
  title     = {Cryogenic contrast-enhanced microCT enables nondestructive 3D quantitative histopathology of soft biological tissues},
  year      = {2022},
  issn      = {2041-1723},
  month     = oct,
  number    = {1},
  volume    = {13},
  doi       = {10.1038/s41467-022-34048-4},
  publisher = {Springer Science and Business Media LLC},
}

@Article{jafari_bayesian_2025,
  author   = {Jafari, Abbas and Vlachas, Konstantinos and Chatzi, Eleni and Unger, J{\"{o}}rg F.},
  journal  = {Computer Methods in Applied Mechanics and Engineering},
  title    = {A {Bayesian} framework for constitutive model identification via use of full field measurements, with application to heterogeneous materials},
  year     = {2025},
  issn     = {0045-7825},
  month    = jan,
  pages    = {117489},
  volume   = {433},
  doi      = {10.1016/j.cma.2024.117489},
  urldate  = {2024-11-11},
}

@Article{rahmaniNewApproachInverse2013,
  author     = {Rahmani, B. and Mortazavi, F. and Villemure, I. and Levesque, M.},
  journal    = {Composite Structures},
  title      = {A New Approach to Inverse Identification of Mechanical Properties of Composite Materials: {{Regularized}} Model Updating},
  year       = {2013},
  issn       = {0263-8223},
  month      = nov,
  pages      = {116--125},
  volume     = {105},
  doi        = {10.1016/j.compstruct.2013.04.025},
  shorttitle = {A New Approach to Inverse Identification of Mechanical Properties of Composite Materials},
  urldate    = {2025-05-16},
}

@Article{Marek2019,
  author   = {Marek, Aleksander and Davis, Frances M. and Rossi, Marco and Pierron, Fabrice},
  journal  = {Int J Mater Form},
  title    = {Extension of the Sensitivity-Based Virtual Fields to Large Deformation Anisotropic Plasticity},
  year     = {2019},
  issn     = {1960-6214},
  month    = may,
  number   = {3},
  pages    = {457--476},
  volume   = {12},
  doi      = {10.1007/s12289-018-1428-1},
  langid   = {english},
  urldate  = {2025-05-20},
}

@Article{Linka2023,
  author   = {Linka, Kevin and Kuhl, Ellen},
  journal  = {Computer Methods in Applied Mechanics and Engineering},
  title    = {A New Family of {{Constitutive Artificial Neural Networks}} towards Automated Model Discovery},
  year     = {2023},
  issn     = {0045-7825},
  month    = jan,
  pages    = {115731},
  volume   = {403},
  doi      = {10.1016/j.cma.2022.115731},
  urldate  = {2024-02-05},
}

@Article{avazmohammadi_integrated_2018,
  author     = {Avazmohammadi, Reza and Li, David S. and Leahy, Thomas and Shih, Elizabeth and Soares, Jo{\~{a}}o S. and Gorman, Joseph H. and Gorman, Robert C. and Sacks, Michael S.},
  journal    = {Biomechanics and Modeling in Mechanobiology},
  title      = {An integrated inverse model-experimental approach to determine soft tissue three-dimensional constitutive parameters: application to post-infarcted myocardium},
  year       = {2018},
  issn       = {1617-7940},
  month      = feb,
  number     = {1},
  pages      = {31--53},
  volume     = {17},
  doi        = {10.1007/s10237-017-0943-1},
  language   = {en},
  shorttitle = {An integrated inverse model-experimental approach to determine soft tissue three-dimensional constitutive parameters},
  urldate    = {2023-01-05},
}

@Article{Thorat2025,
  author   = {Thorat, Rohan Vittal and Anas, Mohammad and Nayek, Rajdip and Chatterjee, Sabyasachi},
  journal  = {Probabilistic Engineering Mechanics},
  title    = {System Identification and Reliability Assessment of Hyperelastic Materials via an Efficient Sparsity-Promoting Variational {{Bayesian}} Approach},
  year     = {2025},
  issn     = {0266-8920},
  month    = apr,
  pages    = {103763},
  volume   = {80},
  doi      = {10.1016/j.probengmech.2025.103763},
  urldate  = {2025-06-11},
}

@Book{Abaqus,
  author    = {{Dassault Syst{\`{e}}mes Simulia Corp.}},
  publisher = {Dassault Syst{\`{e}}mes Simulia Corp.},
  title     = {Abaqus Analysis User{\textquoteright}s guide},
  year      = {2025},
  address   = {Providence, RI, USA},
}

@article{Navy2025,
    title = {Three-Dimensional Tissue Strain Measurement Using a Row–Column Array During Biaxial Testing of Excised Ventricular Porcine Myocardium},
    journal = {Ultrasound in Medicine \& Biology},
    volume = {51},
    number = {9},
    pages = {1622-1626},
    year = {2025},
    issn = {0301-5629},
    doi = {https://doi.org/10.1016/j.ultrasmedbio.2025.05.007},
    author = {Xavier Navy and Zhiyu Sheng and Kang Kim and John M. Cormack},
}

@article{Davis2024,
    title = {Comparison of two contrast-enhancing staining agents for use in X-ray imaging and digital volume correlation measurements across the cartilage-bone interface},
    journal = {Journal of the Mechanical Behavior of Biomedical Materials},
    volume = {152},
    pages = {106414},
    year = {2024},
    issn = {1751-6161},
    doi = {https://doi.org/10.1016/j.jmbbm.2024.106414},
    author = {Sarah Davis and Aikaterina Karali and Tim Balcaen and Jurgita Zekonyte and Maïté Pétré and Marta Roldo and Greet Kerckhofs and Gordon Blunn},
}

@article{Ghouli2025,
  title = {A topology optimisation framework to design test specimens for one-shot identification or discovery of material models},
  volume = {203},
  ISSN = {0022-5096},
  DOI = {10.1016/j.jmps.2025.106210},
  journal = {Journal of the Mechanics and Physics of Solids},
  publisher = {Elsevier BV},
  author = {Ghouli,  Saeid and Flaschel,  Moritz and Kumar,  Siddhant and De Lorenzis,  Laura},
  year = {2025},
  month = oct,
  pages = {106210}
}

@Article{Higham1987,
  author    = {Higham, Nicholas J.},
  journal   = {Linear Algebra and its Applications},
  title     = {Computing real square roots of a real matrix},
  year      = {1987},
  issn      = {0024-3795},
  month     = apr,
  pages     = {405--430},
  volume    = {88–89},
  doi       = {10.1016/0024-3795(87)90118-2},
  publisher = {Elsevier BV},
}

@Article{Niederer2019,
  author    = {Niederer, Steven A. and Campbell, Kenneth S. and Campbell, Stuart G.},
  journal   = {Journal of Molecular and Cellular Cardiology},
  title     = {A short history of the development of mathematical models of cardiac mechanics},
  year      = {2019},
  doi       = {10.1016/j.yjmcc.2018.11.015},
  publisher = {Elsevier BV},
}

@Article{Famaey2026,
  author    = {Famaey, Nele and Fehervary, Heleen and Lafon, Yoann and Akyildiz, Ali and Dreesen, Silke and Bruy{\`{e}}re-Garnier, Karine and Allain, Jean-Marc and Alloisio, Marta and Aparici-Gil, Alejandro and Catalano, Chiara and Chassagne, Fanette and Chokhandre, Snehal and Crevits, Kimberly and Crielaard, Hanneke and Cunnane, Eoghan and Cunnane, Connor and De Leener, Karen and Desai, Amisha and Driessen, Rob and Erdemir, Ahmet and Eskandari, Mona and Evans, Sam and Gasser, Christian and Gebhardt, Marc and Glasmacher, Birgit and Holzapfel, Gerhard A. and Isasi, Mikel and Jennings, Louise and Kurz, Sascha and Leal-Marin, Sara and Lecomte, Pauline and Morch, Annie and Mulvihill, John and Nemavhola, Fulufhelo and Pandelani, Thanyani and Pasta, Salvatore and Pe{\~{n}}a, Estefania and Pierrat, Baptiste and Ploeg, Heidi-Lynn and Polzer, Stanislav and Rausch, Manuel and Schwarz, David and Screen, Hazel and Sherifova, Selda and Sommer, Gerhard and Wang, Shengzhang and Walsh, Darragh and Yadav, Deepesh and Marchal, Thierry and Geris, Liesbet},
  journal   = {Journal of Biomechanics},
  title     = {Community challenge towards consensus on characterization of biological tissue: C4Bio{\textquoteright}s first findings},
  year      = {2026},
  doi       = {10.1016/j.jbiomech.2025.113021},
  publisher = {Elsevier BV},
}

@Article{Vervenne2026,
  author    = {Vervenne, Thibault and Vermeeren, Nic and Demeersseman, Nele and Fehervary, Heleen and Peirlinck, Mathias and Kuhl, Ellen and Famaey, Nele},
  journal   = {Journal of Biomechanical Engineering},
  title     = {Stretching the Limits: From Planar-Biaxial Stress--Stretch to Arterial Pressure--Diameter},
  year      = {2026},
  doi       = {10.1115/1.4070124},
  publisher = {ASME International},
}

@Article{Lazarus2022,
  author    = {Lazarus, Alan and Dalton, David and Husmeier, Dirk and Gao, Hao},
  journal   = {Biomechanics and Modeling in Mechanobiology},
  title     = {Sensitivity analysis and inverse uncertainty quantification for the left ventricular passive mechanics},
  year      = {2022},
  doi       = {10.1007/s10237-022-01571-8},
  publisher = {Springer Science and Business Media LLC},
}

@Article{Gao2015,
  author    = {Gao, H. and Li, W. G. and Cai, L. and Berry, C. and Luo, X. Y.},
  journal   = {Journal of Engineering Mathematics},
  title     = {Parameter estimation in a Holzapfel–Ogden law for healthy myocardium},
  year      = {2015},
  doi       = {10.1007/s10665-014-9740-3},
  publisher = {Springer Science and Business Media LLC},
}

@Article{Palit2018,
  author    = {Palit, Arnab and Bhudia, Sunil K. and Arvanitis, Theodoros N. and Turley, Glen A. and Williams, Mark A.},
  journal   = {Medical \& Biological Engineering \& Computing},
  title     = {In vivo estimation of passive biomechanical properties of human myocardium},
  year      = {2018},
  doi       = {10.1007/s11517-017-1768-x},
  publisher = {Springer Science and Business Media LLC},
}

@Article{Balaban2016,
  author    = {Balaban, Gabriel and Alnæs, Martin S. and Sundnes, Joakim and Rognes, Marie E.},
  journal   = {Biomechanics and Modeling in Mechanobiology},
  title     = {Adjoint multi-start-based estimation of cardiac hyperelastic material parameters using shear data},
  year      = {2016},
  doi       = {10.1007/s10237-016-0780-7},
  publisher = {Springer Science and Business Media LLC},
}

@Misc{Chaurasiya2025,
  author    = {Chaurasiya, Kanhaiya Lal and Dutta, Saurav and Kumar, Siddhant and Joshi, Akshay},
  title     = {Hetero-EUCLID: Interpretable model discovery for heterogeneous hyperelastic materials using stress-unsupervised learning},
  year      = {2025},
  copyright = {arXiv.org perpetual, non-exclusive license},
  doi       = {10.48550/ARXIV.2509.11784},
  publisher = {arXiv},
}

@Article{Ge2025,
  author    = {Ge, Yuzhang and Husmeier, Dirk and Rabbani, Arash and Gao, Hao},
  title     = {Advanced statistical inference of myocardial stiffness: A time series Gaussian process approach of emulating cardiac mechanics for real-time clinical decision support},
  year      = {2025},
  doi       = {10.1016/j.compbiomed.2024.109381},
  publisher = {Elsevier BV},
}

\clearpage
\begin{appendix}
\newpage
\section{Tissue slab geometry and loading conditions}
\label{ap:appendix_tissueslab}
\begin{figure}[ht]
    \centering
    \includegraphics[width=0.95\textwidth]{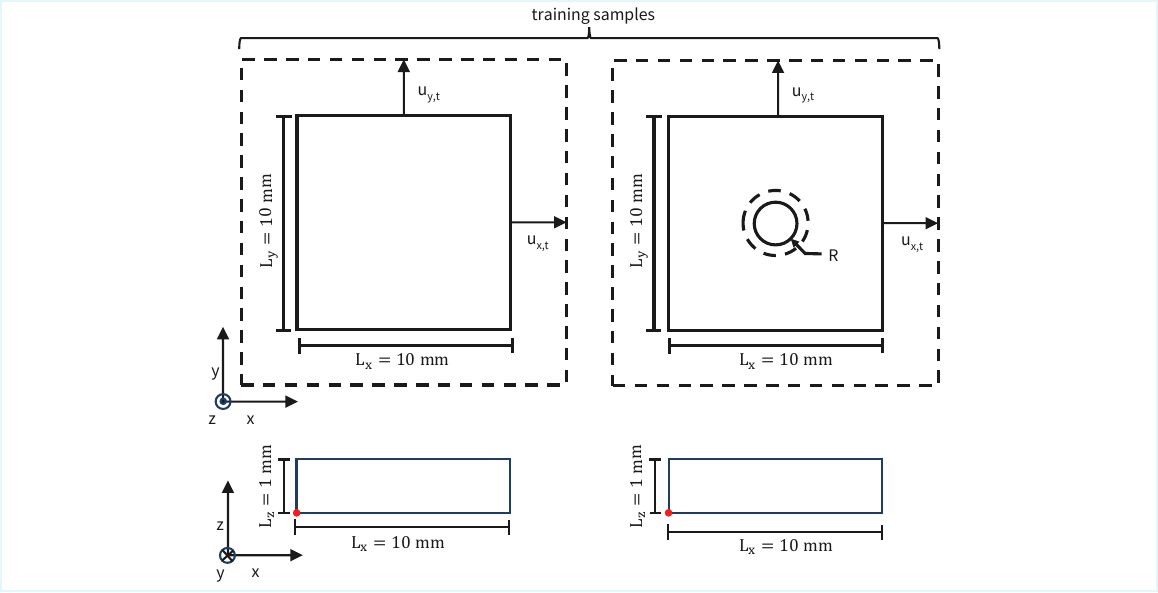}
    \caption[]{\textbf{Synthetic data generation: training sample geometry and loading conditions.}
    Tissue slab geometry and applied boundary conditions for the training samples.
    All specimens are of size $L_x \times L_y \times L_z = 10 \times 10 \times 1~\mathrm{mm}^3$.
    All training samples are subjected to a biaxial loading protocol with loading ratios $\lambda_1:\lambda_2=(1:2)$, $(1:1)$, and $(2:1)$ up to $15\%$ of the initial specimen length.
    The geometrically heterogeneous tissue samples contain a central circular occlusion of radius $R=1~\mathrm{mm}$.}
    \label{fig:geometry_and_bcs}
\end{figure}

We summarize the tissue slab geometry and loading conditions used for synthetic data generation in Figure \ref{fig:geometry_and_bcs}.
All specimens are of size $L_x \times L_y \times L_z = 10 \times 10 \times 1~\mathrm{mm}^3$.
Geometrically heterogeneous specimen have a circular occlusion in the center of the specimen with a radius of $1 \ mm$.
All training samples are subjected to a biaxial loading
protocol with loading ratios $\lambda_1:\lambda_2=(1:2)$, $(1:1)$, and $(2:1)$ up to $15\%$ of the initial specimen length.
These loads are depicted as $u_{x,t}$ and $u_{y,t}$ in the Figure.

\label{sec:appendix_geometries}

{\revision{
\section{Spatially varying and correlated displacement noise}
\label{ap:spatial-noise}
To define a structured displacement perturbation field, we consider a noise model that combines two effects: 
(i) a spatial increase in noise amplitude near specimen boundaries, and 
(ii) spatial correlation between nearby nodes.

We first define, for each Cartesian direction, the distance to the nearest specimen boundary:
\begin{equation}
d_x(X_x) = \min(L_x - X_x, X_x), \qquad
d_y(X_y) = \min(L_y - X_y, X_y), \qquad
d_z(X_z) = \min(L_z - X_z, X_z),
\label{eq:boundary_distance}
\end{equation}
where $(L_x,L_y,L_z)$ denote the specimen dimensions shown in Figure \ref{fig:geometry_and_bcs}.
These distance functions quantify how close a given degree of freedom is to the corresponding specimen boundary.

Next, we use this distance to define a spatially varying local noise amplitude:
\begin{equation}
\sigma^k(X_k) = \sigma_{\mathrm{bulk}}
\left(
1 + \alpha \exp\left[-\frac{d_k(X_k)^2}{2\lambda^2}\right]
\right),
\qquad k \in \{x,y,z\}.
\label{eq:local_noise_amplitude}
\end{equation}
Here, $\sigma_{\mathrm{bulk}}$ denotes the baseline noise level away from the boundaries, $\alpha$ controls the strength of the boundary amplification, and $\lambda$ sets the characteristic decay length scale of this amplification into the bulk.
Thus, $\sigma^k(X_k)$ increases near the specimen boundaries and approaches $\sigma_{\mathrm{bulk}}$ farther away from them.

To introduce spatial correlation, we start from an independent zero-mean Gaussian process in each displacement direction,
\begin{equation}
W_k(X) \sim \mathcal{GP}\!\left(0,\delta^3(\vec{X}-\vec{X}')\right),
\qquad k \in \{x,y,z\},
\label{eq:white_gp}
\end{equation}
where
\begin{equation}
\delta^3(\vec{X}-\vec{X}') = \delta(X_x-X_x')\,\delta(X_y-X_y')\,\delta(X_z-X_z')
\label{eq:dirac3}
\end{equation}
denotes the three-dimensional Dirac delta distribution.
This defines a spatially uncorrelated latent field.
We then correlate this field by convolution with a Gaussian kernel,
\begin{equation}
G_h(r) = \exp\left[-\frac{\|r\|^2}{2h^2}\right],
\label{eq:gaussian_kernel}
\end{equation}
where $h$ is the correlation length scale.
The correlated displacement noise field is then defined as
\begin{equation}
{\sigma}^{k}_u({X}_{}) = 
\sigma^{k}({X}_{k})\int_{\Omega}G_h(\vec{X}_{}-\vec{X}_{}')W_{k}(\vec{X}_{}')d{\vec{X}_{}}',
\qquad k \in \{x,y,z\},
\label{eq:continuous_correlated_noise}
\end{equation}
and we collect the three components into
\[
\vec{\sigma}_u(\vec{X})=\{{\sigma}^{x}_u, {\sigma}^{y}_u, {\sigma}^{z}_u\}.
\]
In this construction, $\sigma^k(X_k)$ determines the local amplitude of the noise, while the convolution with $G_h$ determines its spatial correlation.
The boundary amplification therefore depends on the distance to the nearest boundary, whereas the correlation structure depends on the Euclidean distance between spatial locations.

In this study, we set
\[
h = 0.5, \qquad \lambda = 1, \qquad \alpha = 10,
\]
which yields an approximate order-of-magnitude increase in noise near the specimen boundaries relative to the bulk level noise $\sigma_{\mathrm{bulk}} = 10^{-4}$.
For the present specimen geometry, the amplification decays over a characteristic length scale of approximately 1~mm.
Because the specimen thickness is only 1~mm, all displacement degrees of freedom in the $z$-direction lie within the boundary-affected region and therefore experience elevated noise levels throughout the thickness.

To apply this noise model in our discretized displacement field setting, we consider the discrete nodes in the domain $\vec{X}^i, i\in 1,...,N$ (with $N$ the number of nodes in the mesh) and define the distance metric:

\begin{equation}
    \begin{aligned}
        &d^{i}_{k}({X}^{i}_{k}) = min(L_{k}-{X}^{i}_{k}, {X}^{i}_{k}), \quad k \in \{x, y, z\}\\
    \end{aligned}
\end{equation}

Subsequently, we define the spatial increase in noise due to the distance to the boundaries

\begin{equation}
\sigma_i^{k}({X^i_{k}}) = \sigma_{bulk}\left(1+\alpha \ \mathrm{exp}\left[-\frac{d^i_{k}({X}^i_{k})^2}{2 \lambda^2}\right]\right),
\end{equation}
and the correlation matrix for the spatial noise correlation between nodes
\begin{equation}
C_{ij} = \mathrm{exp}\left(-\frac{||\vec{X^i} - \vec{X^j}||^2}{2h^2}\right).
\end{equation}

Finally, we independently sample realizations of the displacement noise from
\begin{equation}
\vec{\sigma}_{u}^{k}(\vec{X}) \sim \mathcal{N}(\vec{0}, \vec{D}_{k}\mat{C}\vec{D}_{k}), \quad \vec{D}_{k}=\mathrm{diag}(\sigma_1^{k}, ..., \sigma_N^{k}).
\end{equation}
Here $\vec{\sigma}_{u}(\vec{X})=\{\vec{\sigma}_{u}^{x}, \vec{\sigma}_{u}^{y}, \vec{\sigma}_{u}^{z}\} \in \mathbb{R}^{N \times 3}$  denotes a realization of the displacement noise field, not its standard deviation.
}}

\section{Uni-modal validation protocols and stress expressions}
\label{ap:homogeneous_validation_protocols}

This appendix summarizes the homogeneous deformation modes used for validation in Section~\ref{ssec:unimodalvalidation}. 
We reproduce incompressible biaxial tension and triaxial shear protocols following \citep{Dokos2002,Sommer2015} and evaluate the resulting Cauchy stress responses.

We prescribe a spatially homogeneous deformation gradient
\begin{equation}\label{eq:hom_defgrad_app}
\ten{F} 
= \left[ 
  \begin{array}{lll}
  \lambda_{\rm{f}} & \gamma_{\rm{fs}} & \gamma_{\rm{fn}} \\
  \gamma_{\rm{sf}} & \lambda_{\rm{s}} & \gamma_{\rm{sn}} \\
  \gamma_{\rm{nf}} & \gamma_{\rm{ns}} & \lambda_{\rm{n}} \\
  \end{array}
  \right].
\end{equation}

For \textit{biaxial tension}, $\lambda_{\rm f}$ and $\lambda_{\rm n}$ are prescribed, incompressibility implies
$\lambda_{\rm s}=1/(\lambda_{\rm f}\lambda_{\rm n})$, and all shear components are set to zero.
The hydrostatic pressure is obtained by enforcing $\sigma_{\rm ss}=0$, which yields the following non-zero stresses
\beq
  \lambda_{\rm f}\ge 1,\quad
  \lambda_{\rm n}\ge 1,\quad
  \lambda_{\rm s}=\frac{1}{\lambda_{\rm f}\lambda_{\rm n}}\le 1:
  \qquad
  \begin{array}{l@{\hspace*{0.1cm}}c@{\hspace*{0.1cm}}
                l@{\hspace*{0.1cm}}c@{\hspace*{0.1cm}}
                c@{\hspace*{0.1cm}}l@{\hspace*{0.1cm}}
                c@{\hspace*{0.1cm}}c@{\hspace*{0.1cm}}
                l@{\hspace*{0.1cm}}l}
 \D{\sigma_{\rm ff}}
&\D{=}
&\D{2\psi_1}
&\D{[\lambda_{\rm f}^2-\lambda_{\rm s}^2]}
&\D{+}
&\D{2\psi_{\rm 4ff}}
{\lambda_{\rm f}^2}
\\[10.pt]
 \D{\sigma_{\rm nn}}
&\D{=}
&\D{2\psi_1}
&\D{[\lambda_{\rm n}^2-\lambda_{\rm s}^2]}
&\D{+}
&\D{2\psi_{\rm 4nn}}
{\lambda_{\rm n}^2}\,.
  \end{array}
\eeq
Here $\psi_{\star}$ denotes $\psi_{\star}=\partial \psi / \partial I_{\star}$.

For the \textit{triaxial simple shear} cases we consider, $\lambda_{\rm f}=\lambda_{\rm s}=\lambda_{\rm n}\equiv 1$ and a single shear component is activated per test while all others are set to zero.
For $\gamma_{\star}\ge 0$, the non-zero shear stresses read
\beq
\begin{alignedat}{2}
\;&
\sigma_{\rm sf}=2\,\gamma_{\rm fs}\,\psi_1
+ \psi_{\rm 4fs},
&\qquad
&\sigma_{\rm fs}=2\,\gamma_{\rm sf}\!\left[\psi_1
+ \psi_{\rm 4ff}\right]
+ \psi_{\rm 4fs}, \\[6pt]
\;&
\sigma_{\rm nf}=2\,\gamma_{\rm fn}\!\left[\psi_1
+ \psi_{\rm 4nn}\right],
&\qquad
&\sigma_{\rm fn}=2\,\gamma_{\rm nf}\!\left[\psi_1
+ \psi_{\rm 4ff}\right], \\[6pt]
\;&
\sigma_{\rm ns}=2\,\gamma_{\rm sn}\!\left[\psi_1
+ \psi_{\rm 4nn}\right],
&\qquad
&\sigma_{\rm sn}=2\,\gamma_{\rm ns}\psi_1.
\end{alignedat}
\eeq

\section{Mixed-modal invariant sampling and admissibility filtering}
\label{ap:mixed_modal_sampling}

We sample mixed-modal deformation states by drawing invariants uniformly between the extrema induced by the homogeneous deformation modes in Eq.~\eqref{eq:hom_defgrad_app}.
We then reconstruct $\ten{C}$ in the local fiber-sheet-normal basis, compute $\sqrt{\ten{C}}$ via Schur decomposition, and evaluate the maximum principal Cauchy stress used for admissibility filtering.

We take the minimum and maximum values of the invariants induced by Eq.~\eqref{eq:hom_defgrad_app} and uniformly sample in between:
\begin{equation}
    I_{\star} \sim \mathcal{U}(I_{\star}^{\min}, I_{\star}^{\max}), 
    \quad I_\star \in \{I_1, I_{4ff}, I_{4nn}, I_{4fs}\}.
\end{equation}
where $I_{\star}^{\min}$ and $I_{\star}^{\max}$ denote the minimum and maximum deformation invariants seen in the uni-modal biaxial tension and triaxial simple shear experiments studied in Section \ref{ssec:unimodalvalidation}.
We reconstruct the right Cauchy-Green tensor using the identity $I_1=I_{4ff}+I_{4ss}+I_{4nn}$ and assuming the same orientation basis as in the uni-modal spatially homogeneous loading case:
\begin{equation}
    \begin{aligned}
        &C_{11} = I_{4ff}, \qquad && C_{13} = C_{31} = C_{23} = C_{32} = 0,\\
        &C_{22} = I_1 - I_{4ff} - I_{4nn}, \qquad && C_{12} = C_{21} = I_{4fs},\\
        &C_{33} = I_{4nn}. &&
    \end{aligned}
\end{equation}
Using the Schur decomposition \citep{Higham1987}, we compute the right stretch tensor 
$\ten{U}=\sqrt{\ten{C}}$ and use the polar decomposition $\ten{F}=\ten{R}\ten{U}$.
We then evaluate the Cauchy stress and extract the maximum principal stress:
\begin{equation}\label{eq:max_stress_U_identity_app}
    \begin{aligned}
        \sigma_{1} &= \mathcal{F}_{1}( J^{-1}\ten{F}\ten{S}\ten{F}^{T})
        = \mathcal{F}_{1}( J^{-1}\ten{R}\ten{U}\ten{S}\ten{U}\ten{R}^{T})
        = \mathcal{F}_{1}( J^{-1}\ten{U}\ten{S}\ten{U}),
    \end{aligned}
\end{equation}
where $\mathcal{F}_{1}(\ten{\sigma}) = \max \operatorname{eig}(\ten{\sigma})$ and $\ten{S}=\partial \Psi/\partial \ten{C}$.
We accept only physiologically relevant samples satisfying $0.9999 \le J \le 1.0001$, aligned with the tissue's assumed quasi-incompressibility, and $\sigma_{1} \le 20.3~\mathrm{kPa}$, aligned with the maximum principal stresses measured in experimental protocols on human myocardial tissue \citep{Sommer2015} (see Table 3 showing peak Cauchy biaxial tensile stresses in the left ventricle). We repeat sampling until $5{,}000$ states are retained.
%

\section{Noise-free orthotropic constitutive parameter inference results}
\label{ap:noise-free-parameter-results}
%
This appendix summarizes the orthotropic constitutive parameter inference results under noise-free conditions.
{\revision{In Table~\ref{tab:noise-free-hyperparameter-estimates}, we report the optimized hyperparameters at the final epoch.
For the inverse-gamma distribution employed, the mean is defined for values $\alpha > 1$ and can be computed as $\mu_{\mathcal{IG}} := \frac{\beta}{\alpha - 1}$.
The inferred material model parameters and their accompanying standard deviations are reported in Table~\ref{tab:noise-free-parameter-estimates}}}, the relative errors with respect to the ground-truth parameters are reported in Table~\ref{tab:noise-free-relative-error}, and the corresponding unbiased posterior skewness values are reported in Table~\ref{tab:noise-free-skewness}.
\begin{table}[H]
    
\caption{\textbf{Hyperparameter estimates under noise-free conditions.}
Optimized inverse-gamma hyperparameters at the final epoch for the homogeneous and heterogeneous training specimens.
The parameters $\alpha^\star$ and $\beta^\star$ denote the shape and rate parameters of the corresponding inverse-gamma distributions.}
\label{tab:noise-free-hyperparameter-estimates}
\begin{tabular}{lrrrrrr}
\toprule
\textbf{Hyperparameters} & \textbf{circ-long-hom} & \textbf{circ-rad-hom} & \textbf{rad-heli-hom} & \textbf{circ-long-het} & \textbf{circ-rad-het} & \textbf{rad-heli-het} \\
\midrule
$\alpha_{v_s}$ & $1.691\times10^{1}$ & $1.803\times10^{1}$ & $1.800\times10^{1}$ & $1.721\times10^{1}$ & $1.802\times10^{1}$ & $1.799\times10^{1}$ \\
$\beta_{v_s}$ & $7.215\times10^{7}$ & $7.008\times10^{7}$ & $6.978\times10^{7}$ & $7.393\times10^{7}$ & $7.506\times10^{7}$ & $7.491\times10^{7}$ \\
$\alpha_{\sigma^2}$ & $5.204\times10^{6}$ & $5.210\times10^{6}$ & $5.222\times10^{6}$ & $5.404\times10^{6}$ & $5.401\times10^{6}$ & $5.416\times10^{6}$ \\
$\beta_{\sigma^2}$ & $1.920\times10^{1}$ & $1.897\times10^{1}$ & $1.898\times10^{1}$ & $1.850\times10^{1}$ & $1.821\times10^{1}$ & $1.831\times10^{1}$ \\
\bottomrule
\end{tabular}

\end{table}
\begin{table}[H]
    
\caption{\textbf{Orthotropic constitutive parameter estimates under noise-free conditions.}
Inferred constitutive parameters at the final epoch for the homogeneous and heterogeneous training specimens.
Parameters in $\vec{\theta}$ are reported using the notation of the Holzapfel--Ogden model, together with the corresponding posterior standard deviations.}
\label{tab:noise-free-parameter-estimates}
\begin{tabular}{lllllll}
\toprule
\textbf{Parameters} & \textbf{circ-long-hom} & \textbf{circ-rad-hom} & \textbf{rad-heli-hom} & \textbf{circ-long-het} & \textbf{circ-rad-het} & \textbf{rad-heli-het} \\
\midrule
\textit{Inferred $\mu_{\vec{\theta}}$:} &  &  &  &  &  &  \\
$a$ & $8.090\times10^{-4}$ & $8.090\times10^{-4}$ & $8.090\times10^{-4}$ & $8.090\times10^{-4}$ & $8.090\times10^{-4}$ & $8.090\times10^{-4}$ \\
$b$ & $7.474\times10^{0}$ & $7.474\times10^{0}$ & $7.474\times10^{0}$ & $7.474\times10^{0}$ & $7.474\times10^{0}$ & $7.474\times10^{0}$ \\
$a_{f}$ & $1.911\times10^{-3}$ & $1.911\times10^{-3}$ & $1.911\times10^{-3}$ & $1.911\times10^{-3}$ & $1.911\times10^{-3}$ & $1.911\times10^{-3}$ \\
$b_{f}$ & $2.206\times10^{1}$ & $2.206\times10^{1}$ & $2.206\times10^{1}$ & $2.206\times10^{1}$ & $2.206\times10^{1}$ & $2.206\times10^{1}$ \\
$a_{n}$ & $2.270\times10^{-4}$ & $2.270\times10^{-4}$ & $2.270\times10^{-4}$ & $2.270\times10^{-4}$ & $2.270\times10^{-4}$ & $2.270\times10^{-4}$ \\
$b_{n}$ & $3.480\times10^{1}$ & $3.480\times10^{1}$ & $3.480\times10^{1}$ & $3.480\times10^{1}$ & $3.480\times10^{1}$ & $3.480\times10^{1}$ \\
$a_{fs}$ & $3.139\times10^{0}$ & $5.447\times10^{-4}$ & $5.481\times10^{-4}$ & $5.436\times10^{-4}$ & $5.478\times10^{-4}$ & $5.478\times10^{-4}$ \\
$b_{fs}$ & $3.256\times10^{0}$ & $4.921\times10^{0}$ & $4.961\times10^{0}$ & $3.147\times10^{0}$ & $5.695\times10^{0}$ & $5.675\times10^{0}$ \\
$K$ & $1.000\times10^{-1}$ & $10.000\times10^{-2}$ & $1.000\times10^{-1}$ & $1.000\times10^{-1}$ & $1.000\times10^{-1}$ & $1.000\times10^{-1}$ \\
\midrule \textit{Inferred $\sigma_{\vec{\theta}}$:} &  &  &  &  &  &  \\
$\sigma_{a}$ & $1.494\times10^{-8}$ & $1.507\times10^{-8}$ & $1.526\times10^{-8}$ & $1.415\times10^{-8}$ & $1.580\times10^{-8}$ & $1.653\times10^{-8}$ \\
$\sigma_{b}$ & $1.044\times10^{-4}$ & $8.826\times10^{-5}$ & $9.443\times10^{-5}$ & $1.090\times10^{-4}$ & $1.072\times10^{-4}$ & $1.119\times10^{-4}$ \\
$\sigma_{a_{f}}$ & $2.502\times10^{-9}$ & $2.659\times10^{-8}$ & $2.948\times10^{-8}$ & $3.981\times10^{-9}$ & $4.635\times10^{-8}$ & $2.670\times10^{-8}$ \\
$\sigma_{b_{f}}$ & $1.835\times10^{-5}$ & $1.972\times10^{-4}$ & $1.879\times10^{-4}$ & $1.491\times10^{-5}$ & $6.669\times10^{-4}$ & $1.929\times10^{-4}$ \\
$\sigma_{a_{n}}$ & $1.337\times10^{-9}$ & $1.143\times10^{-7}$ & $6.658\times10^{-9}$ & $1.059\times10^{-9}$ & $1.098\times10^{-7}$ & $6.207\times10^{-9}$ \\
$\sigma_{b_{n}}$ & $4.554\times10^{-5}$ & $3.477\times10^{-2}$ & $3.559\times10^{-4}$ & $4.139\times10^{-5}$ & $3.142\times10^{-2}$ & $3.382\times10^{-4}$ \\
$\sigma_{a_{fs}}$ & $1.342\times10^{1}$ & $3.965\times10^{-5}$ & $4.295\times10^{-5}$ & $6.385\times10^{-5}$ & $3.896\times10^{-6}$ & $1.010\times10^{-5}$ \\
$\sigma_{b_{fs}}$ & $1.370\times10^{1}$ & $4.445\times10^{0}$ & $3.318\times10^{0}$ & $1.209\times10^{1}$ & $4.433\times10^{-2}$ & $2.674\times10^{-1}$ \\
$\sigma_{K}$ & $1.965\times10^{-7}$ & $7.520\times10^{-7}$ & $6.438\times10^{-7}$ & $2.033\times10^{-7}$ & $8.810\times10^{-7}$ & $6.560\times10^{-7}$ \\
\bottomrule
\end{tabular}

\end{table}
\begin{table}[H]
    \caption{\textbf{Relative errors of the orthotropic constitutive parameter estimates under noise-free conditions.}
Relative errors at the final epoch between the inferred constitutive parameters and the ground-truth parameter values for the homogeneous and heterogeneous training specimens.}
\label{tab:noise-free-relative-error}
\begin{tabular}{lrrrrrr}
\toprule
\textbf{Errors} & \textbf{circ-long-hom} & \textbf{circ-rad-hom} & \textbf{rad-heli-hom} & \textbf{circ-long-het} & \textbf{circ-rad-het} & \textbf{rad-heli-het} \\
\midrule
$\hat{e}_{a}$ & $-1.022\times10^{-5}$ & $4.572\times10^{-5}$ & $-1.448\times10^{-5}$ & $7.019\times10^{-8}$ & $7.098\times10^{-7}$ & $-1.826\times10^{-5}$ \\
$\hat{e}_{b}$ & $-2.568\times10^{-5}$ & $2.159\times10^{-5}$ & $-3.091\times10^{-5}$ & $-1.492\times10^{-5}$ & $-2.521\times10^{-5}$ & $-3.542\times10^{-5}$ \\
$\hat{e}_{a_{f}}$ & $-8.278\times10^{-5}$ & $9.389\times10^{-6}$ & $-6.745\times10^{-5}$ & $-6.743\times10^{-5}$ & $-9.161\times10^{-5}$ & $-6.393\times10^{-5}$ \\
$\hat{e}_{b_{f}}$ & $-3.188\times10^{-5}$ & $4.899\times10^{-5}$ & $-2.662\times10^{-5}$ & $-1.735\times10^{-5}$ & $-6.930\times10^{-5}$ & $-2.107\times10^{-5}$ \\
$\hat{e}_{a_{n}}$ & $1.158\times10^{-4}$ & $-6.964\times10^{-5}$ & $5.965\times10^{-5}$ & $9.178\times10^{-5}$ & $-8.719\times10^{-5}$ & $7.453\times10^{-5}$ \\
$\hat{e}_{b_{n}}$ & $2.496\times10^{-5}$ & $6.033\times10^{-5}$ & $-2.814\times10^{-5}$ & $1.181\times10^{-6}$ & $3.278\times10^{-5}$ & $-1.752\times10^{-5}$ \\
$\hat{e}_{a_{fs}}$ & $-5.738\times10^{3}$ & $4.131\times10^{-3}$ & $-1.945\times10^{-3}$ & $6.163\times10^{-3}$ & $-1.459\times10^{-3}$ & $-1.505\times10^{-3}$ \\
$\hat{e}_{b_{fs}}$ & $4.279\times10^{-1}$ & $1.354\times10^{-1}$ & $1.283\times10^{-1}$ & $4.471\times10^{-1}$ & $-7.901\times10^{-4}$ & $2.882\times10^{-3}$ \\
$\hat{e}_{K}$ & $-3.005\times10^{-5}$ & $3.963\times10^{-5}$ & $-3.433\times10^{-5}$ & $-1.350\times10^{-5}$ & $-8.489\times10^{-6}$ & $-3.858\times10^{-5}$ \\
\bottomrule
\end{tabular}

\end{table}
\begin{table}[H]
    \begin{revision}
        \caption{\textbf{Unbiased posterior skewness under noise-free conditions.}
Unbiased skewness values of the inferred posterior parameter distributions for the homogeneous and heterogeneous training specimens.
Boldfaced entries indicate parameters for which the absolute skewness exceeds $0.5$, consistent with the colored distributions in the parameter convergence plots.}
\label{tab:noise-free-skewness}
\begin{tabular}{lcccccc}
\toprule
\shortstack{\textbf{Unbiased} \\ \textbf{skewness}} & \textbf{circ-long-hom} & \textbf{circ-rad-hom} & \textbf{rad-heli-hom} & \textbf{circ-long-het} & \textbf{circ-rad-het} & \textbf{rad-heli-het} \\
\midrule
$s_{a}$ & $-0.136$ & $-0.136$ & $-0.136$ & $-0.136$ & $-0.136$ & $-0.136$ \\
$s_{b}$ & $0.088$ & $0.088$ & $0.088$ & $0.088$ & $0.088$ & $0.088$ \\
$s_{a_{f}}$ & $-0.133$ & $-0.133$ & $-0.133$ & $-0.133$ & $-0.133$ & $-0.133$ \\
$s_{b_{f}}$ & $0.003$ & $0.003$ & $0.003$ & $0.003$ & $0.003$ & $0.003$ \\
$s_{a_{n}}$ & $0.007$ & $0.007$ & $0.007$ & $0.007$ & $0.007$ & $0.007$ \\
$s_{b_{n}}$ & $-0.042$ & $-0.042$ & $-0.042$ & $-0.042$ & $-0.042$ & $-0.042$ \\
$s_{a_{fs}}$ & $\mathbf{0.722}$ & $-0.104$ & $-0.104$ & $-0.104$ & $-0.104$ & $-0.104$ \\
$s_{b_{fs}}$ & $\mathbf{0.847}$ & $0.492$ & $0.337$ & $\mathbf{0.838}$ & $-0.033$ & $-0.033$ \\
$s_{K}$ & $0.114$ & $0.114$ & $0.114$ & $0.114$ & $0.114$ & $0.114$ \\
\bottomrule
\end{tabular}

    \end{revision}
\end{table}

\section{Orthotropic constitutive parameter inference results under Gaussian white displacement noise}
\label{ap:noisy-parameter-results}
%
This appendix summarizes the orthotropic constitutive parameter inference results under Gaussian white displacement noise.
In Table~\ref{tab:noisy-hyperparameter-estimates}, we report the optimized hyperparameters at the final epoch.
The inferred material model parameters and their accompanying standard deviations are reported in Table~\ref{tab:noisy-parameter-estimates}, the relative errors with respect to the ground-truth parameters are reported in Table~\ref{tab:noisy-relative-error}, and the corresponding unbiased posterior skewness values are reported in Table~\ref{tab:noisy-skewness}.
\begin{table}[H]
    \caption{\textbf{Hyperparameter estimates under Gaussian white displacement noise.}
Optimized inverse-gamma hyperparameters at the final epoch for the heterogeneous training specimens under low- and high-noise conditions.
The parameters $\alpha^\star$ and $\beta^\star$ denote the shape and rate parameters of the corresponding inverse-gamma distributions.}
\label{tab:noisy-hyperparameter-estimates}
\begin{tabular}{lrrrrrr}
\toprule
\textbf{Hyperparameters} & \shortstack{\textbf{circ-long-het} \\ $\sigma_{u}=10^{-4}$} & \shortstack{\textbf{circ-rad-het} \\ $\sigma_{u}=10^{-4}$} & \shortstack{\textbf{rad-heli-het} \\ $\sigma_{u}=10^{-4}$} & \shortstack{\textbf{circ-long-het} \\ $\sigma_{u}=10^{-3}$} & \shortstack{\textbf{circ-rad-het} \\ $\sigma_{u}=10^{-3}$} & \shortstack{\textbf{rad-heli-het} \\ $\sigma_{u}=10^{-3}$} \\
\midrule
$\alpha_{v_s}$ & $1.837\times10^{1}$ & $1.847\times10^{1}$ & $1.822\times10^{1}$ & $1.923\times10^{1}$ & $1.932\times10^{1}$ & $1.916\times10^{1}$ \\
$\beta_{v_s}$ & $1.957\times10^{7}$ & $2.068\times10^{7}$ & $2.439\times10^{7}$ & $2.245\times10^{6}$ & $1.090\times10^{6}$ & $2.668\times10^{6}$ \\
$\alpha_{\sigma^2}$ & $2.683\times10^{6}$ & $2.890\times10^{6}$ & $2.857\times10^{6}$ & $9.050\times10^{5}$ & $1.080\times10^{6}$ & $9.907\times10^{5}$ \\
$\beta_{\sigma^2}$ & $3.691\times10^{1}$ & $3.411\times10^{1}$ & $3.453\times10^{1}$ & $1.100\times10^{2}$ & $9.200\times10^{1}$ & $1.004\times10^{2}$ \\
\bottomrule
\end{tabular}

\end{table}
\begin{table}[H]
    \caption{\textbf{Orthotropic constitutive parameter estimates under Gaussian white displacement noise.}
Inferred constitutive parameters at the final epoch for the heterogeneous training specimens under low- and high-noise conditions.
Parameters in $\vec{\theta}$ are reported using the notation of the Holzapfel--Ogden model, together with the corresponding posterior standard deviations.}
\label{tab:noisy-parameter-estimates}
\begin{tabular}{lllllll}
\toprule
\textbf{Parameters} & \shortstack{\textbf{circ-long-het} \\ $\sigma_{u}=10^{-4}$} & \shortstack{\textbf{circ-rad-het} \\ $\sigma_{u}=10^{-4}$} & \shortstack{\textbf{rad-heli-het} \\ $\sigma_{u}=10^{-4}$} & \shortstack{\textbf{circ-long-het} \\ $\sigma_{u}=10^{-3}$} & \shortstack{\textbf{circ-rad-het} \\ $\sigma_{u}=10^{-3}$} & \shortstack{\textbf{rad-heli-het} \\ $\sigma_{u}=10^{-3}$} \\
\midrule
\textit{Inferred $\mu_{\vec{\theta}}$:} &  &  &  &  &  &  \\
$a$ & $8.086\times10^{-4}$ & $8.051\times10^{-4}$ & $8.055\times10^{-4}$ & $8.102\times10^{-4}$ & $9.911\times10^{-4}$ & $9.240\times10^{-4}$ \\
$b$ & $7.475\times10^{0}$ & $7.535\times10^{0}$ & $7.491\times10^{0}$ & $7.825\times10^{0}$ & $8.775\times10^{0}$ & $7.922\times10^{0}$ \\
$a_{f}$ & $1.919\times10^{-3}$ & $1.989\times10^{-3}$ & $1.946\times10^{-3}$ & $2.004\times10^{-3}$ & $3.669\times10^{-3}$ & $2.042\times10^{-3}$ \\
$b_{f}$ & $2.202\times10^{1}$ & $2.135\times10^{1}$ & $2.183\times10^{1}$ & $2.129\times10^{1}$ & $9.787\times10^{0}$ & $2.087\times10^{1}$ \\
$a_{n}$ & $2.259\times10^{-4}$ & $2.367\times10^{-4}$ & $2.288\times10^{-4}$ & $2.326\times10^{-4}$ & $1.279\times10^{-8}$ & $2.498\times10^{-4}$ \\
$b_{n}$ & $3.484\times10^{1}$ & $3.136\times10^{1}$ & $3.479\times10^{1}$ & $3.378\times10^{1}$ & $7.070\times10^{-1}$ & $3.406\times10^{1}$ \\
$a_{fs}$ & $2.901\times10^{-8}$ & $2.590\times10^{-4}$ & $4.273\times10^{-5}$ & $2.185\times10^{-8}$ & $1.324\times10^{-8}$ & $1.582\times10^{-8}$ \\
$b_{fs}$ & $3.047\times10^{0}$ & $9.088\times10^{0}$ & $2.492\times10^{1}$ & $2.926\times10^{0}$ & $8.505\times10^{-1}$ & $3.055\times10^{0}$ \\
$K$ & $1.000\times10^{-1}$ & $9.952\times10^{-2}$ & $1.001\times10^{-1}$ & $9.924\times10^{-2}$ & $7.005\times10^{-2}$ & $9.157\times10^{-2}$ \\
\midrule \textit{Inferred $\sigma_{\vec{\theta}}$:} &  &  &  &  &  &  \\
$\sigma_{a}$ & $6.286\times10^{-8}$ & $5.740\times10^{-8}$ & $5.530\times10^{-8}$ & $5.143\times10^{-7}$ & $3.082\times10^{-7}$ & $4.174\times10^{-7}$ \\
$\sigma_{b}$ & $4.969\times10^{-4}$ & $3.975\times10^{-4}$ & $3.900\times10^{-4}$ & $4.260\times10^{-3}$ & $1.882\times10^{-3}$ & $2.899\times10^{-3}$ \\
$\sigma_{a_{f}}$ & $2.011\times10^{-8}$ & $1.834\times10^{-7}$ & $8.448\times10^{-8}$ & $1.811\times10^{-7}$ & $1.930\times10^{-6}$ & $7.679\times10^{-7}$ \\
$\sigma_{b_{f}}$ & $1.139\times10^{-4}$ & $2.209\times10^{-3}$ & $6.359\times10^{-4}$ & $1.074\times10^{-3}$ & $1.491\times10^{-2}$ & $5.549\times10^{-3}$ \\
$\sigma_{a_{n}}$ & $5.345\times10^{-9}$ & $3.721\times10^{-7}$ & $2.430\times10^{-8}$ & $4.829\times10^{-8}$ & $4.575\times10^{-8}$ & $2.148\times10^{-7}$ \\
$\sigma_{b_{n}}$ & $2.732\times10^{-4}$ & $1.098\times10^{-1}$ & $1.305\times10^{-3}$ & $2.254\times10^{-3}$ & $4.072\times10^{0}$ & $1.052\times10^{-2}$ \\
$\sigma_{a_{fs}}$ & $3.916\times10^{-6}$ & $7.839\times10^{-6}$ & $5.946\times10^{-6}$ & $3.022\times10^{-6}$ & $1.813\times10^{-6}$ & $2.142\times10^{-6}$ \\
$\sigma_{b_{fs}}$ & $1.295\times10^{1}$ & $1.624\times10^{-1}$ & $1.060\times10^{0}$ & $1.258\times10^{1}$ & $3.901\times10^{0}$ & $1.148\times10^{1}$ \\
$\sigma_{K}$ & $9.225\times10^{-7}$ & $2.781\times10^{-6}$ & $2.089\times10^{-6}$ & $8.517\times10^{-6}$ & $2.025\times10^{-5}$ & $1.810\times10^{-5}$ \\
\bottomrule
\end{tabular}

\end{table}
\begin{table}[H]
    \caption{\textbf{Relative errors of the orthotropic constitutive parameter estimates under Gaussian white displacement noise.}
Relative errors at the final epoch between the inferred constitutive parameters and the ground-truth parameter values for the heterogeneous training specimens under low- and high-noise conditions.}
\label{tab:noisy-relative-error}
\begin{tabular}{lrrrrrr}
\toprule
\textbf{Errors} & \shortstack{\textbf{circ-long-het} \\ $\sigma_{u}=10^{-4}$} & \shortstack{\textbf{circ-rad-het} \\ $\sigma_{u}=10^{-4}$} & \shortstack{\textbf{rad-heli-het} \\ $\sigma_{u}=10^{-4}$} & \shortstack{\textbf{circ-long-het} \\ $\sigma_{u}=10^{-3}$} & \shortstack{\textbf{circ-rad-het} \\ $\sigma_{u}=10^{-3}$} & \shortstack{\textbf{rad-heli-het} \\ $\sigma_{u}=10^{-3}$} \\
\midrule
$\hat{e}_{a}$ & $4.826\times10^{-4}$ & $4.764\times10^{-3}$ & $4.286\times10^{-3}$ & $-1.526\times10^{-3}$ & $-2.251\times10^{-1}$ & $-1.421\times10^{-1}$ \\
$\hat{e}_{b}$ & $-1.843\times10^{-4}$ & $-8.210\times10^{-3}$ & $-2.226\times10^{-3}$ & $-4.697\times10^{-2}$ & $-1.741\times10^{-1}$ & $-5.992\times10^{-2}$ \\
$\hat{e}_{a_{f}}$ & $-4.237\times10^{-3}$ & $-4.060\times10^{-2}$ & $-1.806\times10^{-2}$ & $-4.859\times10^{-2}$ & $-9.201\times10^{-1}$ & $-6.874\times10^{-2}$ \\
$\hat{e}_{b_{f}}$ & $1.938\times10^{-3}$ & $3.238\times10^{-2}$ & $1.070\times10^{-2}$ & $3.506\times10^{-2}$ & $5.564\times10^{-1}$ & $5.408\times10^{-2}$ \\
$\hat{e}_{a_{n}}$ & $4.971\times10^{-3}$ & $-4.290\times10^{-2}$ & $-7.797\times10^{-3}$ & $-2.456\times10^{-2}$ & $9.999\times10^{-1}$ & $-1.006\times10^{-1}$ \\
$\hat{e}_{b_{n}}$ & $-1.187\times10^{-3}$ & $9.900\times10^{-2}$ & $2.294\times10^{-4}$ & $2.947\times10^{-2}$ & $9.797\times10^{-1}$ & $2.124\times10^{-2}$ \\
$\hat{e}_{a_{fs}}$ & $9.999\times10^{-1}$ & $5.264\times10^{-1}$ & $9.219\times10^{-1}$ & $10.000\times10^{-1}$ & $10.000\times10^{-1}$ & $10.000\times10^{-1}$ \\
$\hat{e}_{b_{fs}}$ & $4.645\times10^{-1}$ & $-5.970\times10^{-1}$ & $-3.379\times10^{0}$ & $4.859\times10^{-1}$ & $8.506\times10^{-1}$ & $4.632\times10^{-1}$ \\
$\hat{e}_{K}$ & $-1.080\times10^{-4}$ & $4.751\times10^{-3}$ & $-1.271\times10^{-3}$ & $7.621\times10^{-3}$ & $2.995\times10^{-1}$ & $8.432\times10^{-2}$ \\
\bottomrule
\end{tabular}

\end{table}
\begin{table}[H]
    \begin{revision}
        \caption{\textbf{Unbiased posterior skewness under Gaussian white displacement noise.}
Unbiased skewness values of the inferred posterior parameter distributions for the heterogeneous training specimens under low- and high-noise conditions.
Boldfaced entries indicate parameters for which the absolute skewness exceeds $0.5$.}
\label{tab:noisy-skewness}
\begin{tabular}{lcccccc}
\toprule
\shortstack{\textbf{Unbiased} \\ \textbf{skewness}} & \shortstack{\textbf{circ-long-het} \\ $\sigma_{u}=10^{-4}$} & \shortstack{\textbf{circ-rad-het} \\ $\sigma_{u}=10^{-4}$} & \shortstack{\textbf{rad-heli-het} \\ $\sigma_{u}=10^{-4}$} & \shortstack{\textbf{circ-long-het} \\ $\sigma_{u}=10^{-3}$} & \shortstack{\textbf{circ-rad-het} \\ $\sigma_{u}=10^{-3}$} & \shortstack{\textbf{rad-heli-het} \\ $\sigma_{u}=10^{-3}$} \\
\midrule
$s_{a}$ & $-0.136$ & $-0.136$ & $-0.136$ & $-0.136$ & $-0.136$ & $-0.136$ \\
$s_{b}$ & $0.088$ & $0.088$ & $0.088$ & $0.088$ & $0.088$ & $0.088$ \\
$s_{a_{f}}$ & $-0.133$ & $-0.133$ & $-0.133$ & $-0.133$ & $-0.133$ & $-0.133$ \\
$s_{b_{f}}$ & $0.003$ & $0.003$ & $0.003$ & $0.003$ & $0.003$ & $0.003$ \\
$s_{a_{n}}$ & $0.007$ & $0.007$ & $0.007$ & $0.007$ & $\mathbf{0.933}$ & $0.007$ \\
$s_{b_{n}}$ & $-0.042$ & $-0.042$ & $-0.042$ & $-0.042$ & $\mathbf{0.878}$ & $-0.042$ \\
$s_{a_{fs}}$ & $\mathbf{0.799}$ & $-0.104$ & $-0.104$ & $\mathbf{0.799}$ & $\mathbf{0.799}$ & $\mathbf{0.799}$ \\
$s_{b_{fs}}$ & $\mathbf{0.848}$ & $-0.033$ & $-0.033$ & $\mathbf{0.849}$ & $\mathbf{0.855}$ & $\mathbf{0.836}$ \\
$s_{K}$ & $0.114$ & $0.114$ & $0.114$ & $0.114$ & $0.114$ & $0.114$ \\
\bottomrule
\end{tabular}

    \end{revision}
\end{table}

{\revision{
\section{Orthotropic constitutive parameter inference results under spatially correlated displacement noise}
\label{ap:spat-parameter-results}
%
This appendix summarizes the orthotropic constitutive parameter inference results under spatially correlated displacement noise.
In Table~\ref{tab:spat-hyperparameter-estimates}, we report the optimized hyperparameters at the final epoch.
The inferred material model parameters and their accompanying standard deviations are reported in Table~\ref{tab:spat-parameter-estimates}, the relative errors with respect to the ground-truth parameters are reported in Table~\ref{tab:spat-relative-error}, and the corresponding unbiased posterior skewness values are reported in Table~\ref{tab:spat-skewness}.
\begin{table}[H]
    \begin{revision}
        \caption{\textbf{Hyperparameter estimates under spatially correlated displacement noise.}
Optimized inverse-gamma hyperparameters at the final epoch for the heterogeneous training specimens under spatially correlated displacement noise.
The parameters $\alpha^\star$ and $\beta^\star$ denote the shape and rate parameters of the corresponding inverse-gamma distributions.}
\label{tab:spat-hyperparameter-estimates}
\begin{tabular}{lrrr}
\toprule
\textbf{Hyperparameters} & \shortstack{\textbf{circ-long-het} \\ $\sigma_{u}(\mathbf{X})$} & \shortstack{\textbf{circ-rad-het} \\ $\sigma_{u}(\mathbf{X})$} & \shortstack{\textbf{rad-heli-het} \\ $\sigma_{u}(\mathbf{X})$} \\
\midrule
$\alpha_{v_s}$ & $1.938\times10^{1}$ & $1.944\times10^{1}$ & $1.906\times10^{1}$ \\
$\beta_{v_s}$ & $1.515\times10^{6}$ & $7.690\times10^{5}$ & $3.695\times10^{6}$ \\
$\alpha_{\sigma^2}$ & $7.540\times10^{5}$ & $1.017\times10^{6}$ & $1.355\times10^{6}$ \\
$\beta_{\sigma^2}$ & $1.323\times10^{2}$ & $9.776\times10^{1}$ & $7.322\times10^{1}$ \\
\bottomrule
\end{tabular}

    \end{revision}
\end{table}
\begin{table}[H]
    \begin{revision}
        \caption{\textbf{Orthotropic constitutive parameter estimates under spatially correlated displacement noise.}
Inferred constitutive parameters at the final epoch for the heterogeneous training specimens under spatially correlated displacement noise.
Parameters in $\vec{\theta}$ are reported using the notation of the Holzapfel--Ogden model, together with the corresponding posterior standard deviations.}
\label{tab:spat-parameter-estimates}
\begin{tabular}{llll}
\toprule
\textbf{Parameters} & \shortstack{\textbf{circ-long-het} \\ $\sigma_{u}(\mathbf{X})$} & \shortstack{\textbf{circ-rad-het} \\ $\sigma_{u}(\mathbf{X})$} & \shortstack{\textbf{rad-heli-het} \\ $\sigma_{u}(\mathbf{X})$} \\
\midrule
\textit{Inferred $\mu_{\vec{\theta}}$:} &  &  &  \\
$a$ & $4.348\times10^{-4}$ & $8.135\times10^{-4}$ & $5.520\times10^{-4}$ \\
$b$ & $1.076\times10^{1}$ & $1.053\times10^{1}$ & $9.863\times10^{0}$ \\
$a_{f}$ & $2.580\times10^{-3}$ & $6.905\times10^{-3}$ & $1.262\times10^{-3}$ \\
$b_{f}$ & $1.920\times10^{1}$ & $1.265\times10^{-2}$ & $2.893\times10^{1}$ \\
$a_{n}$ & $2.541\times10^{-4}$ & $1.714\times10^{-8}$ & $3.668\times10^{-3}$ \\
$b_{n}$ & $3.257\times10^{1}$ & $5.721\times10^{-1}$ & $1.187\times10^{-4}$ \\
$a_{fs}$ & $5.688\times10^{-5}$ & $8.329\times10^{-8}$ & $1.889\times10^{-7}$ \\
$b_{fs}$ & $2.856\times10^{0}$ & $4.417\times10^{-1}$ & $1.867\times10^{0}$ \\
$K$ & $9.678\times10^{-2}$ & $5.278\times10^{-2}$ & $9.391\times10^{-2}$ \\
\midrule \textit{Inferred $\sigma_{\vec{\theta}}$:} &  &  &  \\
$\sigma_{a}$ & $4.725\times10^{-7}$ & $2.584\times10^{-7}$ & $1.626\times10^{-7}$ \\
$\sigma_{b}$ & $7.082\times10^{-3}$ & $1.884\times10^{-3}$ & $1.828\times10^{-3}$ \\
$\sigma_{a_{f}}$ & $3.194\times10^{-7}$ & $3.074\times10^{-6}$ & $2.374\times10^{-7}$ \\
$\sigma_{b_{f}}$ & $1.470\times10^{-3}$ & $1.399\times10^{-2}$ & $2.462\times10^{-3}$ \\
$\sigma_{a_{n}}$ & $7.436\times10^{-8}$ & $6.089\times10^{-8}$ & $1.222\times10^{-6}$ \\
$\sigma_{b_{n}}$ & $3.217\times10^{-3}$ & $3.227\times10^{0}$ & $1.810\times10^{-4}$ \\
$\sigma_{a_{fs}}$ & $2.880\times10^{-4}$ & $8.836\times10^{-6}$ & $1.078\times10^{-5}$ \\
$\sigma_{b_{fs}}$ & $1.218\times10^{1}$ & $2.678\times10^{0}$ & $7.454\times10^{0}$ \\
$\sigma_{K}$ & $1.194\times10^{-5}$ & $2.176\times10^{-5}$ & $9.835\times10^{-6}$ \\
\bottomrule
\end{tabular}

    \end{revision}
\end{table}
\begin{table}[H]
    \begin{revision}
        \caption{\textbf{Relative errors of the orthotropic constitutive parameter estimates under spatially correlated displacement noise.}
Relative errors at the final epoch between the inferred constitutive parameters and the ground-truth parameter values for the heterogeneous training specimens under spatially correlated displacement noise.}
\label{tab:spat-relative-error}
\begin{tabular}{lrrr}
\toprule
\textbf{Errors} & \shortstack{\textbf{circ-long-het} \\ $\sigma_{u}(\mathbf{X})$} & \shortstack{\textbf{circ-rad-het} \\ $\sigma_{u}(\mathbf{X})$} & \shortstack{\textbf{rad-heli-het} \\ $\sigma_{u}(\mathbf{X})$} \\
\midrule
$\hat{e}_{a}$ & $4.626\times10^{-1}$ & $-5.595\times10^{-3}$ & $3.177\times10^{-1}$ \\
$\hat{e}_{b}$ & $-4.398\times10^{-1}$ & $-4.090\times10^{-1}$ & $-3.196\times10^{-1}$ \\
$\hat{e}_{a_{f}}$ & $-3.501\times10^{-1}$ & $-2.613\times10^{0}$ & $3.399\times10^{-1}$ \\
$\hat{e}_{b_{f}}$ & $1.296\times10^{-1}$ & $9.994\times10^{-1}$ & $-3.111\times10^{-1}$ \\
$\hat{e}_{a_{n}}$ & $-1.194\times10^{-1}$ & $9.999\times10^{-1}$ & $-1.516\times10^{1}$ \\
$\hat{e}_{b_{n}}$ & $6.421\times10^{-2}$ & $9.836\times10^{-1}$ & $10.000\times10^{-1}$ \\
$\hat{e}_{a_{fs}}$ & $8.960\times10^{-1}$ & $9.998\times10^{-1}$ & $9.997\times10^{-1}$ \\
$\hat{e}_{b_{fs}}$ & $4.981\times10^{-1}$ & $9.224\times10^{-1}$ & $6.719\times10^{-1}$ \\
$\hat{e}_{K}$ & $3.217\times10^{-2}$ & $4.722\times10^{-1}$ & $6.093\times10^{-2}$ \\
\bottomrule
\end{tabular}

    \end{revision}
\end{table}
\begin{table}[H]
    \begin{revision}
        \caption{\textbf{Unbiased posterior skewness under spatially correlated displacement noise.}
Unbiased skewness values of the inferred posterior parameter distributions for the heterogeneous training specimens under spatially correlated displacement noise.
Boldfaced entries indicate parameters for which the absolute skewness exceeds $0.5$.}
\label{tab:spat-skewness}
\begin{tabular}{lcccc}
\toprule
\shortstack{\textbf{Unbiased} \\ \textbf{skewness}} & \shortstack{\textbf{circ-long-het} \\ $\sigma_{u}(\mathbf{X})$} & \shortstack{\textbf{circ-rad-het} \\ $\sigma_{u}(\mathbf{X})$} & \shortstack{\textbf{rad-heli-het} \\ $\sigma_{u}(\mathbf{X})$} \\
\midrule
$s_{a}$ & $-0.136$ & $-0.136$ & $-0.136$ \\
$s_{b}$ & $0.088$ & $0.088$ & $0.088$ \\
$s_{a_{f}}$ & $-0.133$ & $-0.133$ & $-0.133$ \\
$s_{b_{f}}$ & $0.003$ & $\mathbf{0.611}$ & $0.003$ \\
$s_{a_{n}}$ & $0.007$ & $\mathbf{0.932}$ & $0.007$ \\
$s_{b_{n}}$ & $-0.042$ & $\mathbf{0.877}$ & $\mathbf{0.688}$ \\
$s_{a_{fs}}$ & $\mathbf{0.735}$ & $\mathbf{0.798}$ & $\mathbf{0.796}$ \\
$s_{b_{fs}}$ & $\mathbf{0.848}$ & $\mathbf{0.875}$ & $\mathbf{0.842}$ \\
$s_{K}$ & $0.114$ & $0.114$ & $0.114$ \\
\bottomrule
\end{tabular}

    \end{revision}
\end{table}
}}

{\revision{
\section{Sensitivity to load-step selection}
\label{ap:loadstep-sensitivity}
%
This appendix summarizes a sensitivity analysis with respect to the selected biaxial load-step set.
Using the \textit{rad-heli-het} specimen, we compare mixed-modal validation performance across datasets constructed from different sets of prescribed boundary displacements{\secondrevision{ as well as mean relative parameter performance in the noiseless setting}}.
Each load-step set shown on the x-axis denotes the applied equibiaxial displacement levels, relative to the original tissue slab dimensions, included in the corresponding dataset.
\begin{figure}[!ht]
    \centering
    \includegraphics[width=0.99\linewidth]{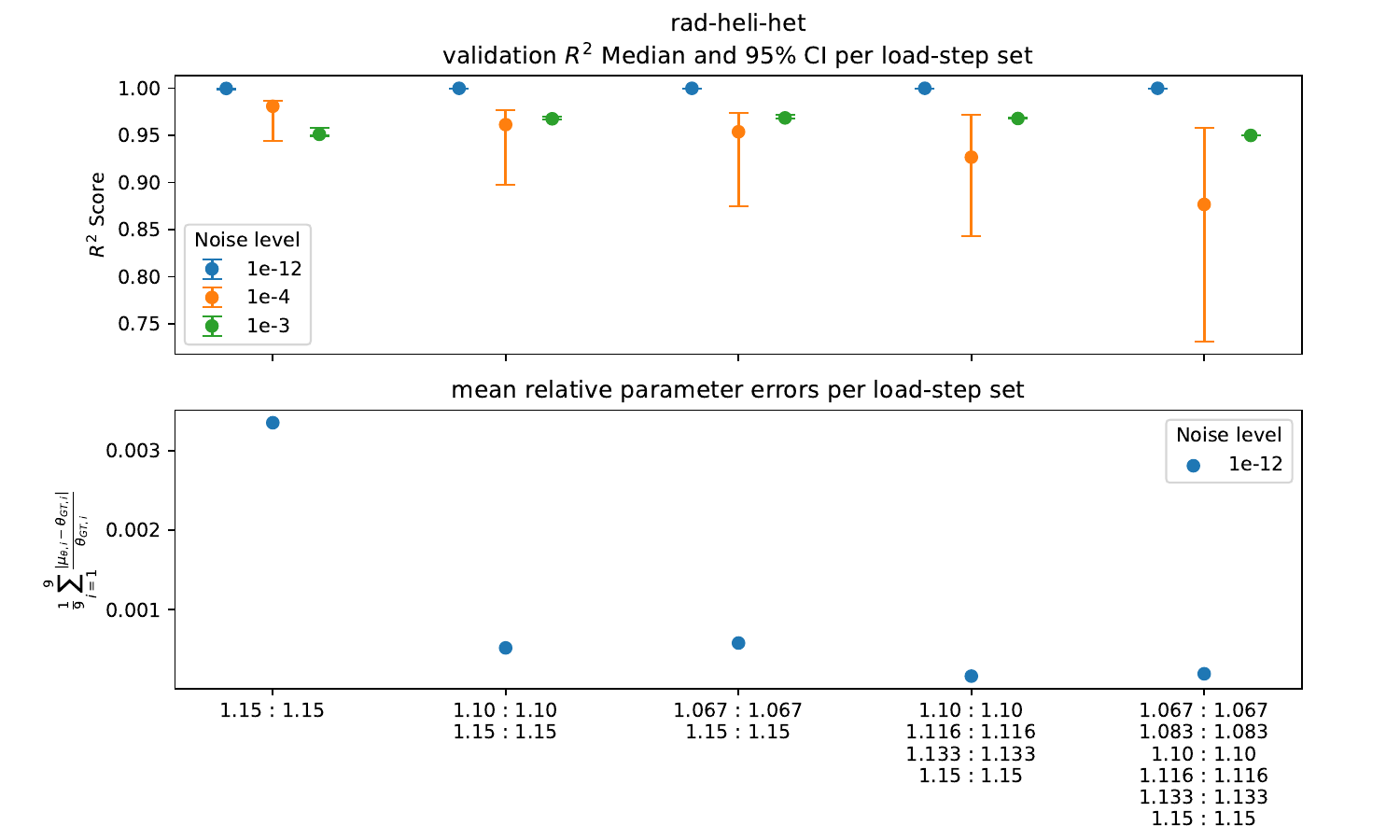}
    \caption{\textbf{Sensitivity of mixed-modal validation strain-energy $R^2$ scores and mean relative parameter error to load-step selection.}
    Validation strain-energy $R^2$ scores are reported for different biaxial load-step sets under noise-free, low-noise, and high-noise conditions, with median values and corresponding 95\% confidence intervals.
    For the noise-free case, the mean relative parameter error is additionally provided.
    Each load-step set represents a sequence of prescribed biaxial loading steps applied at fixed stretch ratios, as indicated on the x-axis ($\lambda_1:\lambda_2$).}
    \label{fig:loadstep-sensitivity}
\end{figure}

We consider load-step sets with $n_t=2$ up to $n_t=6$, corresponding to different sets of prescribed biaxial boundary displacement levels.
For example, we consider both $(1.10:1.10; 1.15:1.15)$ and $(1.067:1.067; 1.15:1.15)$ where, hypothetically, larger gaps in strain invariants could be observed.
In principle, varying the number and spacing of these load steps may enrich the range of induced strain invariants and thereby provide a more informative dataset for constitutive inference.

Overall, increasing the number of considered load steps does not provide a clear benefit.
Instead, using more load steps appears to increase sensitivity to low measurement noise.
Among the tested configurations, the load-step set $(1.10:1.10; 1.15:1.15)$ provides a reasonable balance between mixed-modal validation performance, the number of load steps included in the dataset{\secondrevision{, and the capability of estimating the ground-truth parameter set accurately}}.
Reducing the minimum prescribed biaxial stretch level tends to degrade the resulting $R^2$ scores, especially in the low-noise setting where numerically non-zero estimates of $a_{fs}$ and $b_{fs}$ are still obtained.
We interpret this trend as sensitivity to low-valued induced strain invariants entering the loss function.
We further observe that in the noiseless setting, including multiple load-steps in beneficial in estimating the ground-truth parameter set.
}}
\end{appendix}
\end{document}